\documentclass{aa}  
\usepackage{lipsum}
\usepackage{color}
\newcommand{\om}[1]{{\color{black} #1}}

\usepackage{float}
\usepackage{graphicx}
\usepackage{epstopdf}
\usepackage{txfonts}
\begin{document}

   \title{The Leo-I group: new dwarf galaxy and UDG candidates}

   \author{Oliver M\"uller
          \inst{1}
          \and
           Helmut Jerjen\inst{2} \and Bruno Binggeli\inst{1}
          }

\institute{Departement Physik, Universit\"at Basel, 
 Klingelbergstr. 82, CH-4056 Basel, Switzerland\\
              \email{oliver89.mueller@unibas.ch}
         \and
			Research School of Astronomy and Astrophysics, Australian National University, Canberra,
ACT 2611, Australia
             }

   \date{Received XX, 2018; accepted TBD}

% \abstract{}{}{}{}{} 
% 5 {} token are mandatory
 
  \abstract
     {The study of dwarf galaxies and their environments provides crucial testbeds for predictions of cosmological models and insights on the structure formation on small cosmological scales. In recent years, many problems on the scale of groups of galaxies challenged the current standard model of cosmology.}
   %{bla} %leave it empty if necessary  
   %{bla}
   {We aim to increase the sample of known galaxies in the Leo-I group, containing the M\,96 subgroup and the Leo Triplet. This galaxy aggregate is located at the edge of the Local Volume at a mean distance of 10.7\,Mpc.}
   {We employ image enhancing techniques to search for low-surface brightness objects in publicly available $gr$ images taken by the Sloan Digital Sky Survey within 500 square degrees around the Leo-I group. Once detected, we perform surface photometry and compare their structural parameters to other known dwarf galaxies in the nearby universe. }
   {We found 36 new dwarf galaxy candidates within the search area. Their morphology and structural parameters resemble known dwarfs in other groups. Among the candidates 5 to 6 galaxies are considered as ultra diffuse galaxies candidates. If confirmed, they would be some of the closest examples of this galaxy type. We assessed the luminosity function of the Leo-I group and find it to be considerably rich in dwarf galaxies, with twice the number of galaxies as the Local Group at a limiting magnitude of $M_V=-10$ \om{and a steeper faint-end slope}.}
{}

   \maketitle
%
%________________________________________________________________

\section{Introduction}
In a sphere of 11\,Mpc radius around the Milky Way reside more than one thousand galaxies, mostly of the type of dwarf galaxies (M$_B>-17.7$\,mag). 
This so-called Local Volume \citep{1979AN....300..181K,2004AJ....127.2031K,2013AJ....145..101K} contains many prominent galaxy aggregates, e.g. our own Local Group (LG), the Sculptor filament, the Centaurus group, the M\,81 group, the Canes-Venatici cloud, the M\,101 group complex, and the Leo I group \citep{1988ang..book.....T}. In recent years many teams have taken up the challenge to search for new dwarf galaxies in the local universe and measure their distances \citep{2009AJ....137.3009C,2013AJ....146..126C,2014ApJ...787L..37M,2014MNRAS.441.2124B,2014ApJ...795L..35C,2016ApJ...823...19C,2015ApJ...804L..44K,2015A&A...583A..79M,2017A&A...597A...7M,2017A&A...602A.119M,2016ApJ...828L...5C,2016A&A...588A..89J,2017ApJ...837..136D,2017MNRAS.465.5026C,2017A&A...603A..18H,2017ApJ...848...19P,2018MNRAS.474.3221M}. 
%, e.g. around M\,81 (Chiboucas et al. 2009, 2013), M\,83 (M\"uller et al. 2015, Mueller et al.), Cen\,A (Crnojevic et al 2014, 2016, M\"uller et al. 2017a), M\,101 (Merritt et al. 2014, Javanmardi et al. 2016, Danieli et al. 2017, M\"uller et al. 2017b), NGC\,253 \citep{2016ApJ...816L...5T}, and NGC\,2784 \citep{2017ApJ...848...19P}. 
These studies can be used to test the theoretical predictions from the standard model of cosmology ($\Lambda$CDM). For the LG, there is indeed a serious tension between observation and theory, i.e. the long-standing missing satellite problem \citep{1999ApJ...524L..19M}; the too-big-too-fail (TBTF) problem \citep{2010A&A...523A..32K,2011MNRAS.415L..40B}; and the plane-of-satellites problem \citep{2005A&A...431..517K,2012MNRAS.423.1109P,2013Natur.493...62I,2018arXiv180202579P}, see \citet{2017ARA&A..55..343B} for a recent review on small-scale challenges.
Such studies are now extended to other nearby galaxy groups, e.g. for Cen\,A \citep{2015ApJ...802L..25T,2016A&A...595A.119M,2018Sci...359..534M}, addressing the plane-of-satellite problem, or M\,101 \citep{2017ApJ...837..136D,2017A&A...602A.119M}, addressing the TBTF and missing satellite problems.

Using public data from the Sloan Digital Sky Survey (SDSS) we have started to systematically search for new, hitherto undetected dwarf galaxies in the Local Volume, beginning with the M\,101 group complex, covering 330\,deg$^2$ around the spiral galaxies M\,101, M\,51, and M\,63. We found 15 new dwarf galaxy candidates \citep{2017A&A...602A.119M}. We now continue our optical search for dwarf galaxies in an area that covers 500\,deg$^2$ around the Leo-I group (Fig.\,1).

The Leo-I group, with a mean distance of 10.7\,Mpc \citep{2004AJ....127.2031K,2013AJ....145..101K}, consists of seven bright galaxies, NGC\,3351 (= M\,95), NGC\,3368 (= M\,96), NGC\,3377, NGC\,3379 (= M\,105), NGC\,3384, NGC\,3412, and NGC\,3489 \citep{2004ARep...48..267K}. Another four bright galaxies, NGC\,3632 (= M\,65), NGC\,3627 (= M\,66), NGC\,3628, and NGC\,3593 -- the Leo Triplet, about six degrees to the east of the main aggregate -- are possibly also part of the group based on their common distances and systemic velocities \citep{2000ApJS..128..431F}. 
Note that about eight degrees to the north-east is another quartet of bright galaxies (NGC\,3599, NGC\,3605, NGC\,3607, and NGC\,3608), which shares the same systemic velocity but is farther behind and is arguably not associated to the group \citep{2000ApJS..128..431F}.

A spectacular feature of the Leo-I group in HI is the so-called Leo ring \citep{1985ApJ...288L..33S} around NGC\,3384/M\,105, one of the largest HI structures in the nearby universe. \citet{2010ApJ...717L.143M} followed this up with a deep optical survey using MegaCam on the CFHT and found no diffuse stellar optical component down to 28 mag\,arcsec$^{-2}$ surface brightness. The authors suggest an origin based on a collision between NGC\,3384 and M\,105 using gas/dark matter simulations and can explain the structure of the ring, together with the absence of apparent light. Deeper images ($\mu_V$ > 29.5 mag\,arcsec$^{-2}$) taken by \citet{2014ApJ...791...38W} still revealed no optical counterpart of the ring, however, they found some stream-like features associated to the ring, which are possibly of tidal origin.
In the Leo Triplet another intriguing feature, this time in the optical, is a stellar stream associated to \om{the} boxy spiral NGC\,3628 \citep{1956ErNW...29..344Z}, which hosts a tidal dwarf galaxy \citep{2014ApJ...786..144N} and an ultra compact dwarf galaxy \citep{2015ApJ...812L..10J}. 

For the central part of the Leo-I group (i.e. the M\,96 subgroup) an initial catalogue of 50 dwarf galaxy candidates was produced by \citet{1990AJ....100....1F}. The
authors argued based on morphological properties that half of them are group members. Another collection of dwarf galaxies were discovered by \citet{2002MNRAS.335..712T} who surveyed a $10\times 10$ \om{deg$^2$} field partially covering the Leo-I group. Using the digitized sky survey \citet{2004ARep...48..267K} refined and extended this list to 50 likely members. For many members HI velocities were derived \citep{2009AJ....138..338S}, making it possible to distinguish between actual Leo-I members and background galaxies belonging to the more distant Leo cloud (see Fig.\,1  in \citealt{2002MNRAS.335..712T} for the difference in velocity space). A  very deep but spatially limited image, based on amateur telescopes, was produced for NGC\,3628 in the Leo Triplet and revealed another faint dwarf galaxy \citep{2016A&A...588A..89J}.

To follow a consistent naming convention in this paper, from now on we use the term M\,96 subgroup to describe the main galaxy aggregate around M\,96, and the term Leo-Triplet (Leo-T) for the aggregate around M\,66. Both subgroups together are called the Leo-I group (see Fig.~\ref{fieldImage}).\\

In this work we present a search for unresolved dwarf galaxies using publicly available data from the Sloan Digital Sky Survey (SDSS) in 500 deg$^2$, covering the extended Leo-I group region. In Section\,\ref{search} we summarize our search strategy and in Section\,\ref{phot} we present the surface photometry performed for all known and newly found members of the Leo-I group. In Section\,\ref{disc} we discuss our candidate list and potential background contamination.
Finally, in Section\,\ref{concl} we draw our conclusions and give a brief outlook.

\section{Discovery of new dwarf \om{galaxy candidates}}
\label{search}
\om{In recent years, different automatic detection approaches have been proposed to search for low surface brightness galaxies \citep[e.g.][]{2014ApJ...787L..37M,2014ApJ...788..188S,2016A&A...590A..20V,2017ApJ...850..109B} with encouraging results. On the other hand, these pipelines were only applied  on small areas of the sky (< 10 deg$^2$) and still have a considerable rate of false detections, or rely on a large number of existing galaxies to study galaxy groups on a statistical basis. It remains to be seen how these methods perform on large-field surveys with areas of several hundreds of square degrees and how time-consuming the task of rejecting false-positives will be. We, as well as other authors \citep[e.g.][]{2017ApJ...848...19P,2017MNRAS.470.1512W}, argue that a visual search on images is still on par with algorithm-based detections.\\}

\om{In this work, we} follow the same methods as described in \citet{2017A&A...602A.119M} to search for dwarf galaxies in an area of $\sim$500\,deg$^2$ around the Leo-I group using data taken from the SDSS. In summary this involves the creation of 1 square degree mosaics of $g$ and $r$ images, the use of several image processing algorithms (e.g. binning and Gaussian convolution) to enhance the low-surface brightness features within the images, and the final visual search for dwarf galaxies in these processed images. Once an object is detected, surface photometry is applied to derive the structural parameters, which are compared to the properties of known dwarf galaxies of the LG and other groups. Based on this morphological comparison, a detection is considered or rejected as dwarf galaxy candidate. To estimate our detection rate we conducted an experiment where we induced artificial galaxies into the SDSS images and derived the recovery rate of these objects (Fig.\,3 in \citealt{2017A&A...602A.119M}). \\

\begin{figure*}[ht]
\centering
\hspace*{-0.7cm}
\includegraphics[width=20cm]{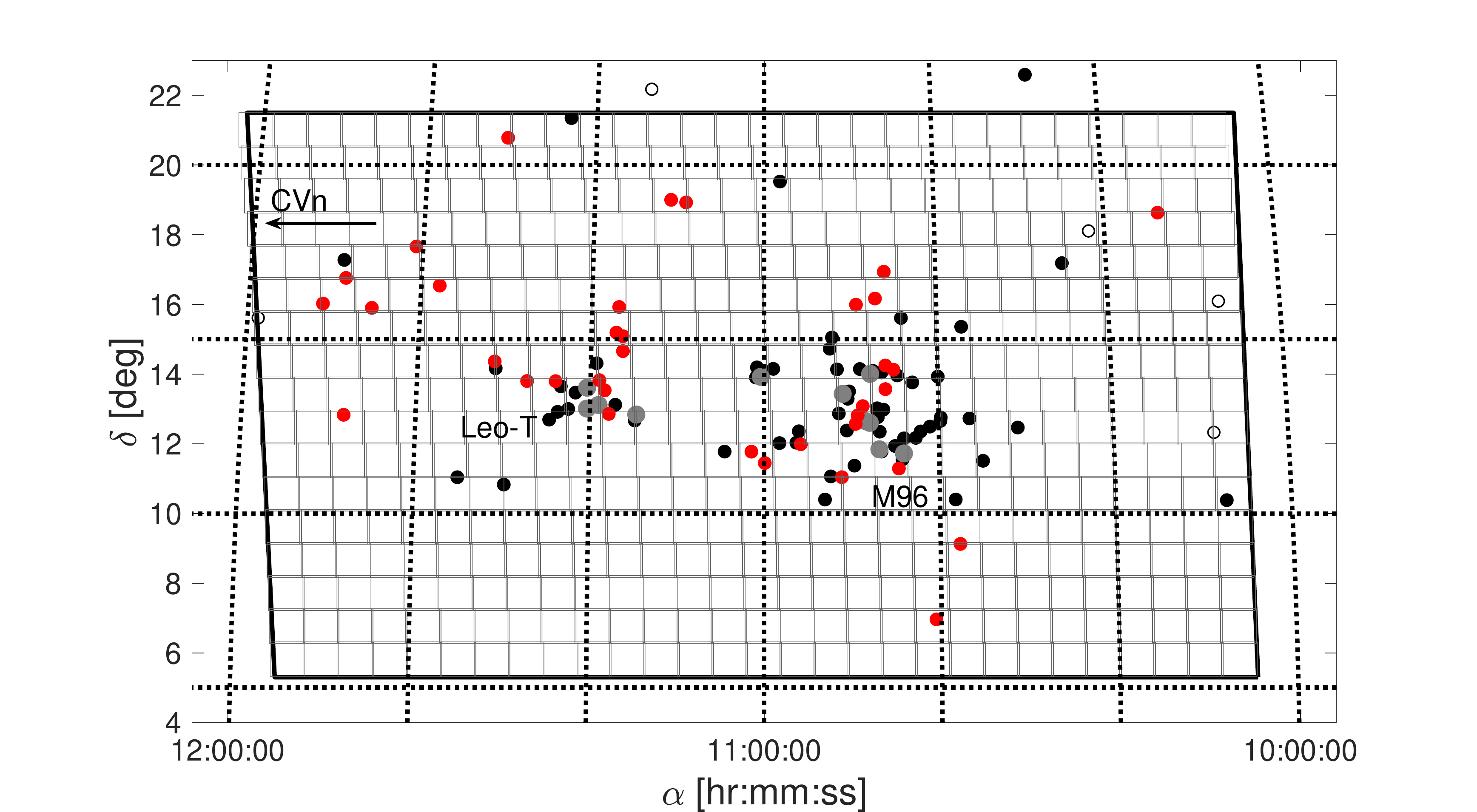}
 \caption{ Survey area of $\approx 500$ \om{deg$^2$} in the Leo-I group region. The squares correspond to the created 1 \om{deg$^2$} mosaics.
The small black dots are previously known members based on their photometric properties, compiled from the Local Volume Catalog \citep{2004AJ....127.2031K,2013AJ....145..101K}. The large gray dots are the major galaxies in the M\,96 subgroup and Leo Triplet. The red dots indicate the positions of the 36 new dwarf candidates. Open circles are confirmed foreground (<7\,Mpc) galaxies taken from the LV Catalog.
 }
 \label{fieldImage}
\end{figure*}

In Fig.\,\ref{fieldImage} we present the survey footprint, the known galaxies in this field (black and gray dots), and the new dwarf galaxy candidates (red dots) found in our search. In the up-to-date online version\footnote{last checked: 19. December 2017.} of the LV catalog 63 dwarf galaxies are listed within our footprint, with four (open circle) having a distance estimate smaller than 7\,Mpc.  In Table\,\ref{table:1} we present the coordinates of the 36 dwarf galaxy candidates found in the survey, together with our galaxy type classification and comments on the objects. We indicate if the objects are found in the vicinity of M\,96, the Leo Triplet, or in the surrounding field.

\begin{table}[ht]
\centering
\setlength{\tabcolsep}{3pt}
\caption{Names, coordinates, and morphological types of the 36 new dwarf galaxy candidates of the Leo-I group.}
\label{table:1}
\begin{tabular}{lccll}
\hline\hline
& $\alpha$ & $\delta$ & \\ 
Name & (J2000) & (J2000) & Type & Notes\\ 
\hline \\[-2mm]
dw1013+18 & 10:13:29 & $+$18:36:44 & dSph & field \\ %380
dw1037+09 & 10:37:40 & $+$09:06:20  & dIrr & M\,96  \\ % 96
dw1040+06 & 10:40:30 & $+$06:56:27  & dSph   & field \\ % 39  unreliable phot, near LEDA31770, UDG? (180Mpc)
dw1044+11 & 10:44:33 & $+$11:16:10 & dSph & M\,96   \\ % 185 nearby Cluster 
dw1045+14a & 10:45:01 & $+$14:06:20 & dSph &   M\,96    \\ %272
dw1045+14b & 10:45:56 & $+$14:13:37 & dSph & M\,96   \\ % 272
dw1045+16 & 10:45:56& $+$16:55:00 & dSph, bg? & M\,96  \\ %359
dw1045+13 & 10:45:58 & $+$13:32:52 & dSph & M\,96  \\ % 243  near NGC 3367 (35Mpc) 
dw1047+16 & 10:47:00 & $+$16:08:50 & dSph,N & M\,96  \\ %330
dw1048+13 & 10:48:36 & $+$13:03:34 & dSph  & M\,96  \\ % 244  near LEDA 32315 (146Mpc)
dw1049+12a& 10:49:11 & $+$12:47:34 & dSph & M\,96  \\ % 215 
dw1049+15 & 10:49:14 &$+$15:58:20&  dSph/dIrr & M\,96 \\ %331
dw1049+12b& 10:49:26 & $+$12:33:08 &  dSph/dIrr? & M\,96   \\ % 215
dw1051+11 & 10:51:03 & $+$11:01:13  & dSph, UDG? & M\,96   \\ %  157 
dw1055+11 & 10:55:43 & $+$11:58:05 &  dSph,N, UDG?  & M\,96 \\ % 188 LEDA 1400089 (216Mpc)
dw1059+11 & 10:59:51 & $+$11:25:38 & dSph & M\,96 \\ %189
dw1101+11 & 11:01:22 & $+$11:45:12 & dSph & M\,96  \\ % former bg189
dw1109+18 & 11:09:08 & $+$18:54:20 & dIrr & field \\ % 422
dw1110+18 & 11:10:55 & $+$18:58:52 & dSph & field  \\ % 423
%dw1113+14 & 11:13:29 & $+$14:20:32 & BCD?  bg?& Leo-T  \\ % 279 near NGC3596 (15Mpc)
dw1116+14 & 11:16:14& $+$14:38:17 & dSph, bg? & Leo-T\\ % 280EXTREMLY UNRELIABLE PHOT
dw1116+15a & 11:16:17 & $+$15:04:02 & dSph, bg? &Leo-T \\ % 309 ex faint, bg NGC 3596? swithed ab
dw1116+15b & 11:16:46 & $+$15:54:19 & dSph, bg? & Leo-T\\ % 338EXTREMLY UNRELIABLE PHOT
dw1117+15 &11:17:02 & $+$15:10:17 & dSph, UDG?, bg? &Leo-T\\ %309 
dw1117+12 & 11:17:44 & $+$12:50:10 & dSph & Leo-T\\ % 222  EXTREMLY UNRELIABLE PHOT
dw1118+13a& 11:18:15 & $+$13:30:53 & dSph & Leo-T \\ % 251/252switched ab
dw1118+13b & 11:18:53 & $+$13:48:18 & dSph & Leo-T \\ % 251 cirrus? 
dw1123+13 & 11:23:56 & $+$13:46:41 & dSph &  Leo-T\\ % 253 ext. of stream
dw1127+13 & 11:27:13 & $+$13:46:50 & dSph &Leo-T \\ %  254
dw1130+20 & 11:30:32 & $+$20:45:41 & dIrr & field \\ %485 
dw1131+14 & 11:31:01 & $+$15:54:52 & dSph  & field \\ %341
dw1137+16 & 11:37:46& $+$16:31:09 &  dSph, UDG?& field \\%343 bg of 	UGC 6594?
dw1140+17 & 11:40:43 & $+$17:38:36 & dSph & field\\ %373  bg? Extr z
dw1145+14 &11:45:32 & $+$15:52:50 & dSph&field \\ %345  bg?
dw1148+12 & 11:48:09 & $+$12:48:43 & dSph  &field\\ %  230 UDG? IC0737 (55Mpc)
dw1148+16 & 11:48:45& $+$16:44:24 & dSph &field \\ %346
dw1151+16 & 11:51:15& $+$16:00:20 & dSph & field\\ %346 
\hline\hline
\end{tabular}
\end{table}

\begin{figure*}
 \includegraphics[width=3.6cm]{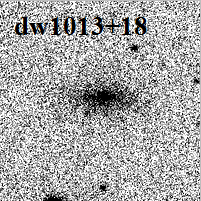}
 \includegraphics[width=3.6cm]{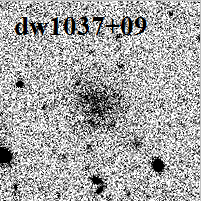}
 \includegraphics[width=3.6cm]{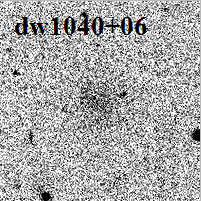}
 \includegraphics[width=3.6cm]{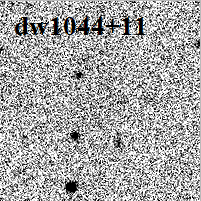}
 \includegraphics[width=3.6cm]{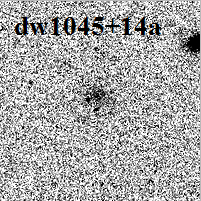}\\
 \includegraphics[width=3.6cm]{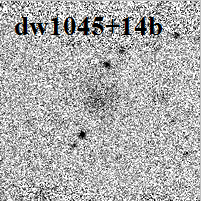}
 \includegraphics[width=3.6cm]{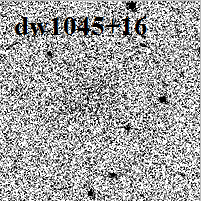}
 \includegraphics[width=3.6cm]{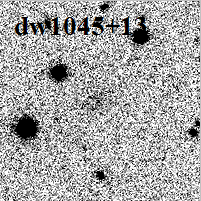}
 \includegraphics[width=3.6cm]{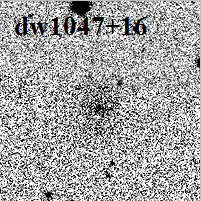}
 \includegraphics[width=3.6cm]{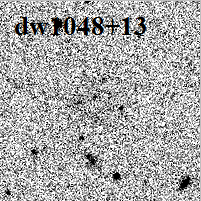}\\
 \includegraphics[width=3.6cm]{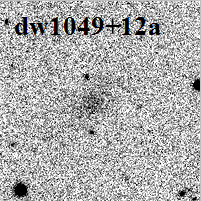}
 \includegraphics[width=3.6cm]{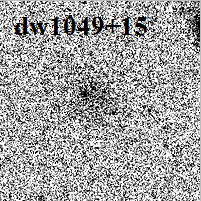}
 \includegraphics[width=3.6cm]{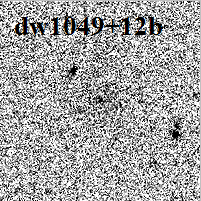}
 \includegraphics[width=3.6cm]{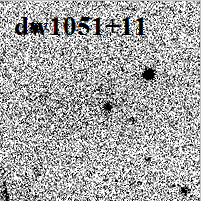}
 \includegraphics[width=3.6cm]{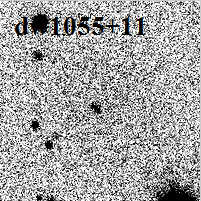}\\
 \includegraphics[width=3.6cm]{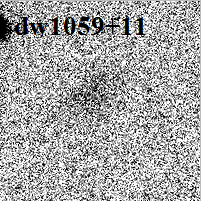}
 \includegraphics[width=3.6cm]{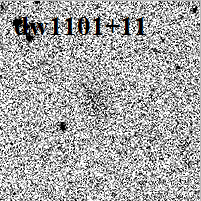}
 \includegraphics[width=3.6cm]{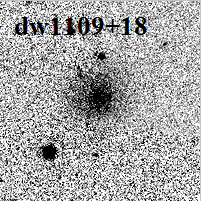}
 \includegraphics[width=3.6cm]{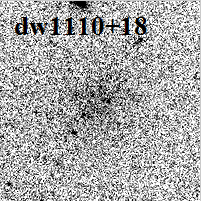}
  \includegraphics[width=3.6cm]{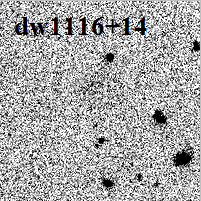}
\caption{Gallery showing SDSS $r$-band images of the new Leo-I group member candidates. 
One side of an image corresponds to 80\,arcsec or 3.88\,kpc at the distance of 10\,Mpc. \om{North is to the top, east to the right.} }
\label{sample0}
\end{figure*}

\begin{figure*} % grayscale: 0.05, -0.002
 \includegraphics[width=3.6cm]{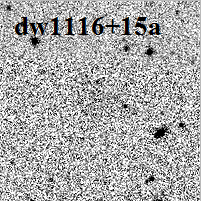}
 \includegraphics[width=3.6cm]{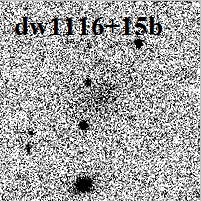}
 \includegraphics[width=3.6cm]{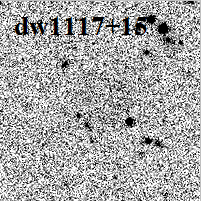}
 \includegraphics[width=3.6cm]{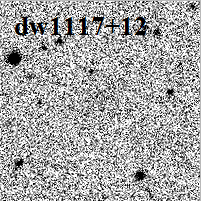}
 \includegraphics[width=3.6cm]{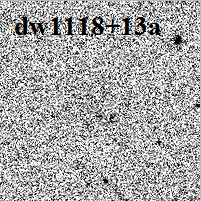}\\
 \includegraphics[width=3.6cm]{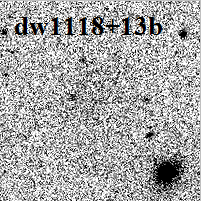}
 \includegraphics[width=3.6cm]{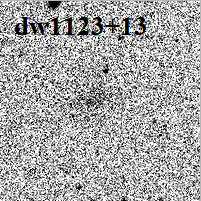}
 \includegraphics[width=3.6cm]{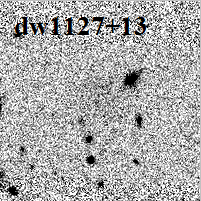}
 \includegraphics[width=3.6cm]{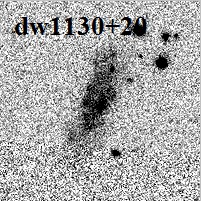}
 \includegraphics[width=3.6cm]{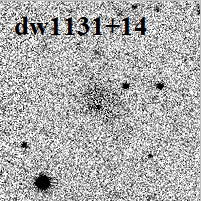}\\
 \includegraphics[width=3.6cm]{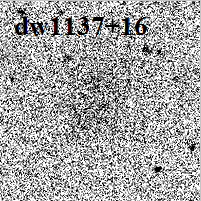}
 \includegraphics[width=3.6cm]{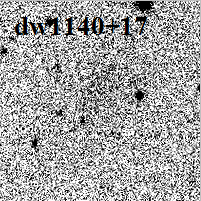}
 \includegraphics[width=3.6cm]{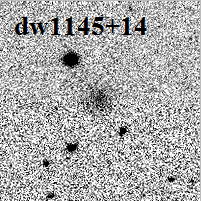}
 \includegraphics[width=3.6cm]{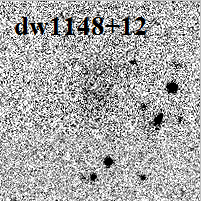}
 \includegraphics[width=3.6cm]{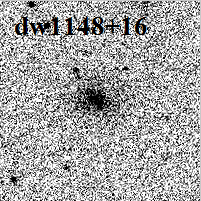}\\
 \includegraphics[width=3.6cm]{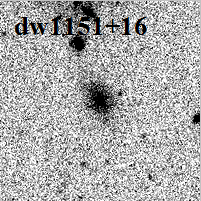}
\caption{Fig.\,\ref{sample0} continued.}
\label{sample1}
\end{figure*}

\section{Surface photometry}
\label{phot}
\begin{figure*}
 \includegraphics[width=3.6cm]{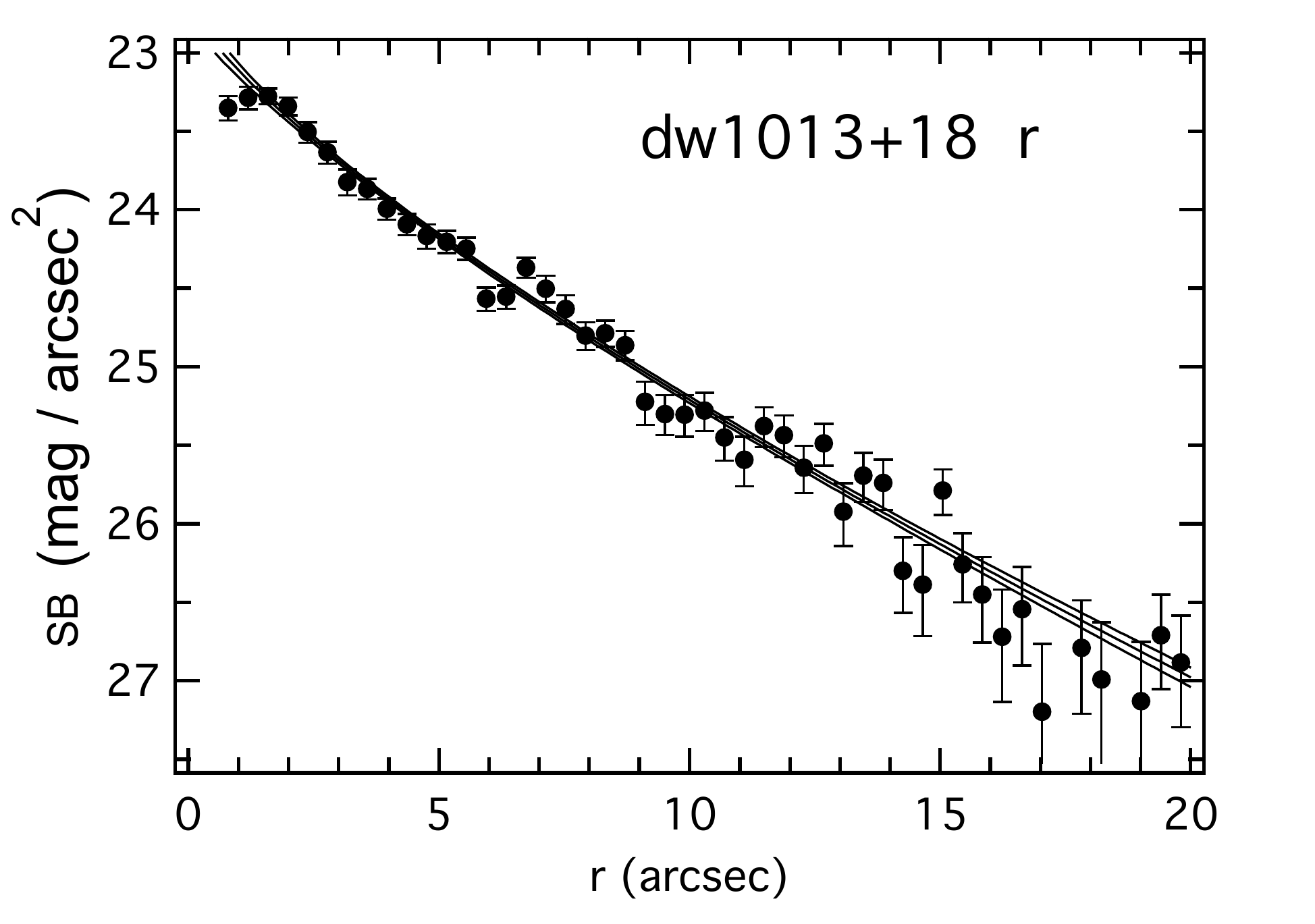}
 \includegraphics[width=3.6cm]{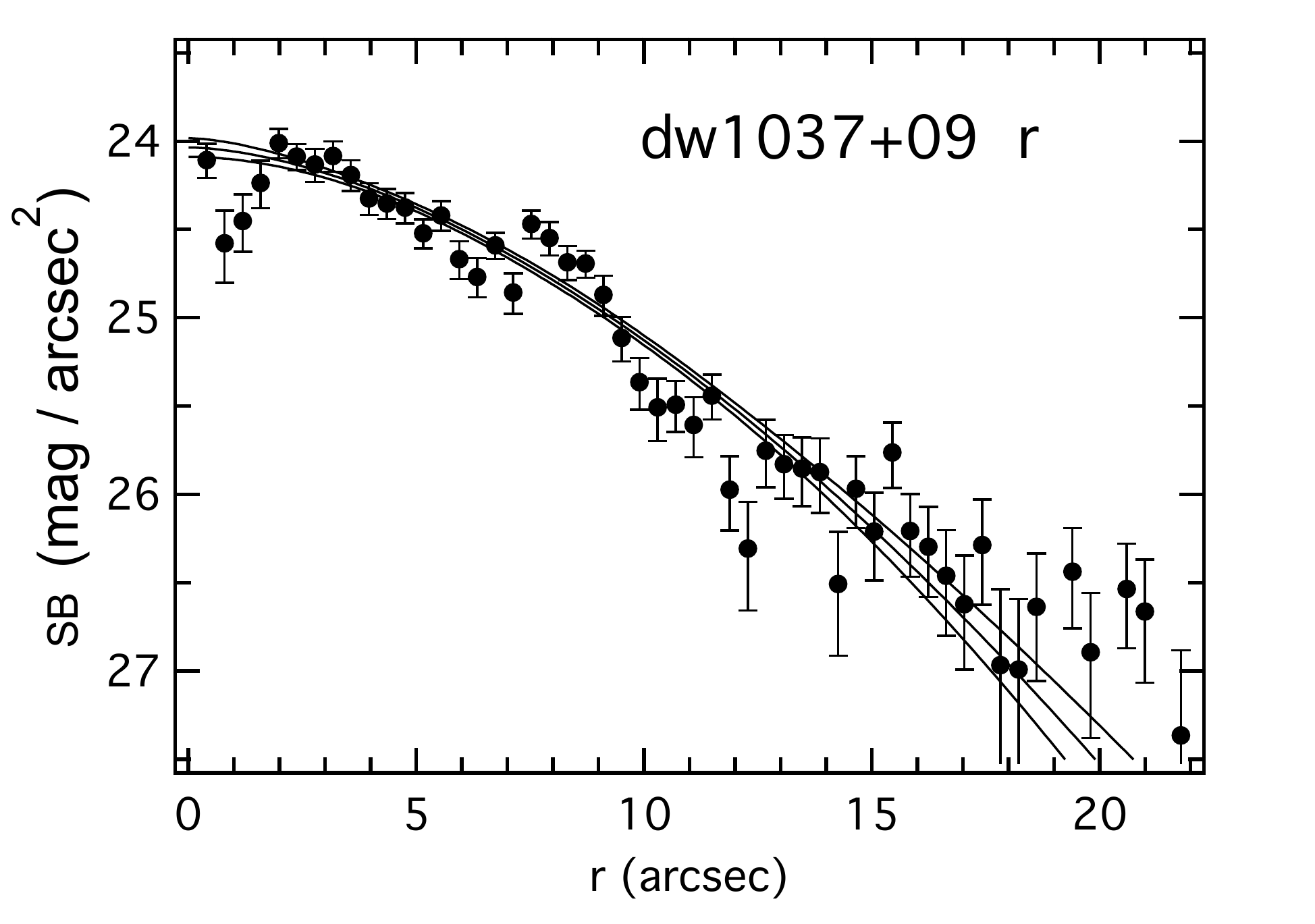}
 \includegraphics[width=3.6cm]{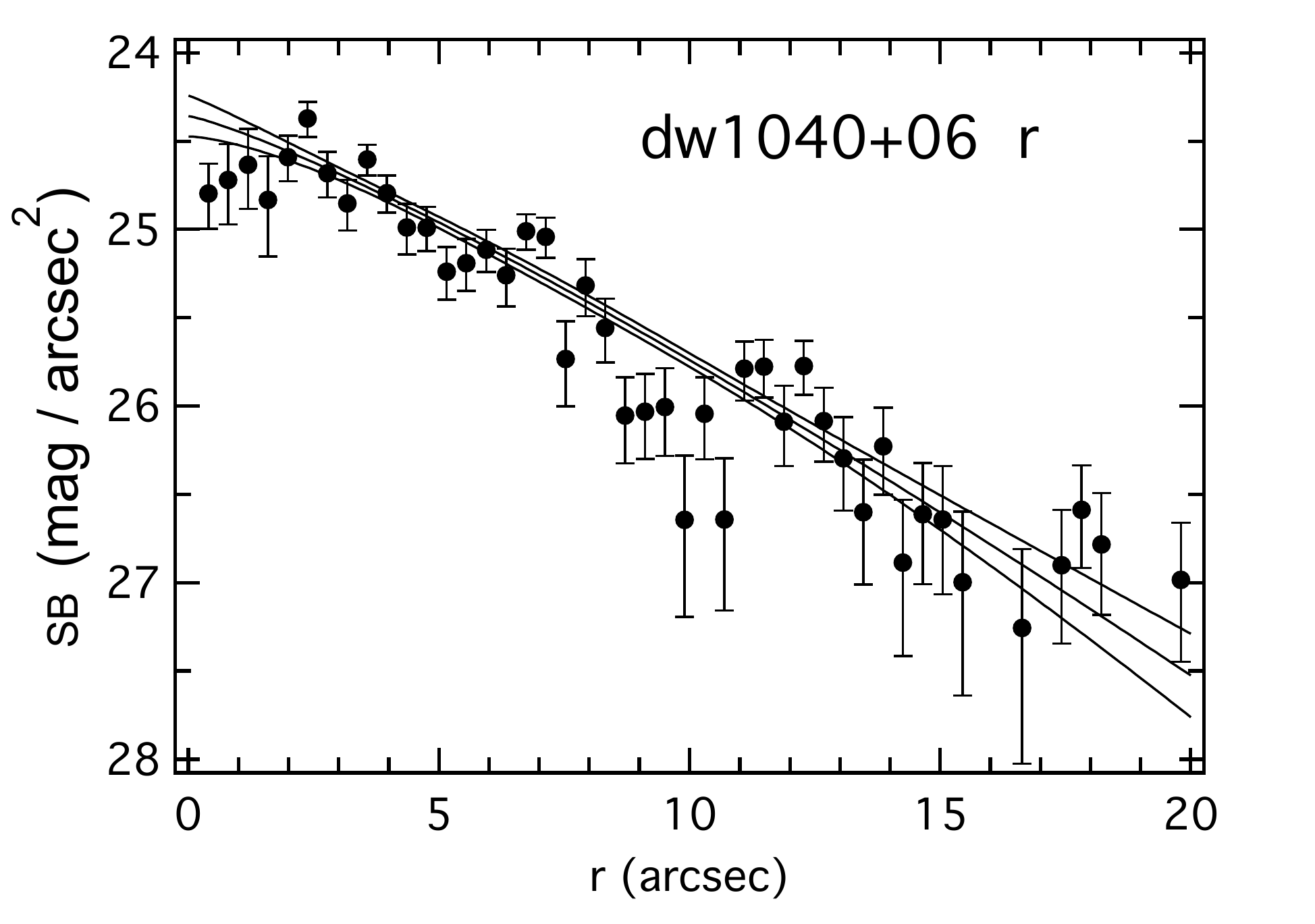}
 \includegraphics[width=3.6cm]{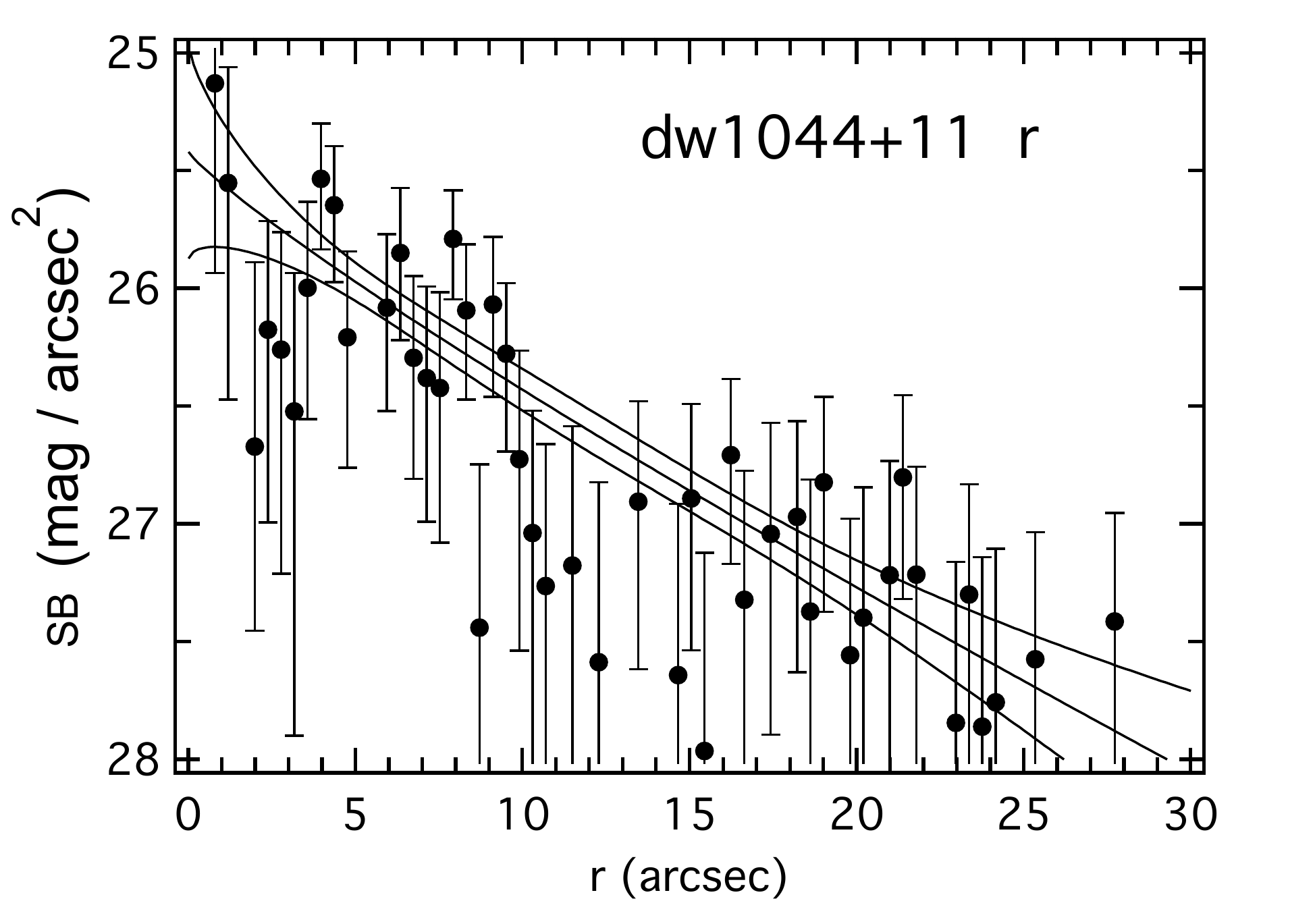}
 \includegraphics[width=3.6cm]{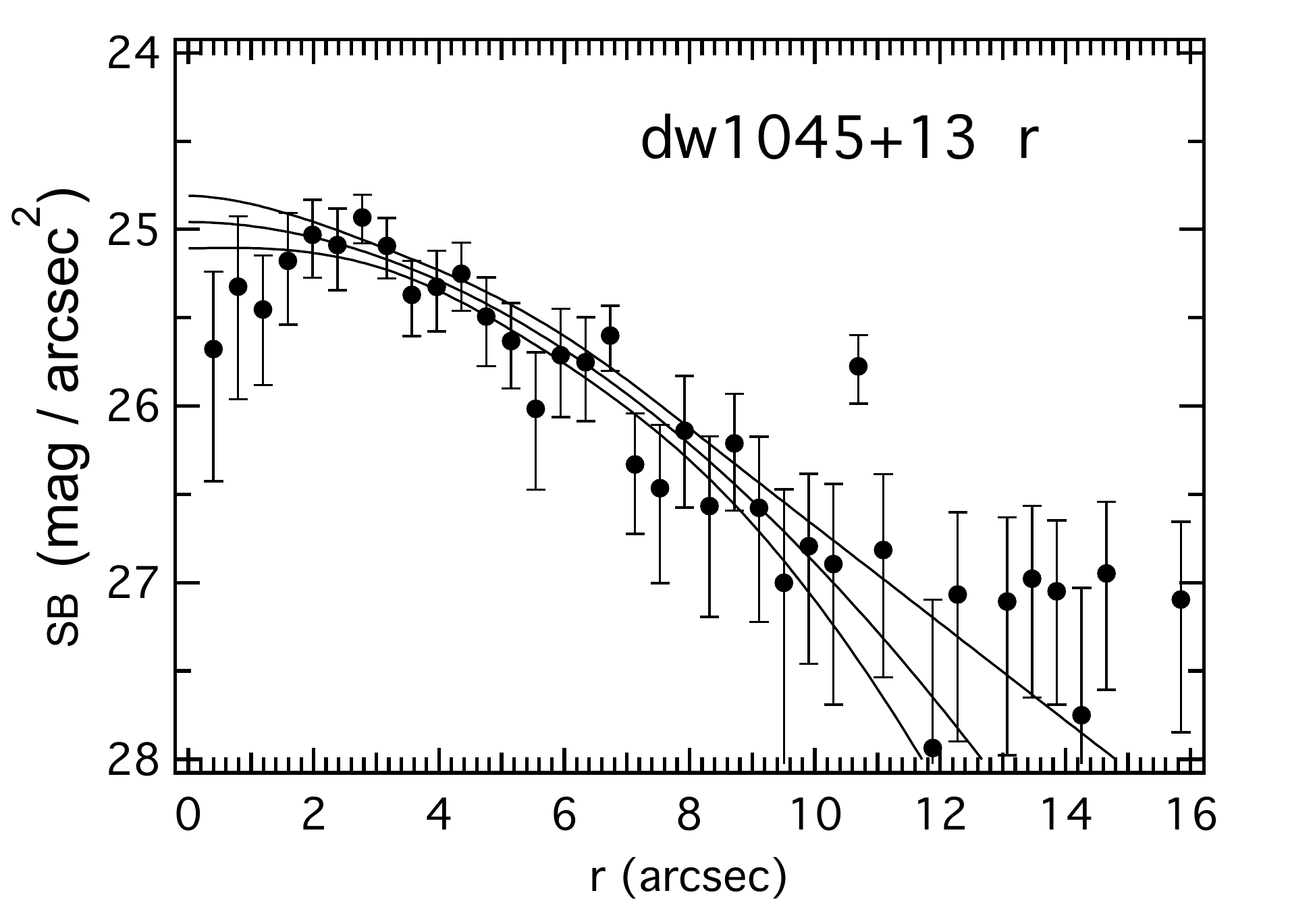}\\
 \includegraphics[width=3.6cm]{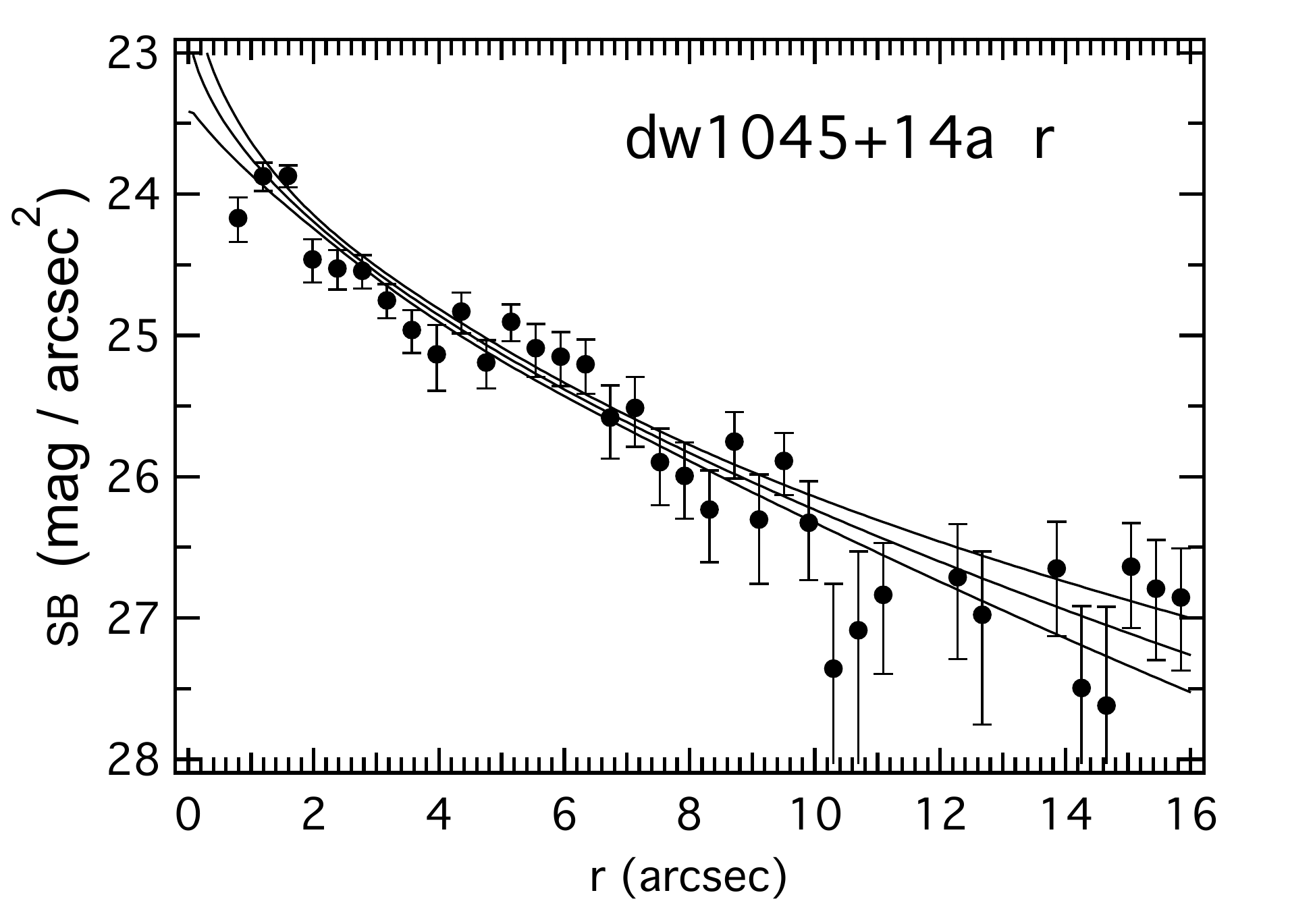}
 \includegraphics[width=3.6cm]{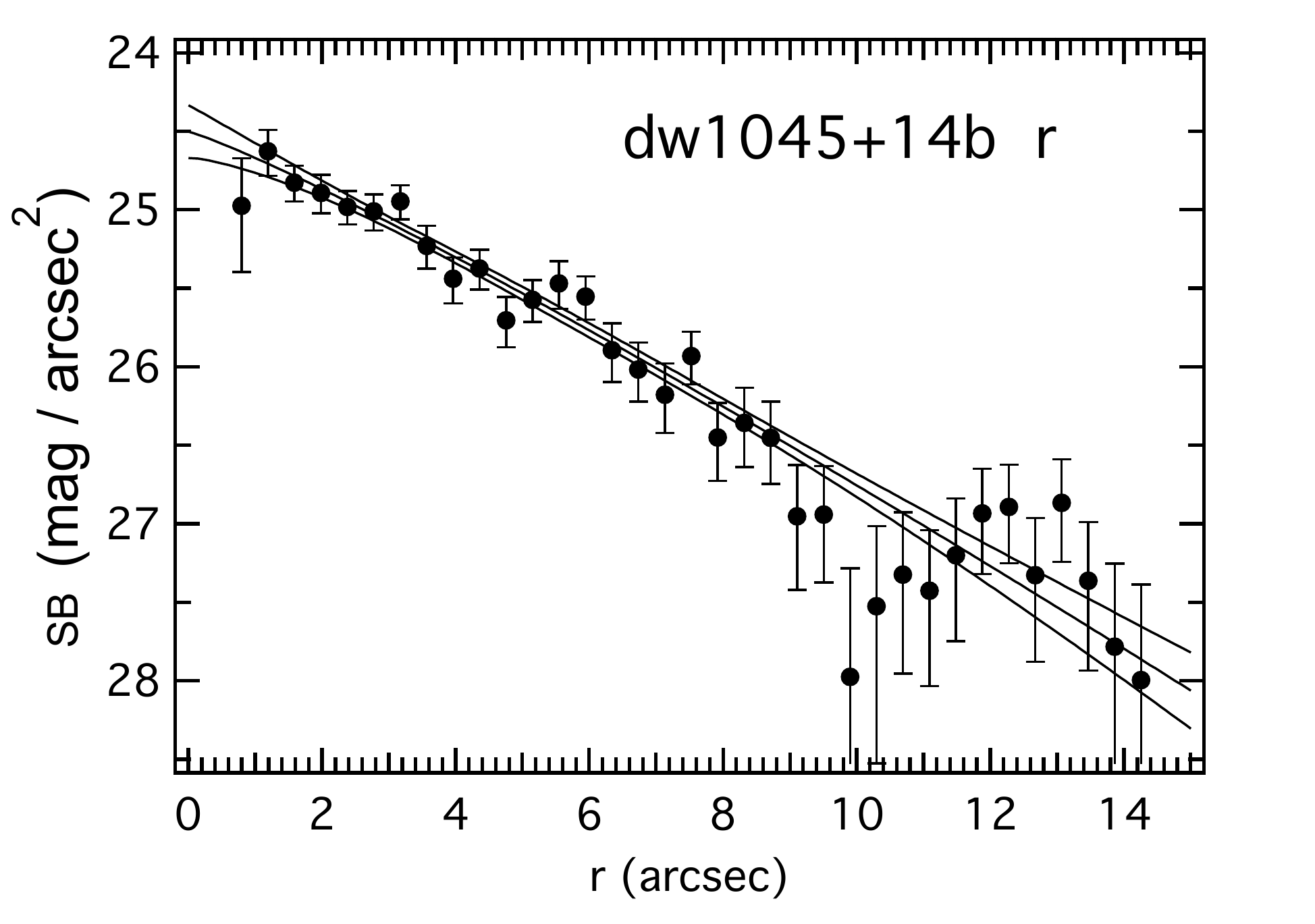}
 \includegraphics[width=3.6cm]{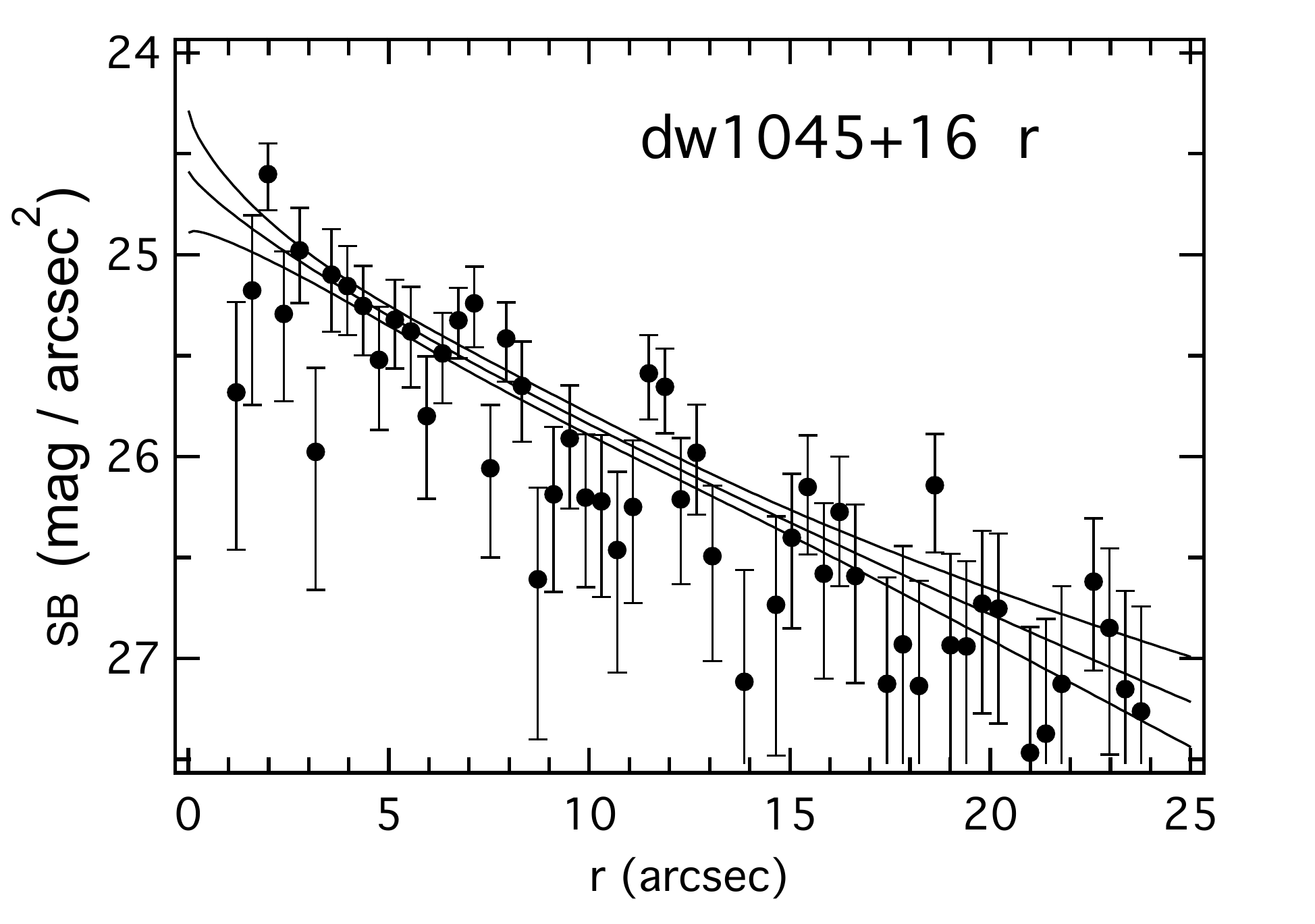}
 \includegraphics[width=3.6cm]{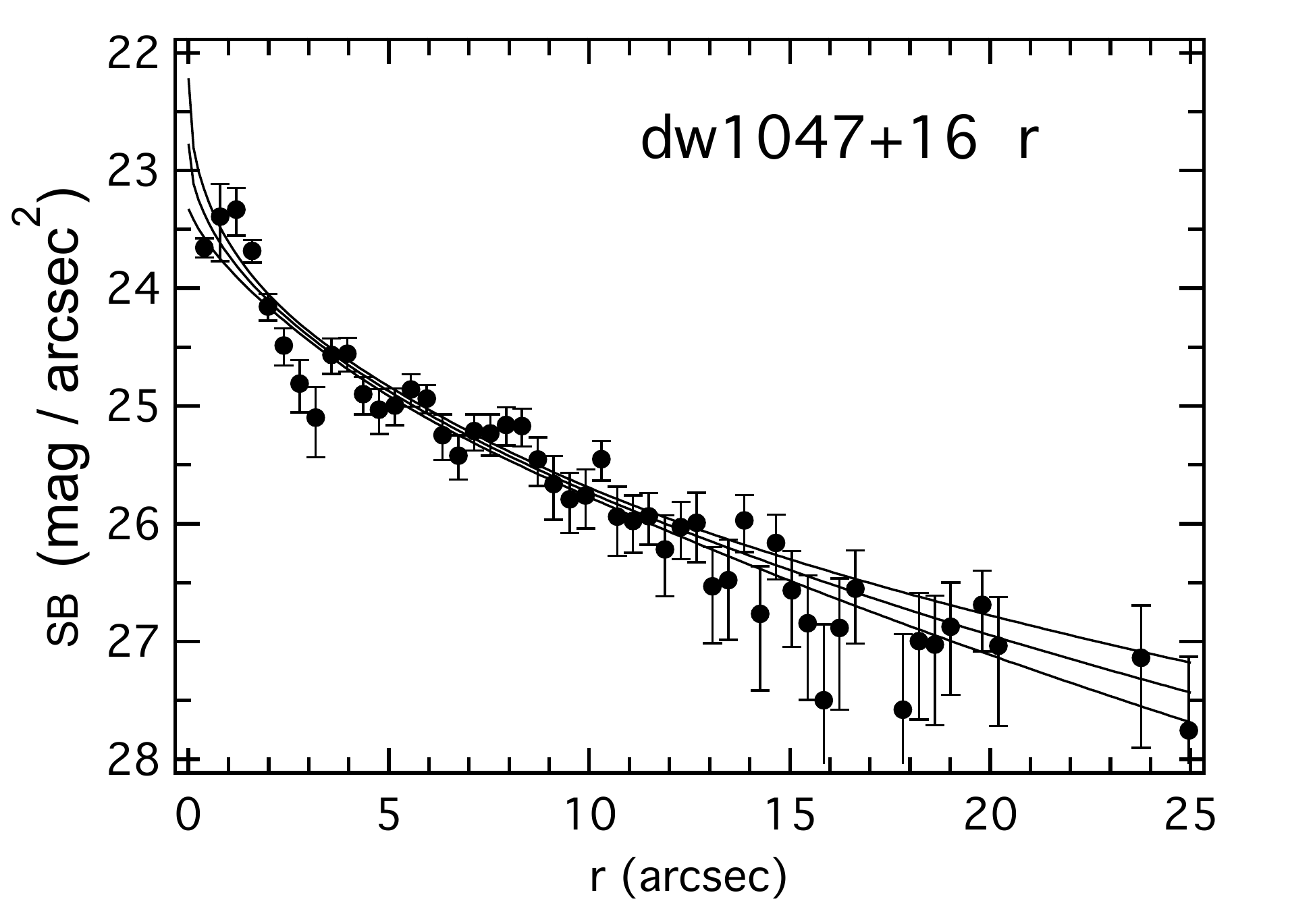}
 \includegraphics[width=3.6cm]{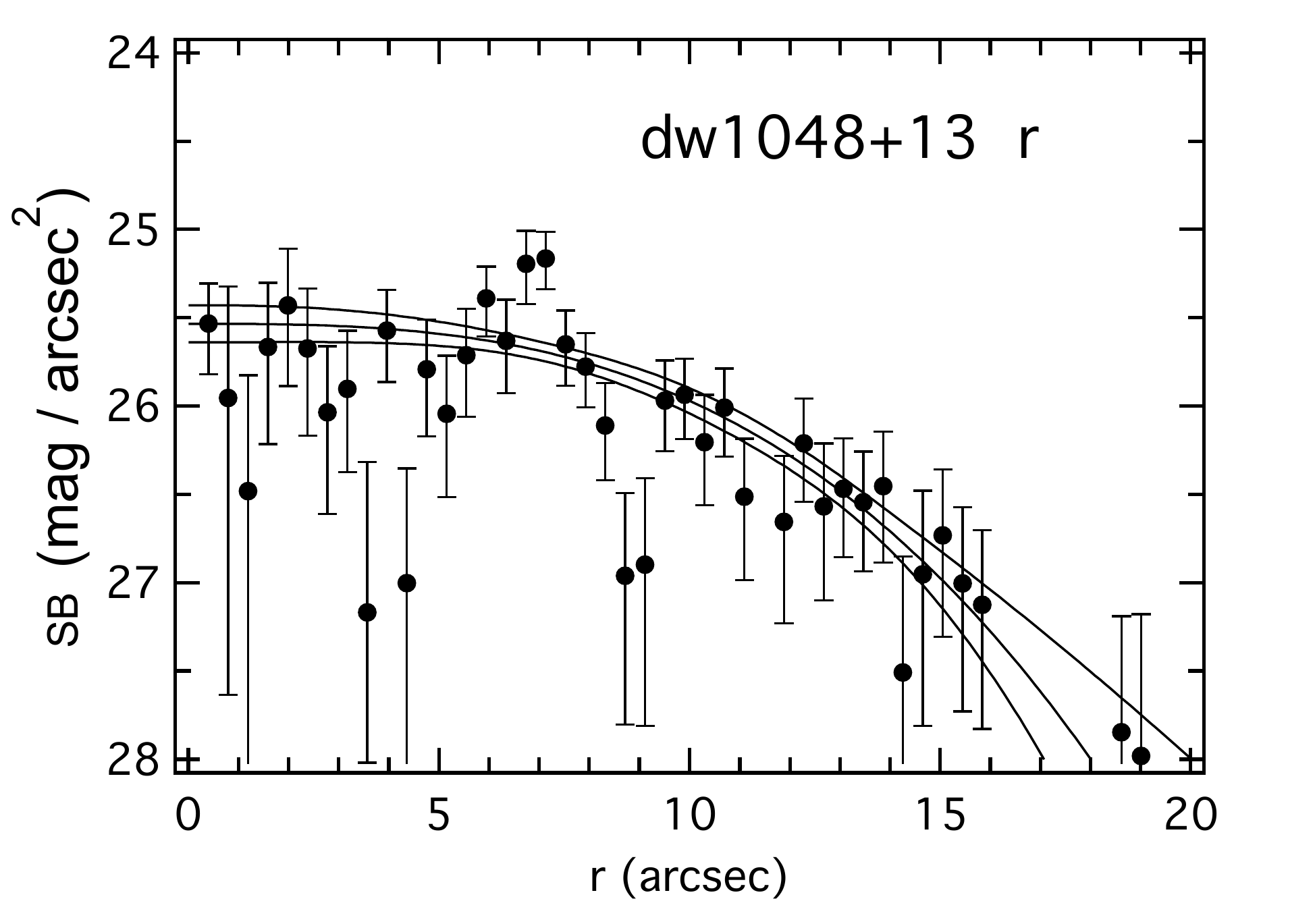}\\
 \includegraphics[width=3.6cm]{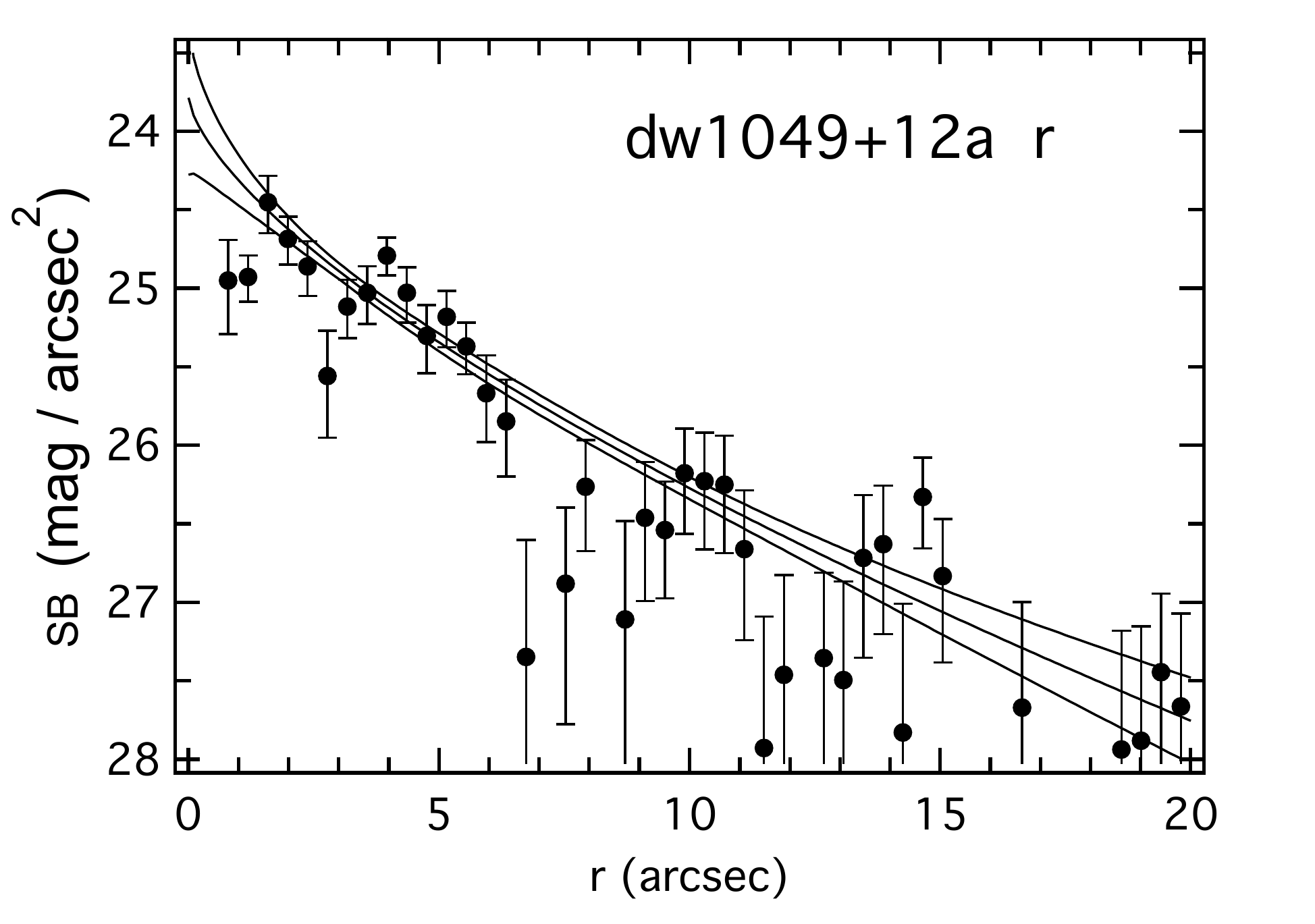}
 \includegraphics[width=3.6cm]{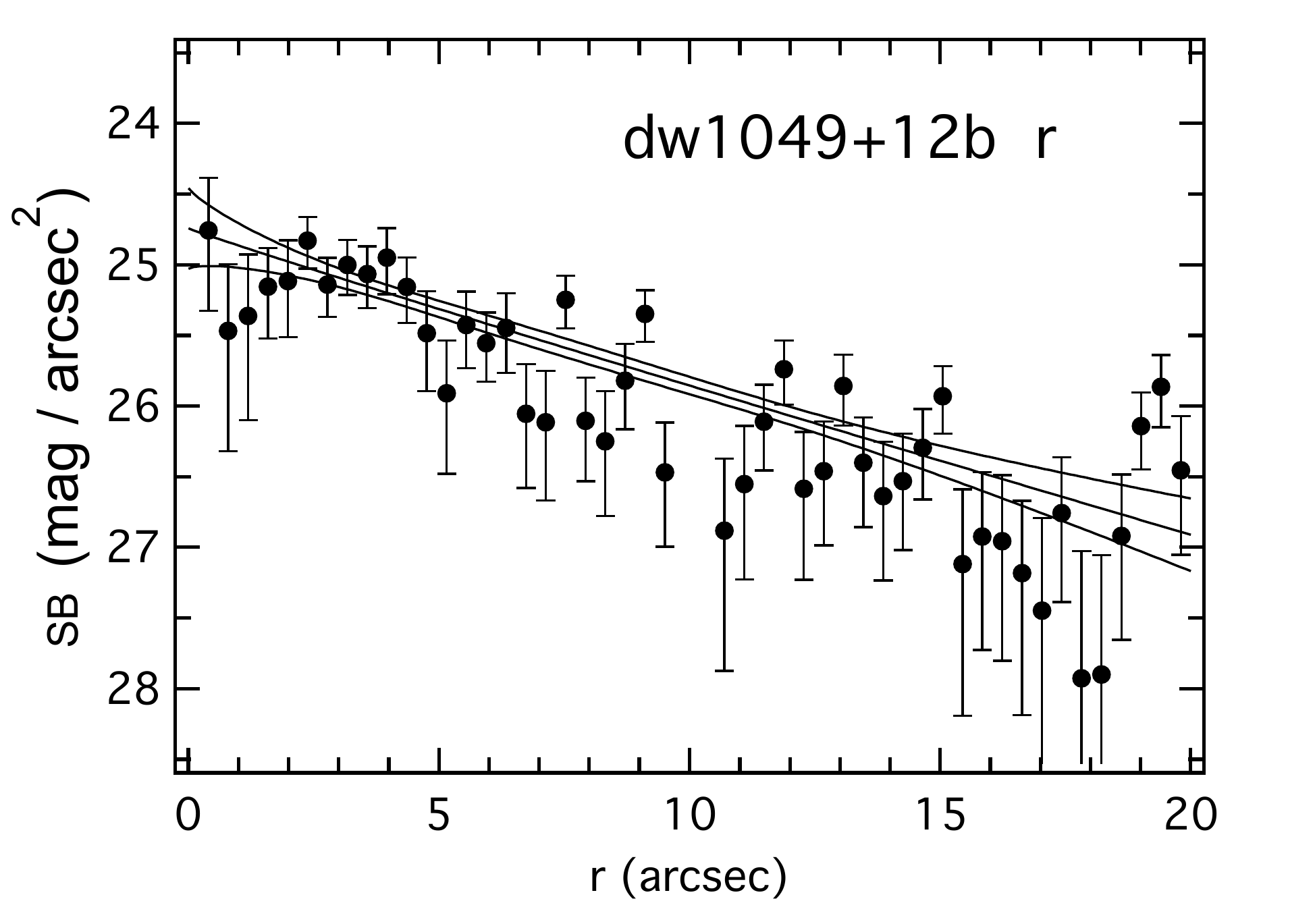}
 \includegraphics[width=3.6cm]{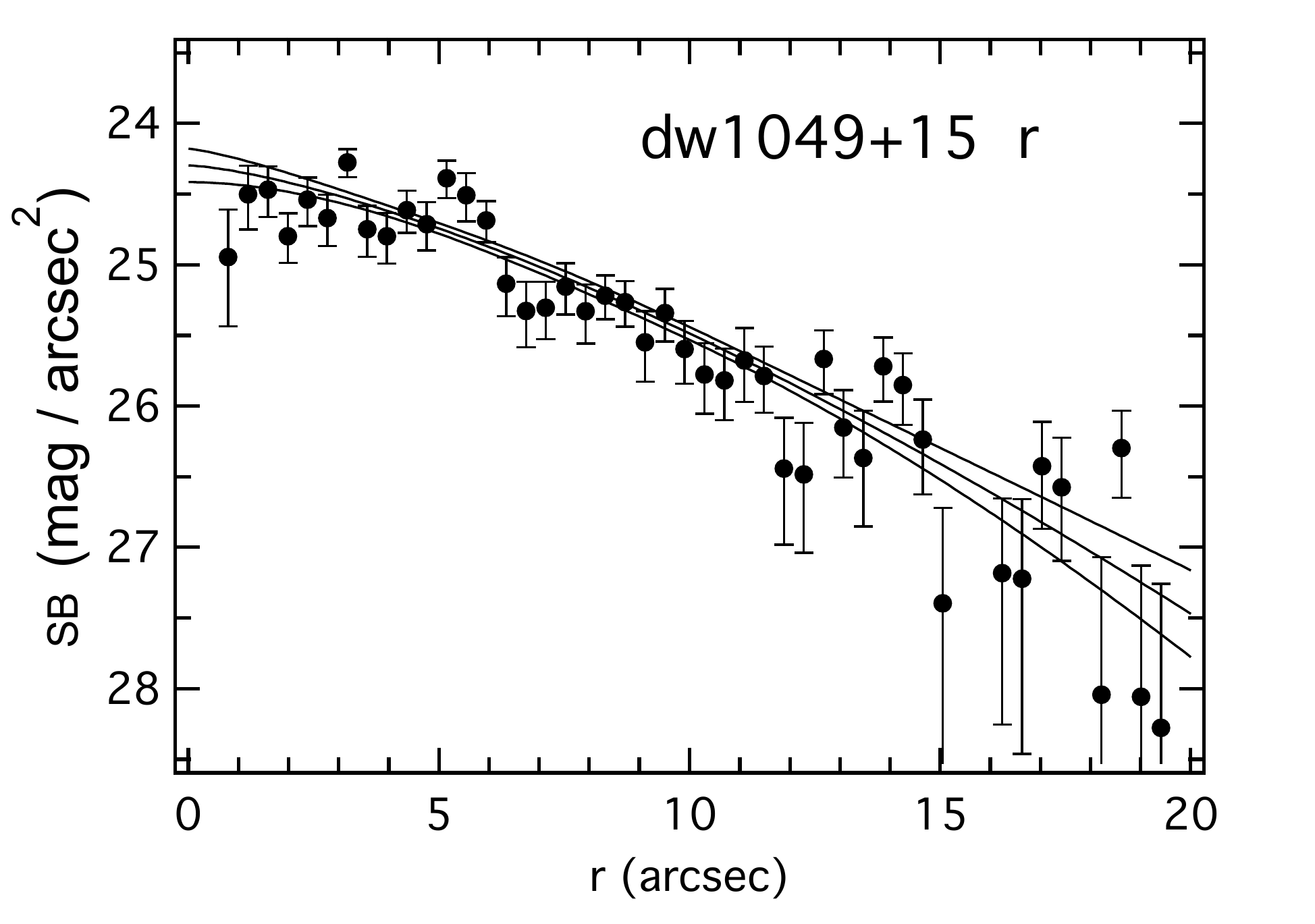}
 \includegraphics[width=3.6cm]{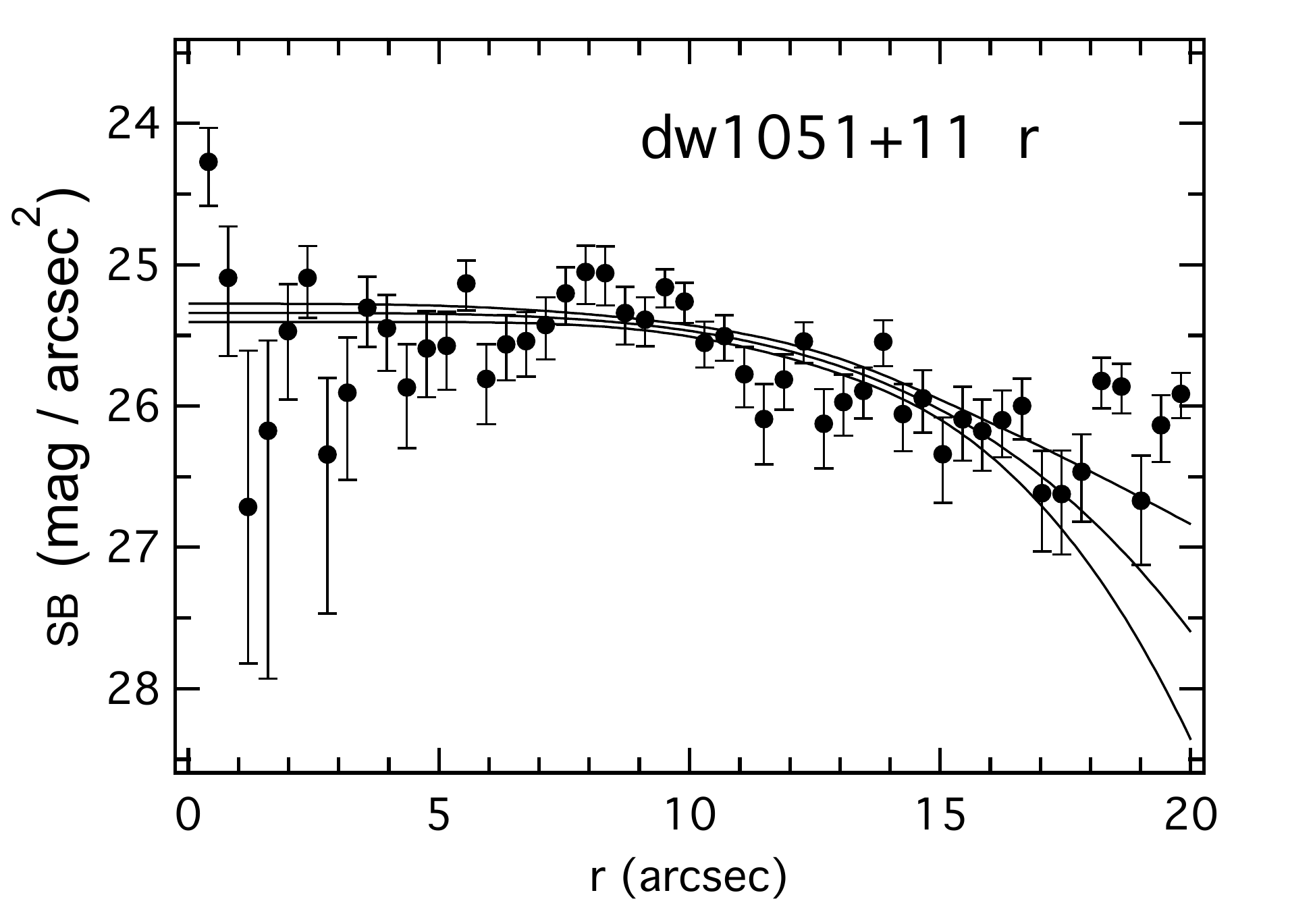}
 \includegraphics[width=3.6cm]{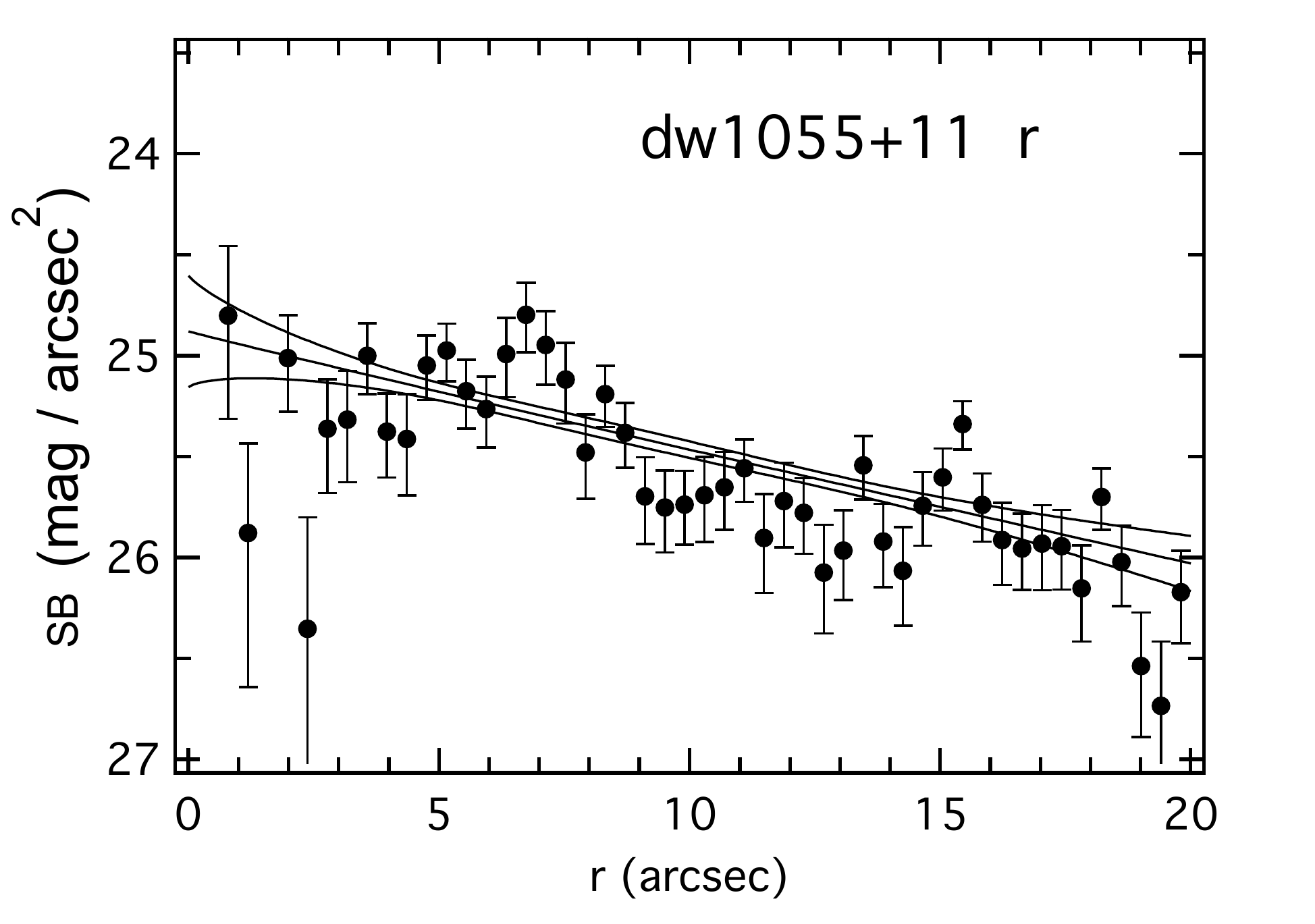}\\
 \includegraphics[width=3.6cm]{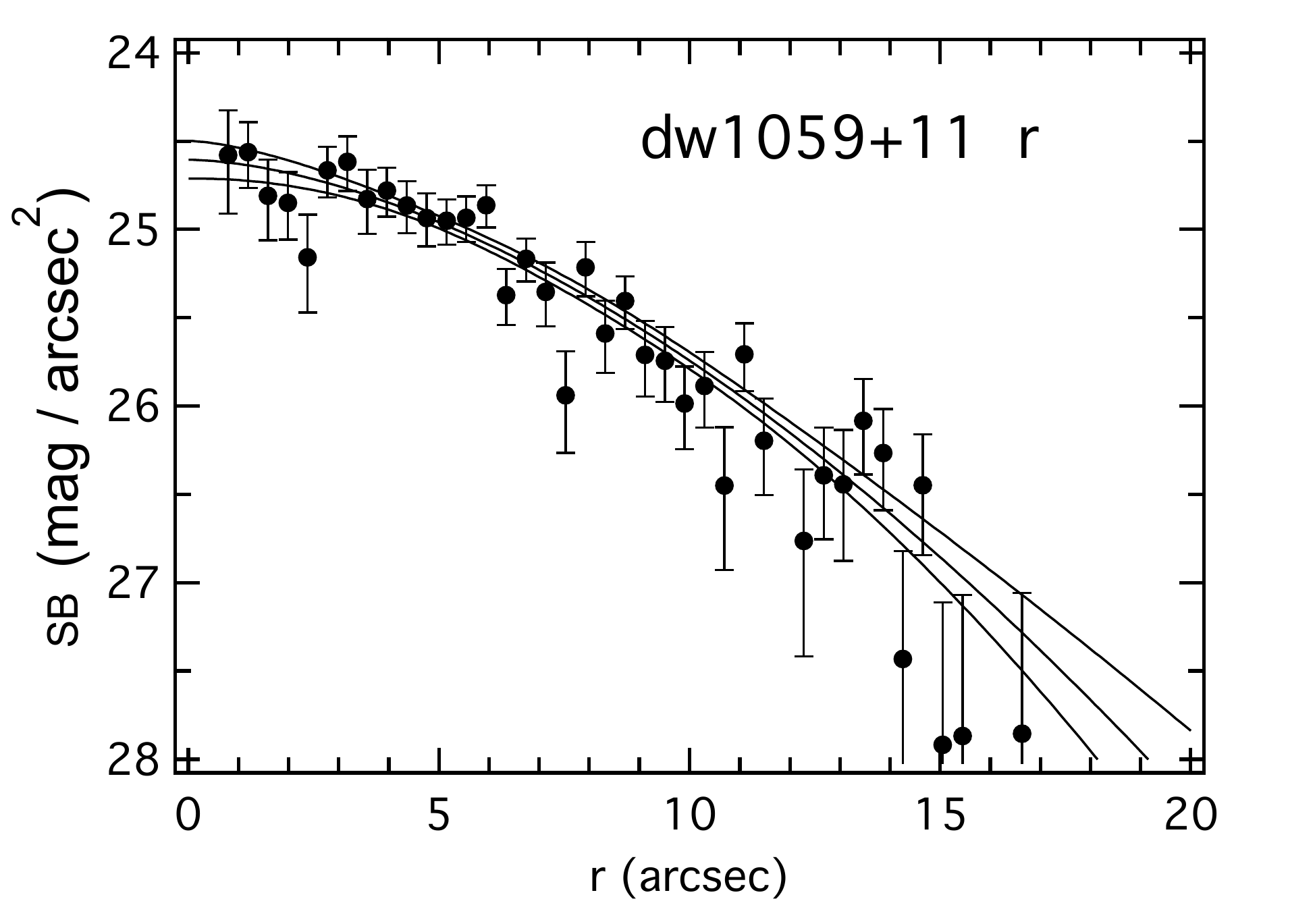}
 \includegraphics[width=3.6cm]{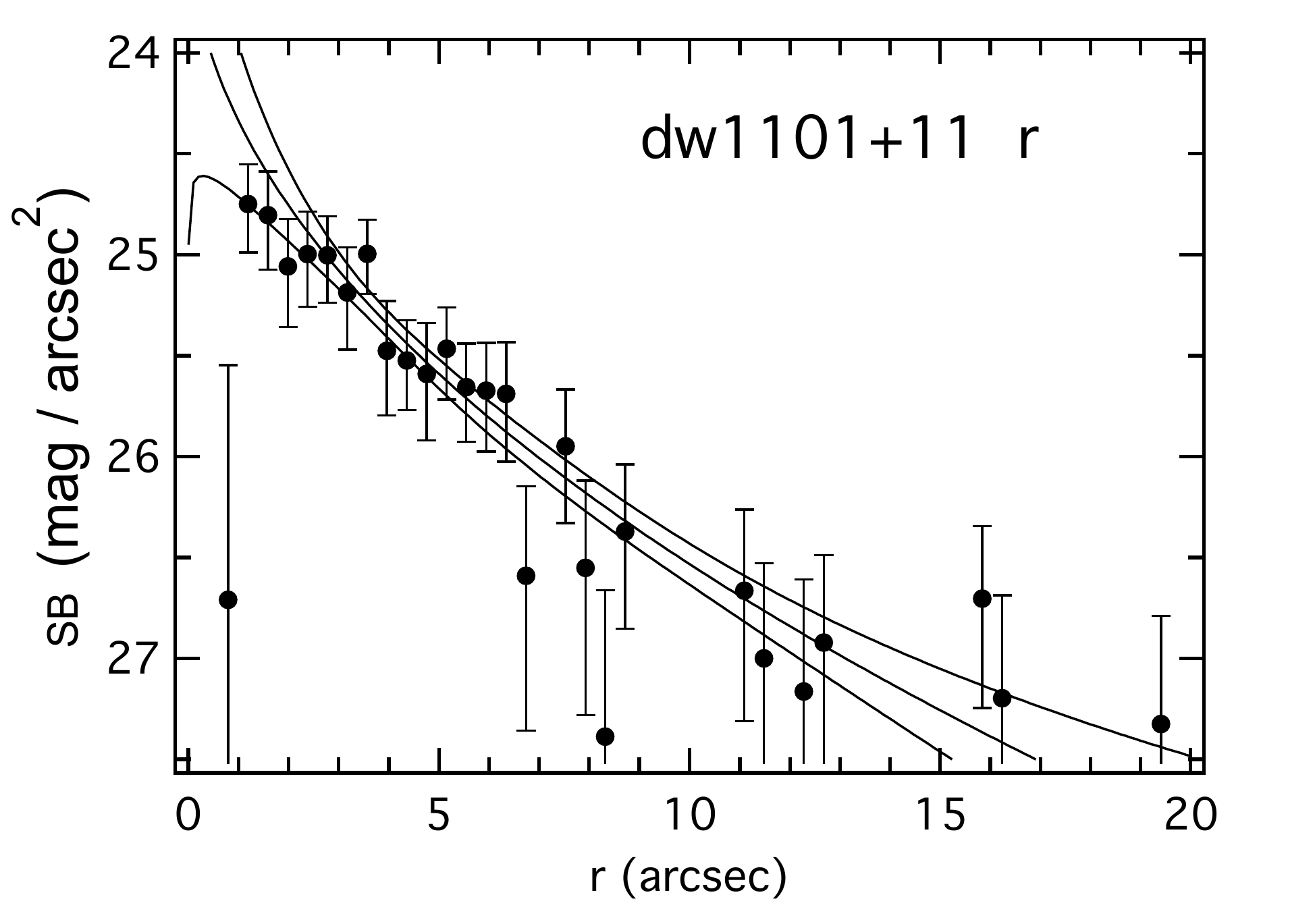}
 \includegraphics[width=3.6cm]{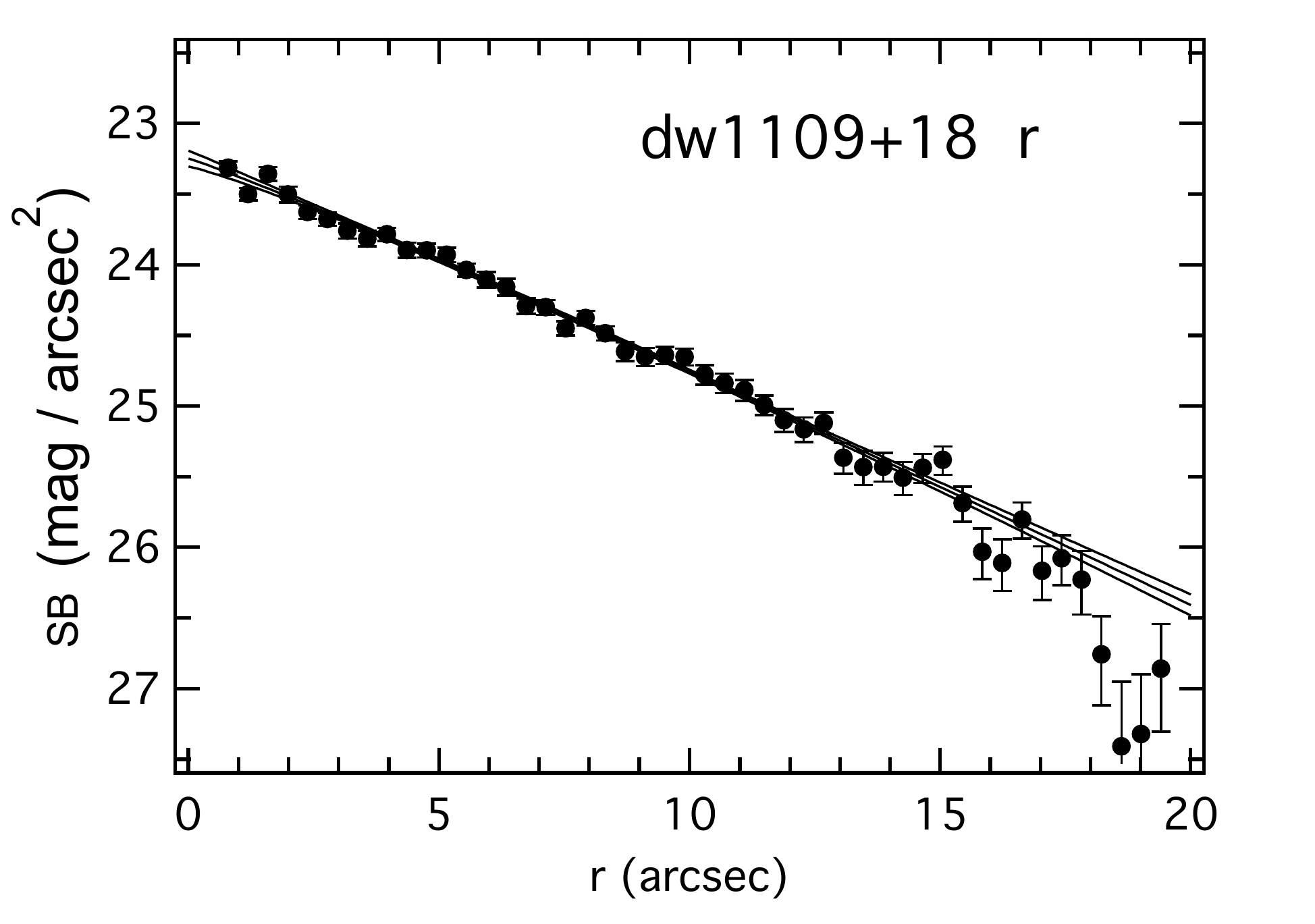}
 \includegraphics[width=3.6cm]{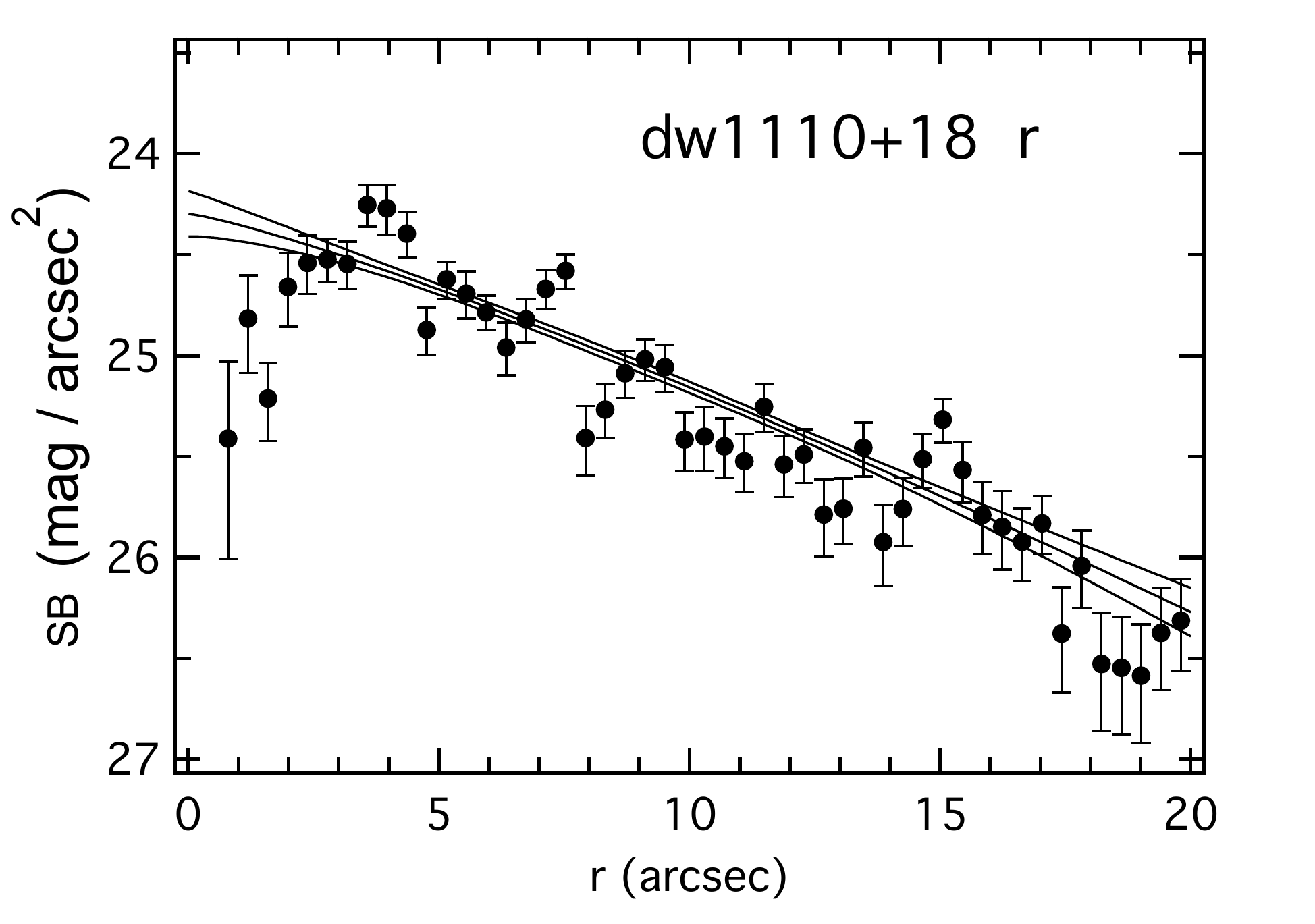}
 \includegraphics[width=3.6cm]{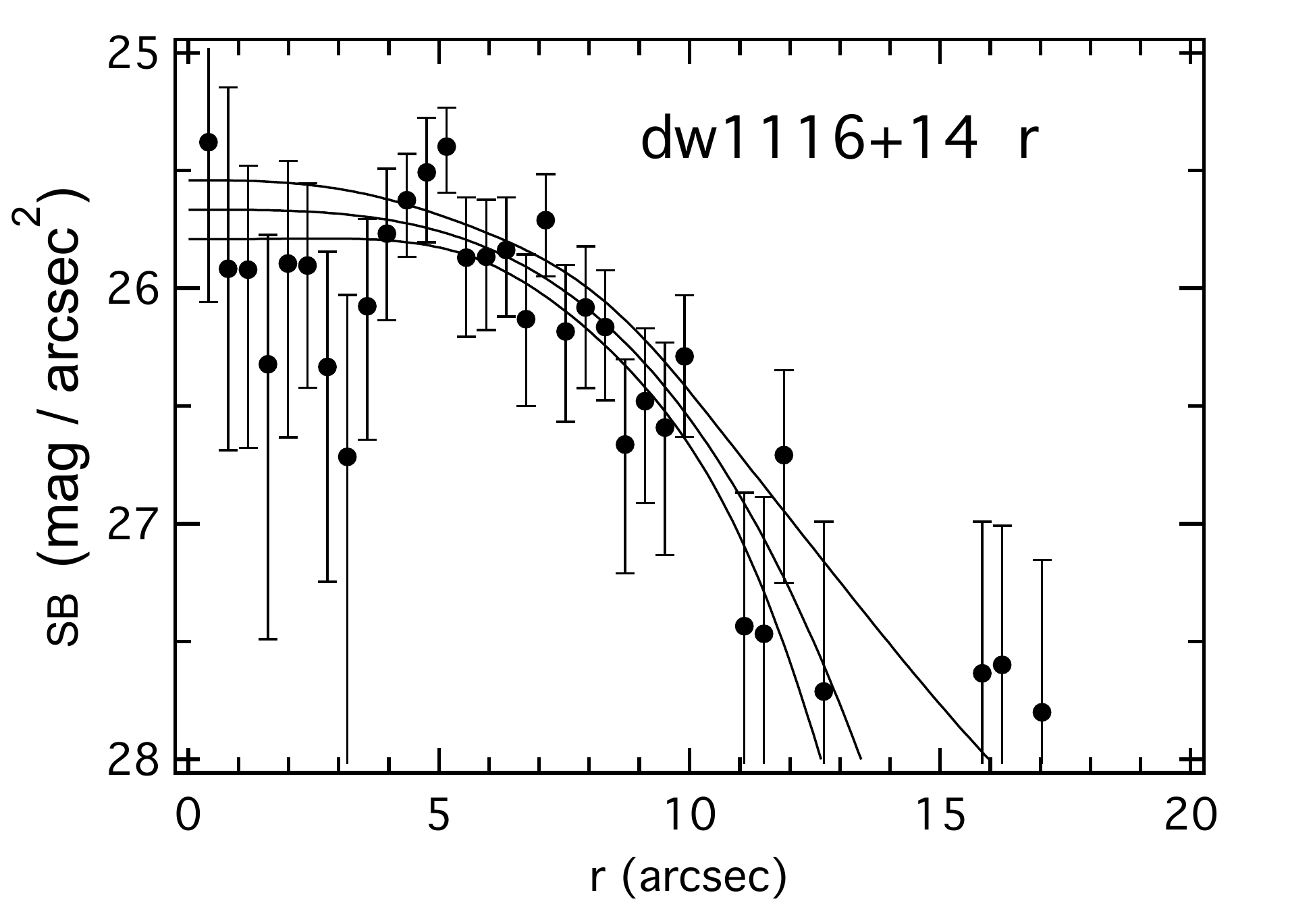}\\
 \includegraphics[width=3.6cm]{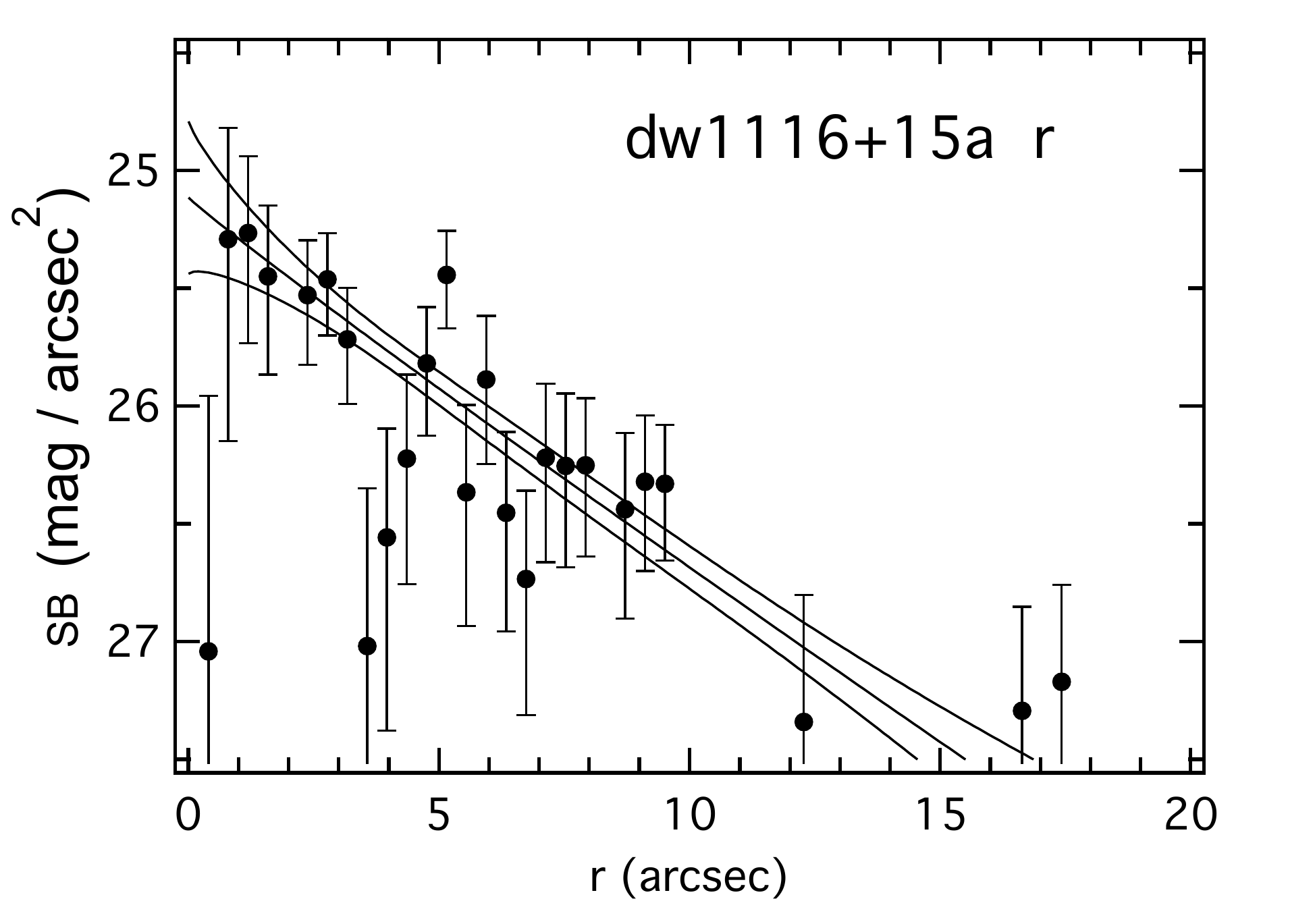}
 \includegraphics[width=3.6cm]{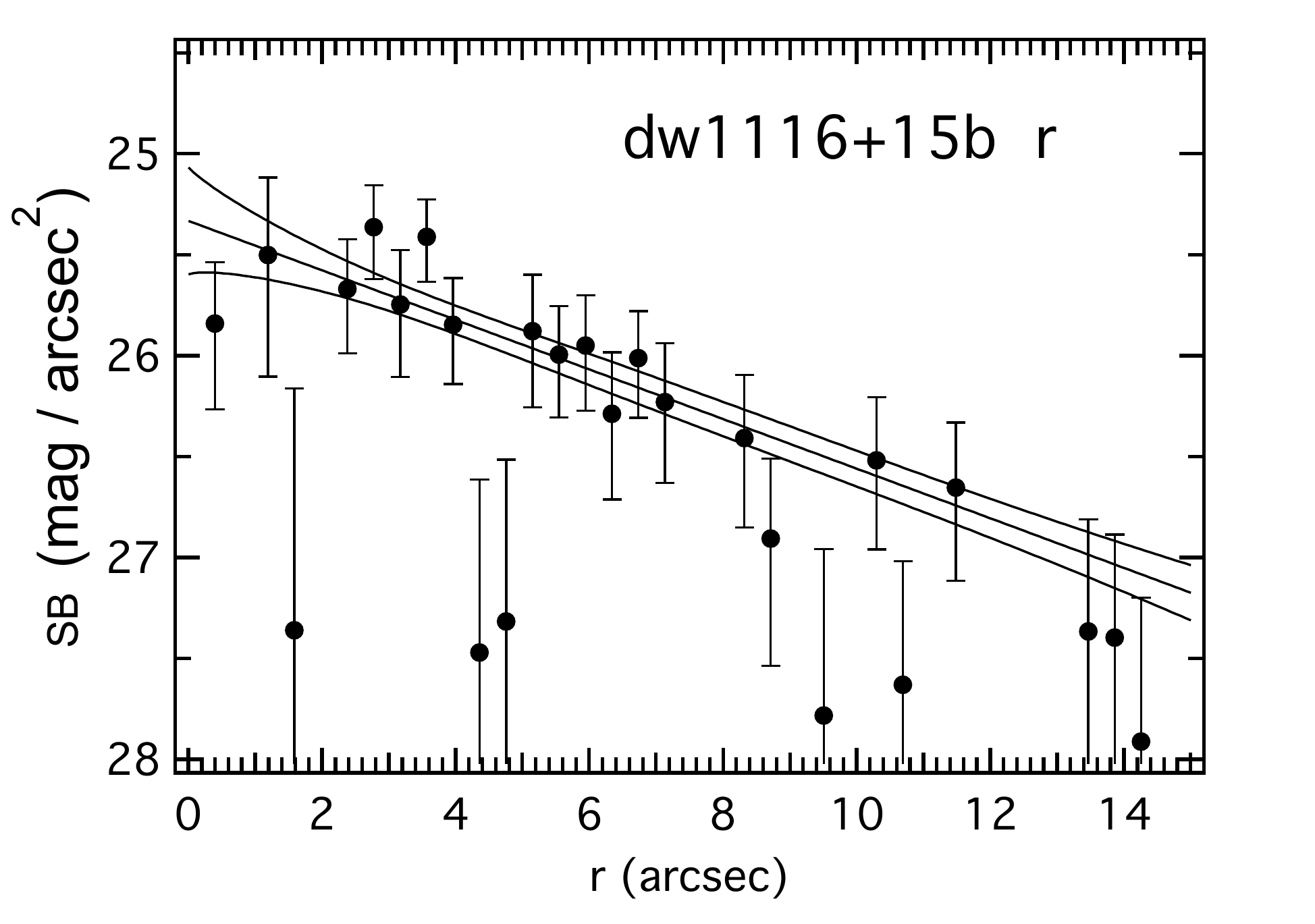}
 \includegraphics[width=3.6cm]{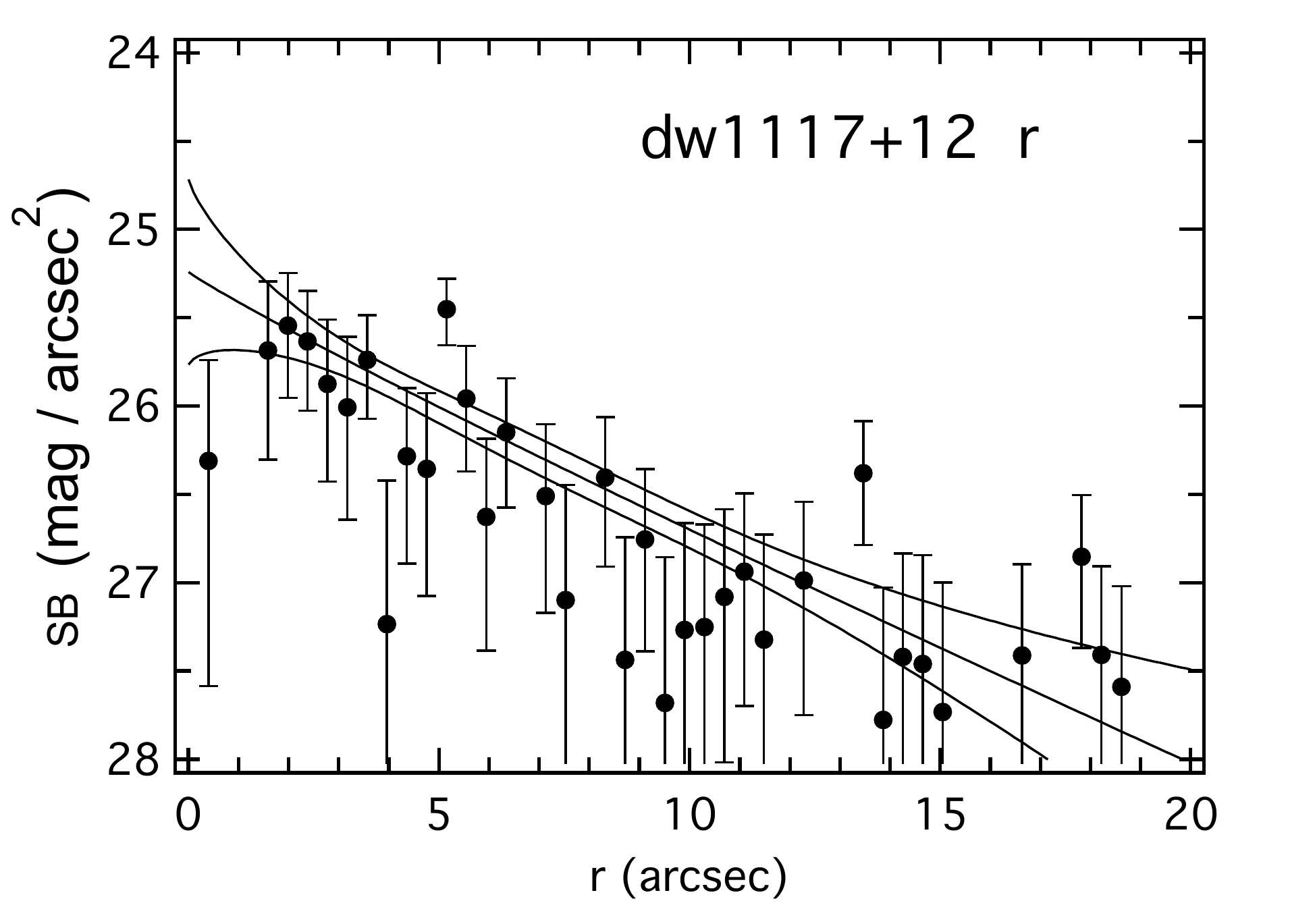}
 \includegraphics[width=3.6cm]{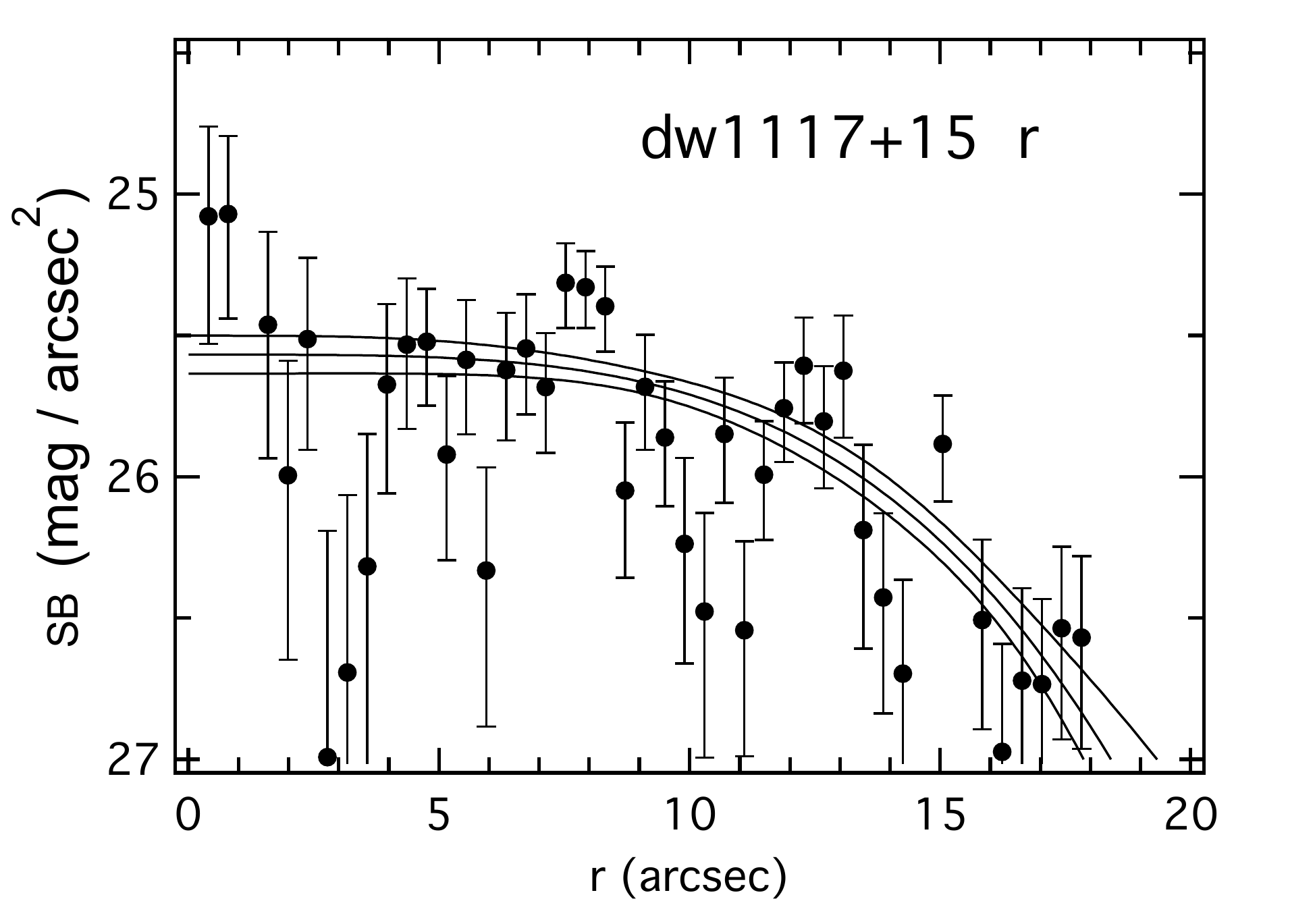}
 \includegraphics[width=3.6cm]{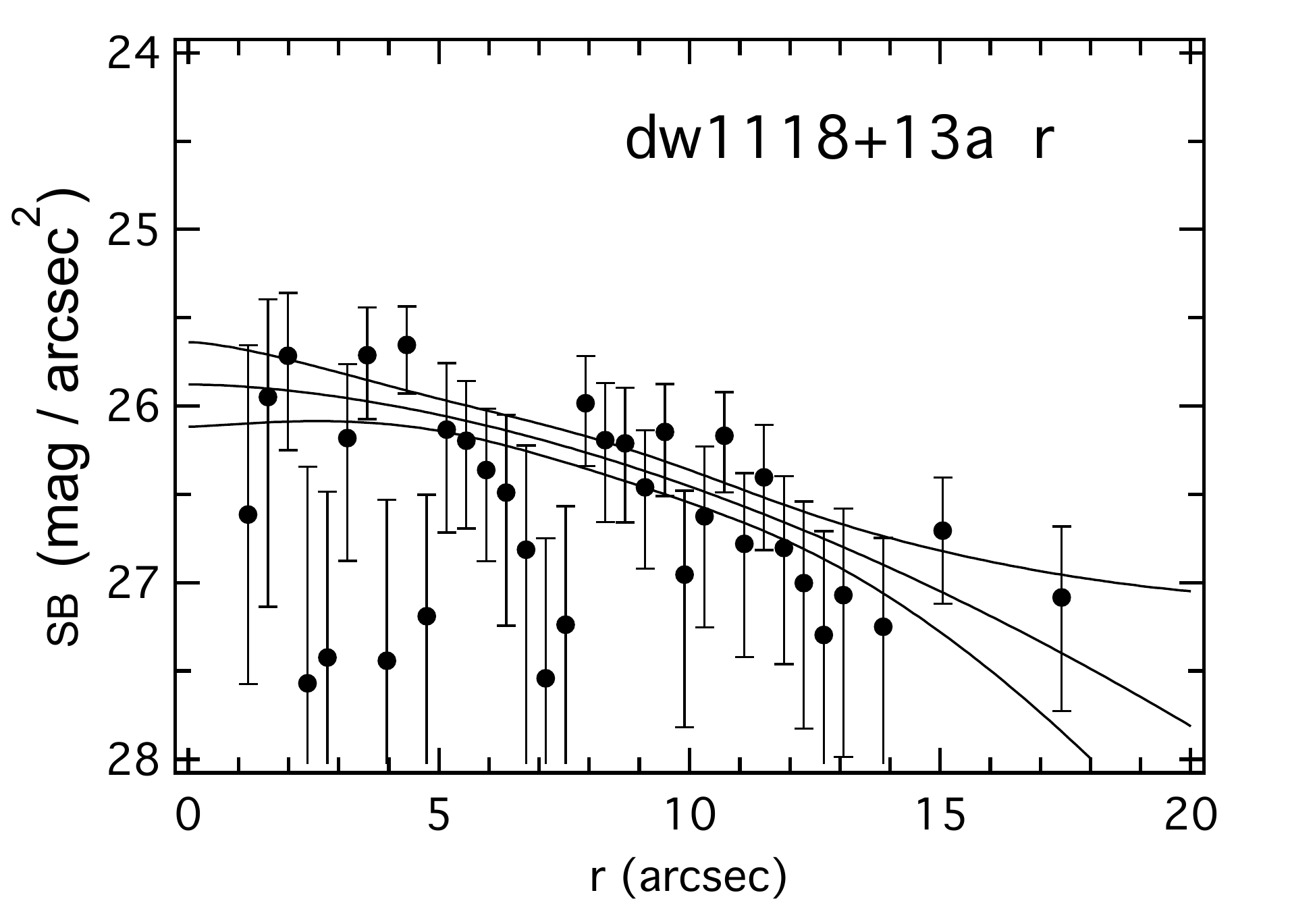}\\
 \includegraphics[width=3.6cm]{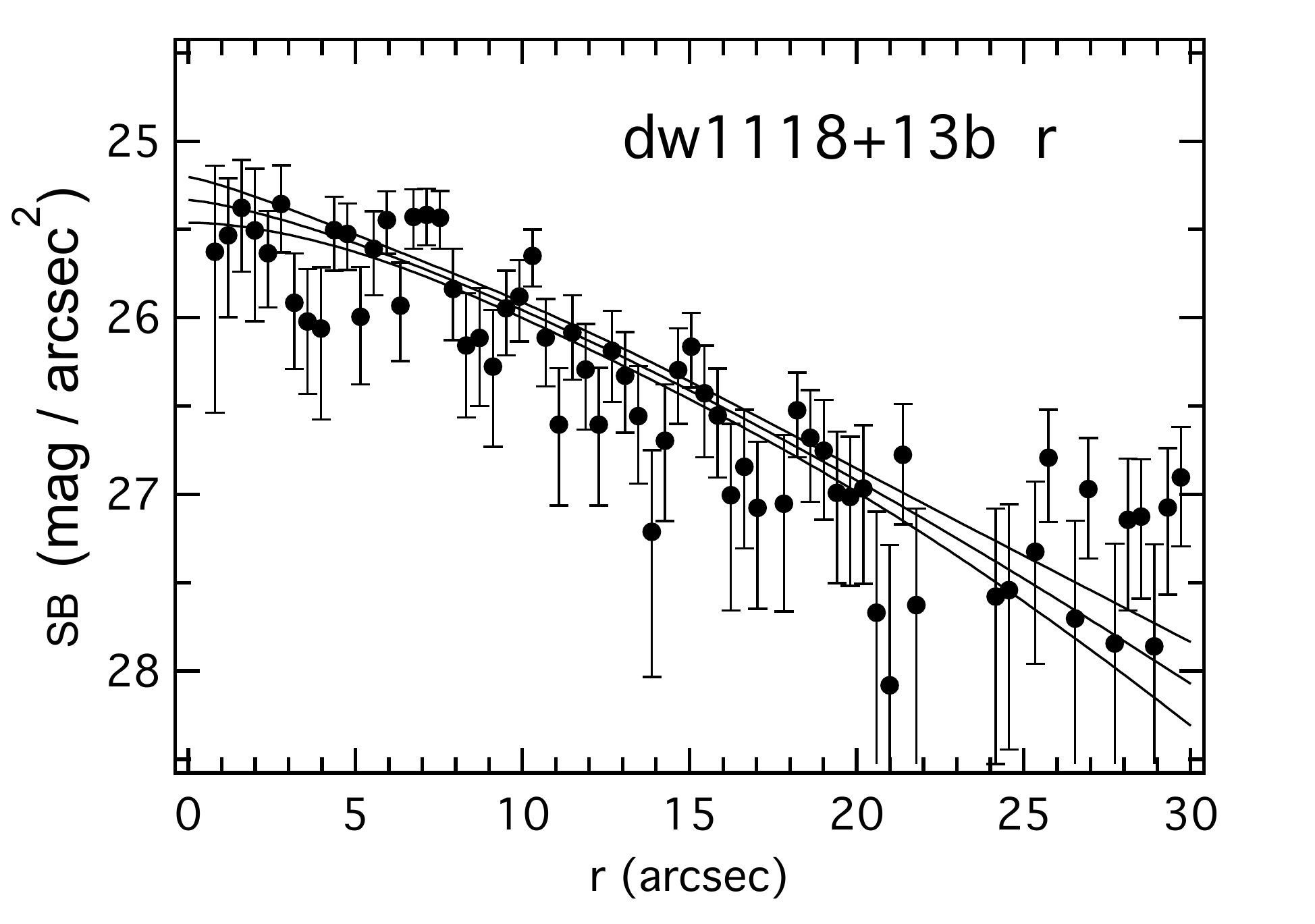}
 \includegraphics[width=3.6cm]{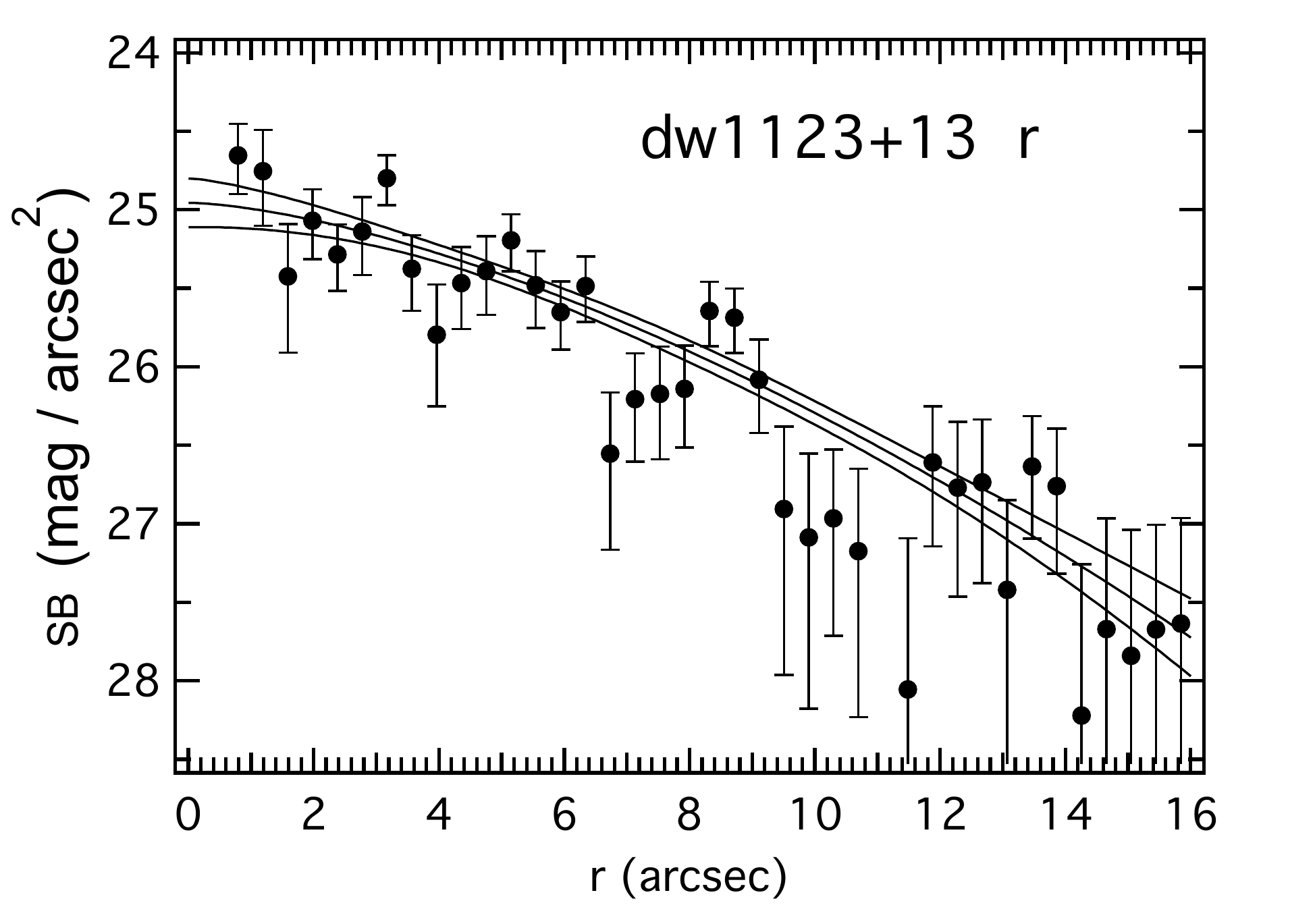}
 \includegraphics[width=3.6cm]{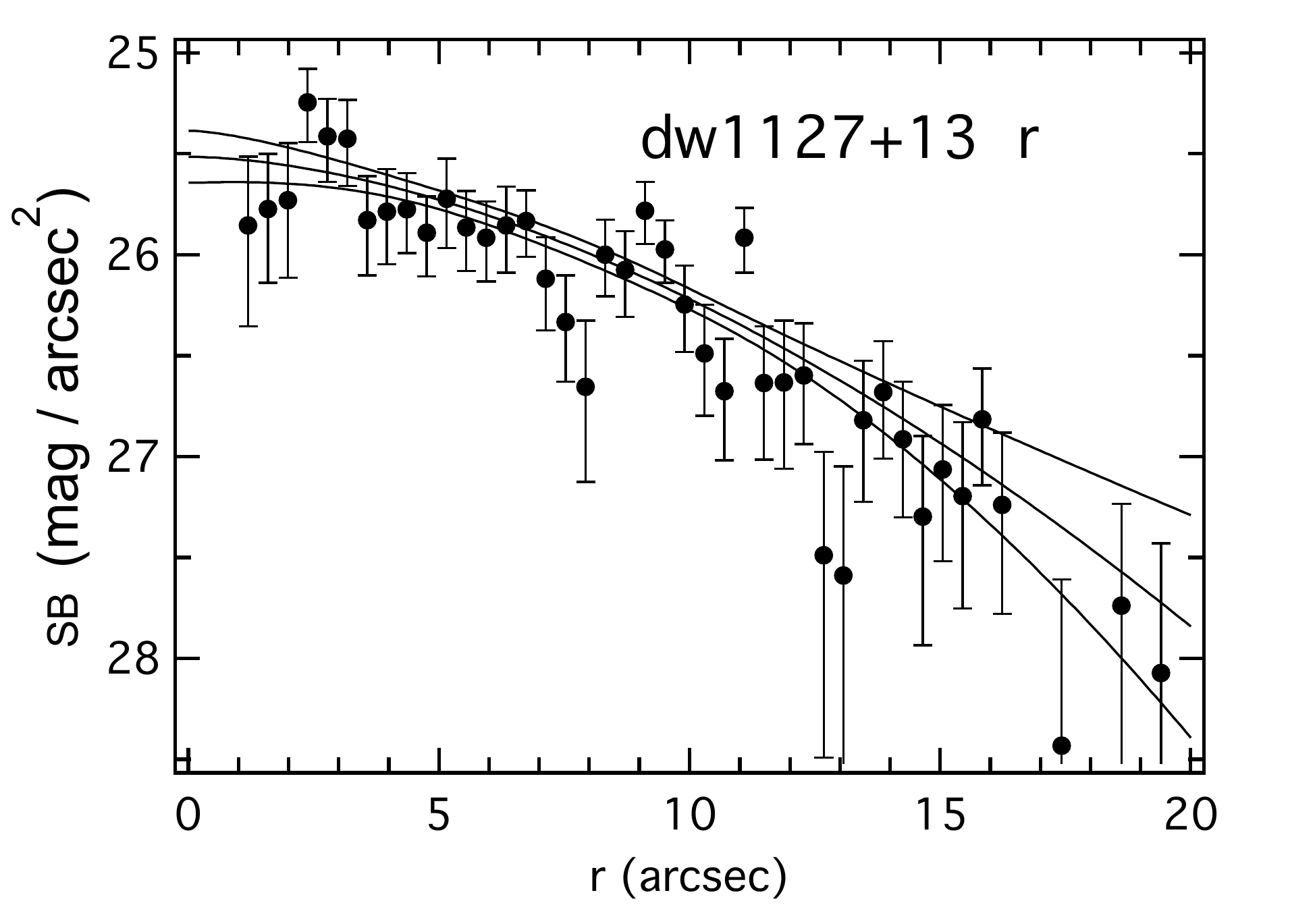}
 \includegraphics[width=3.6cm]{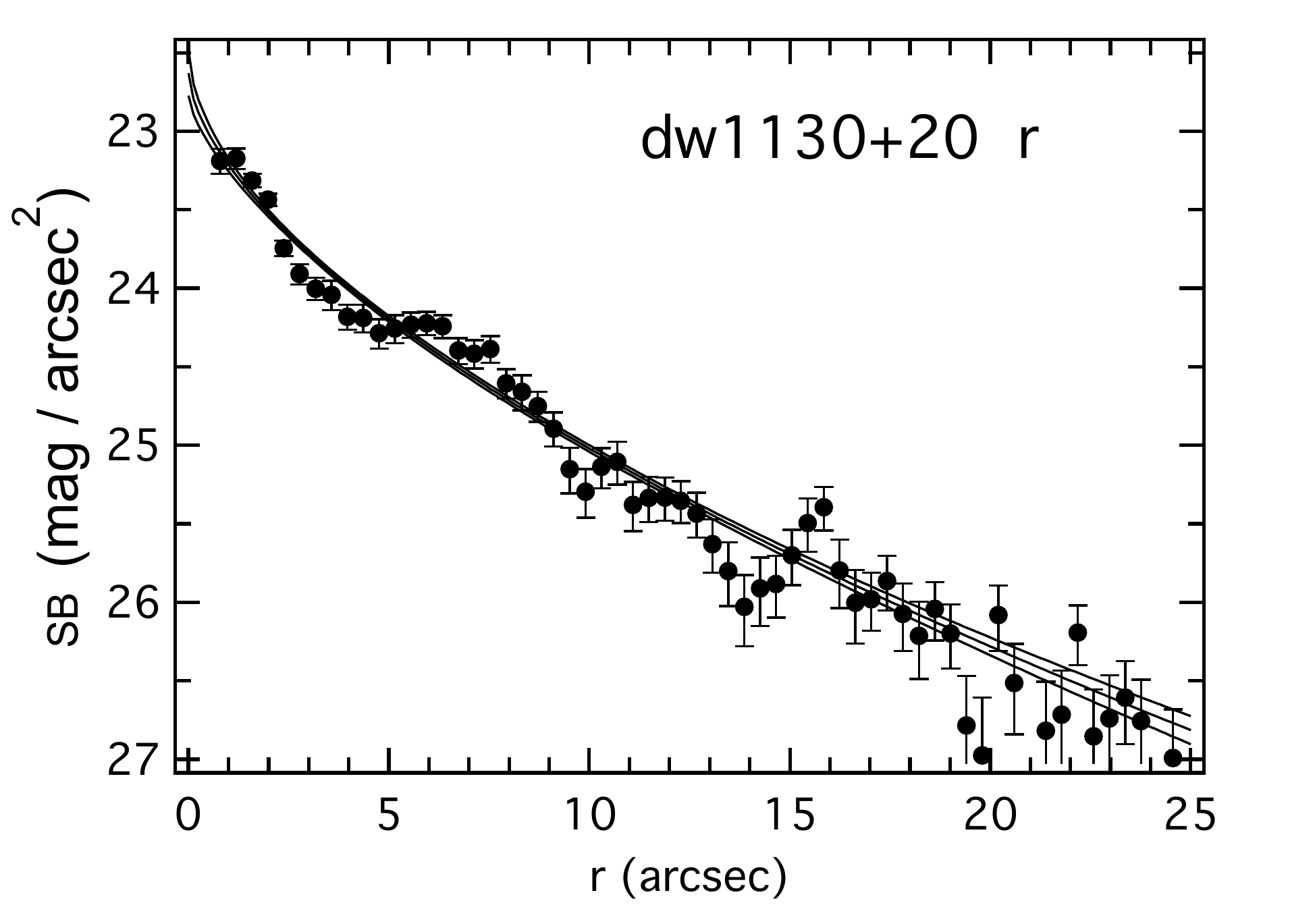}
 \includegraphics[width=3.6cm]{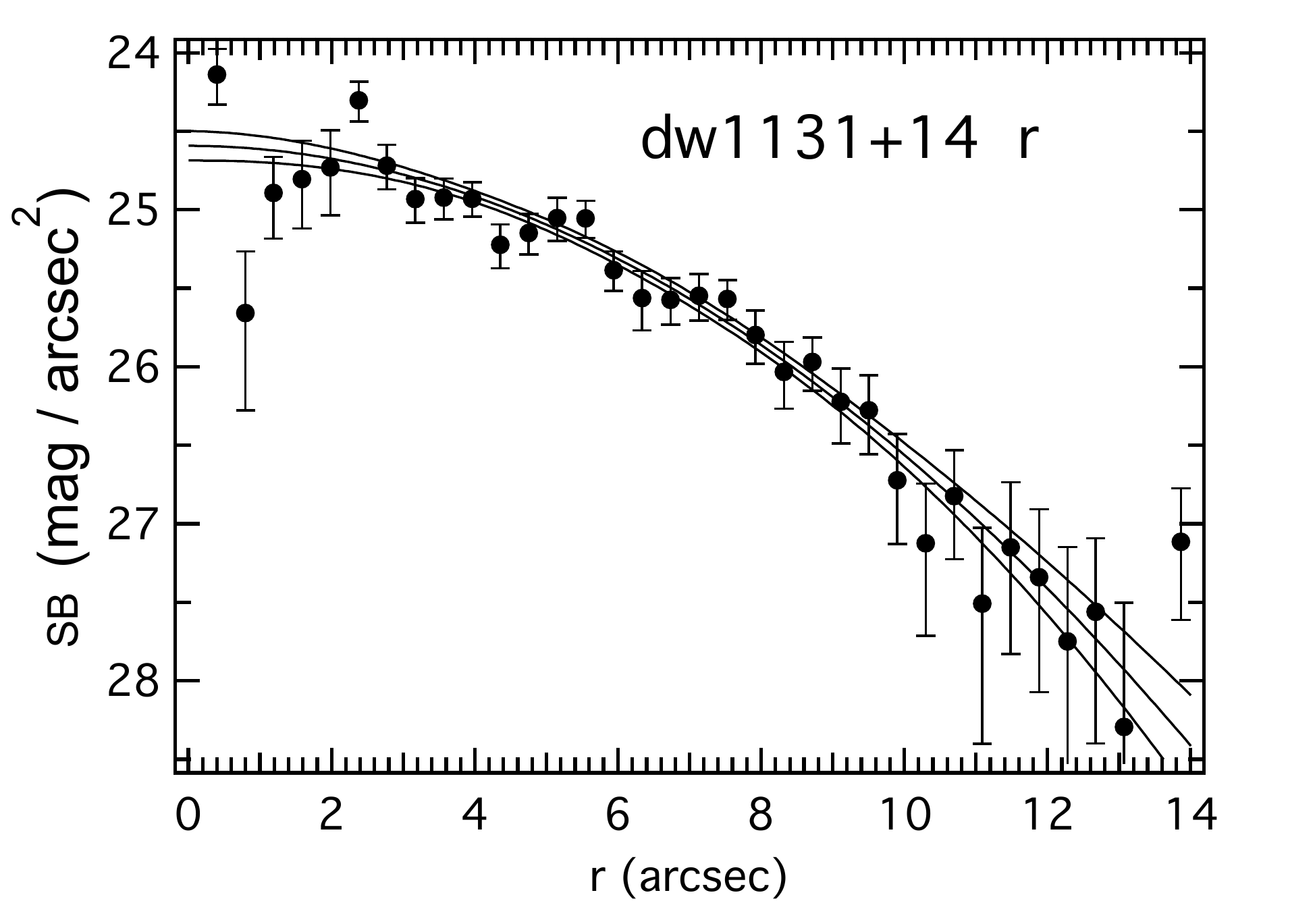}\\
 \includegraphics[width=3.6cm]{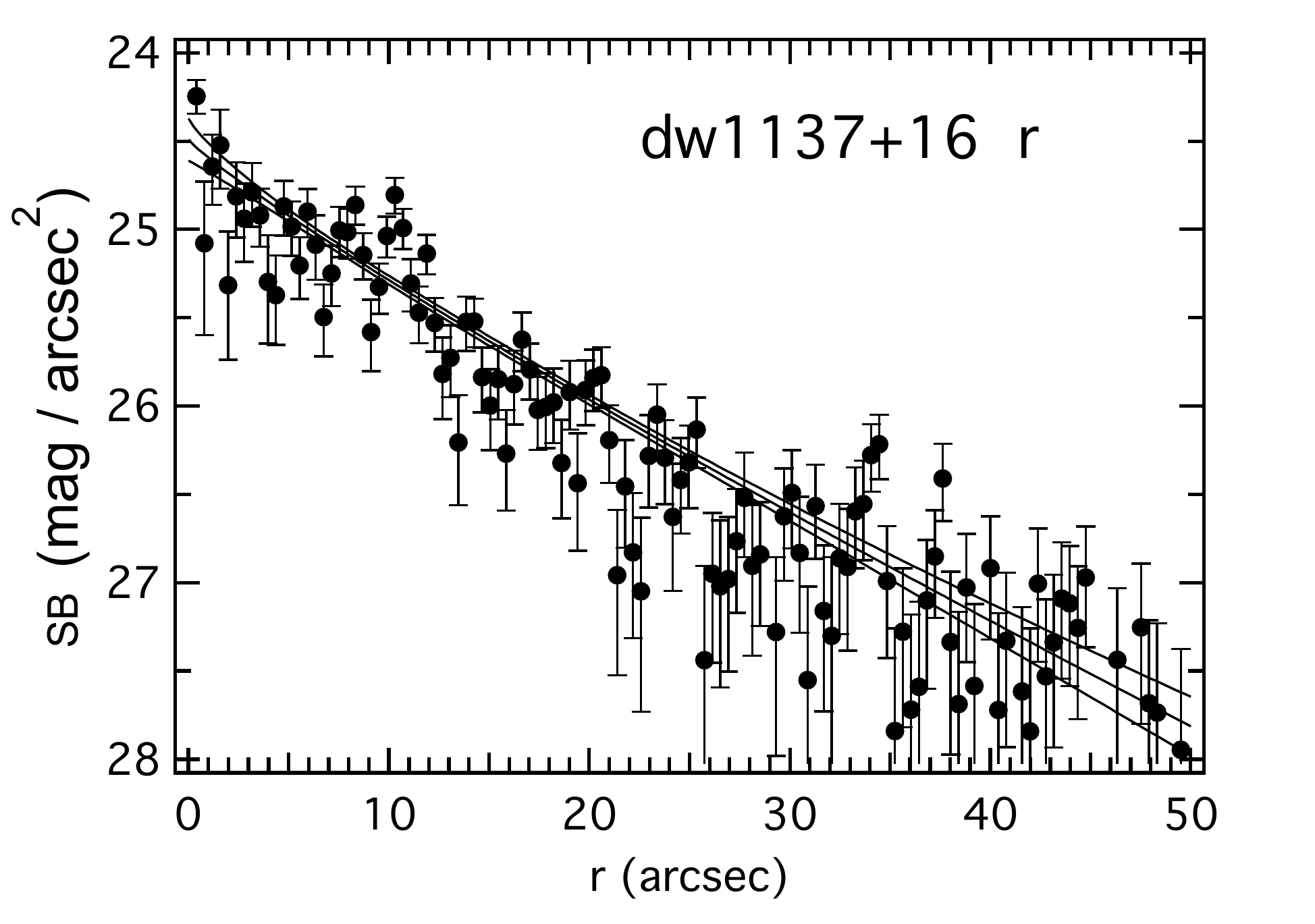}
 \includegraphics[width=3.6cm]{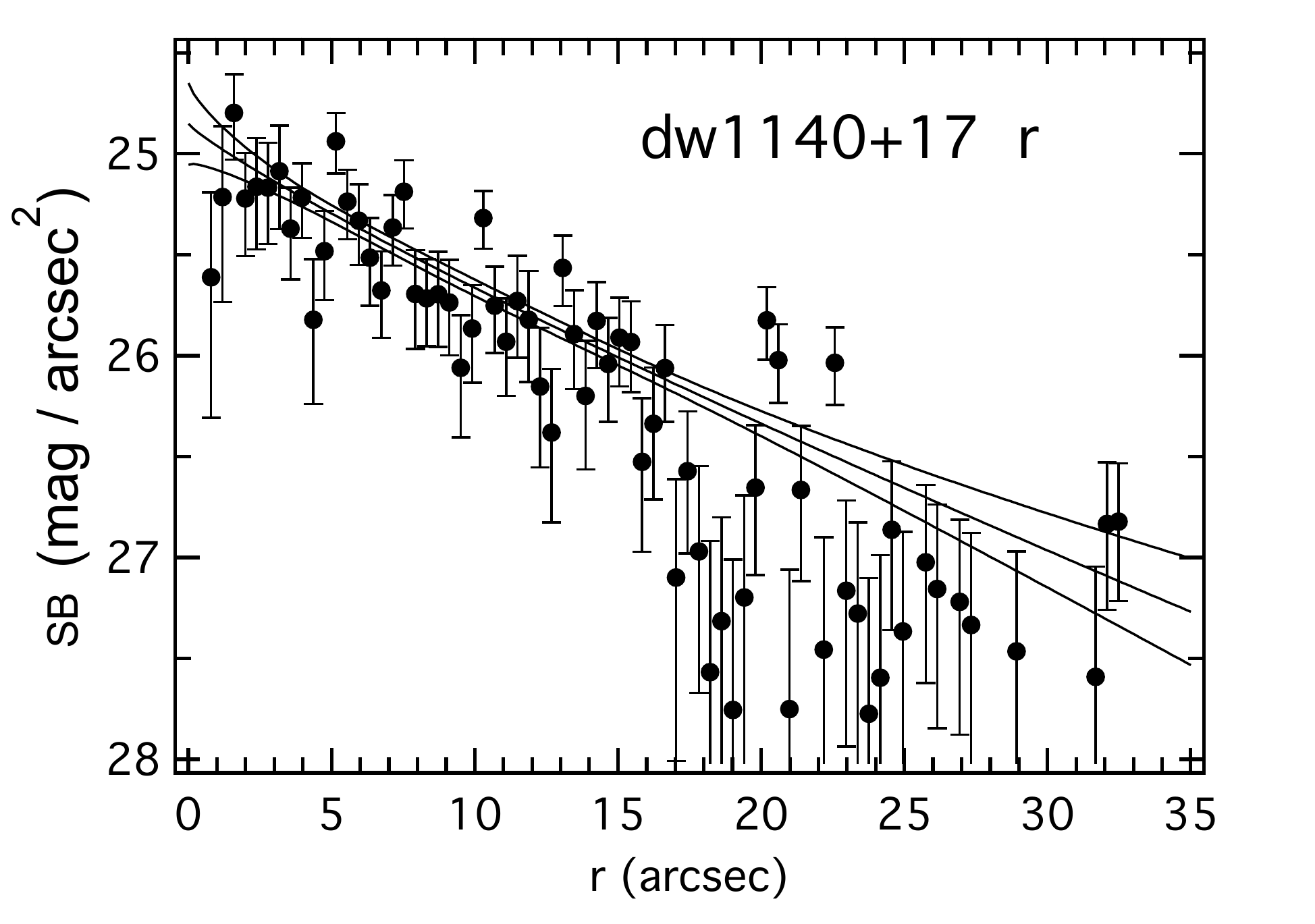}
 \includegraphics[width=3.6cm]{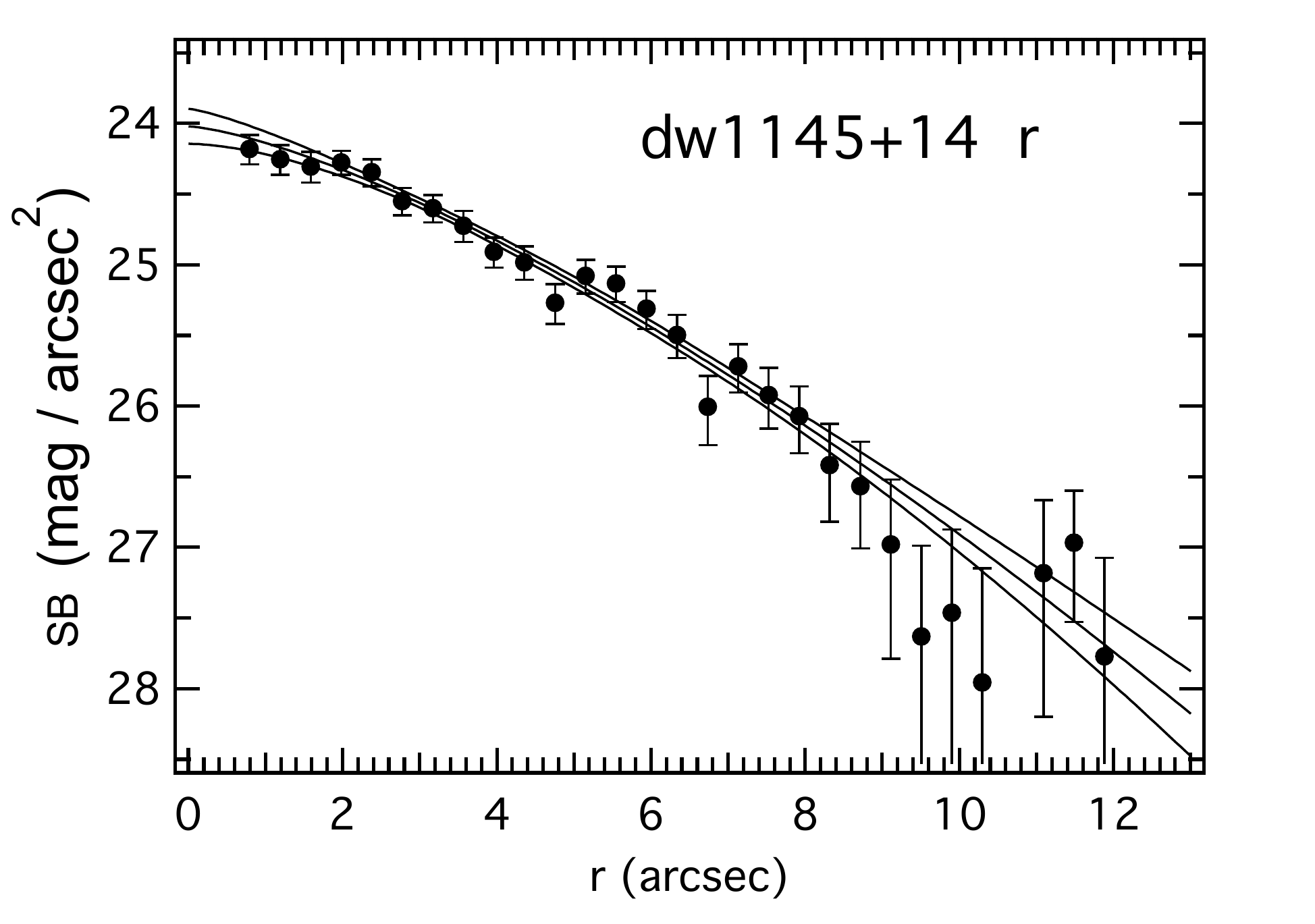}
 \includegraphics[width=3.6cm]{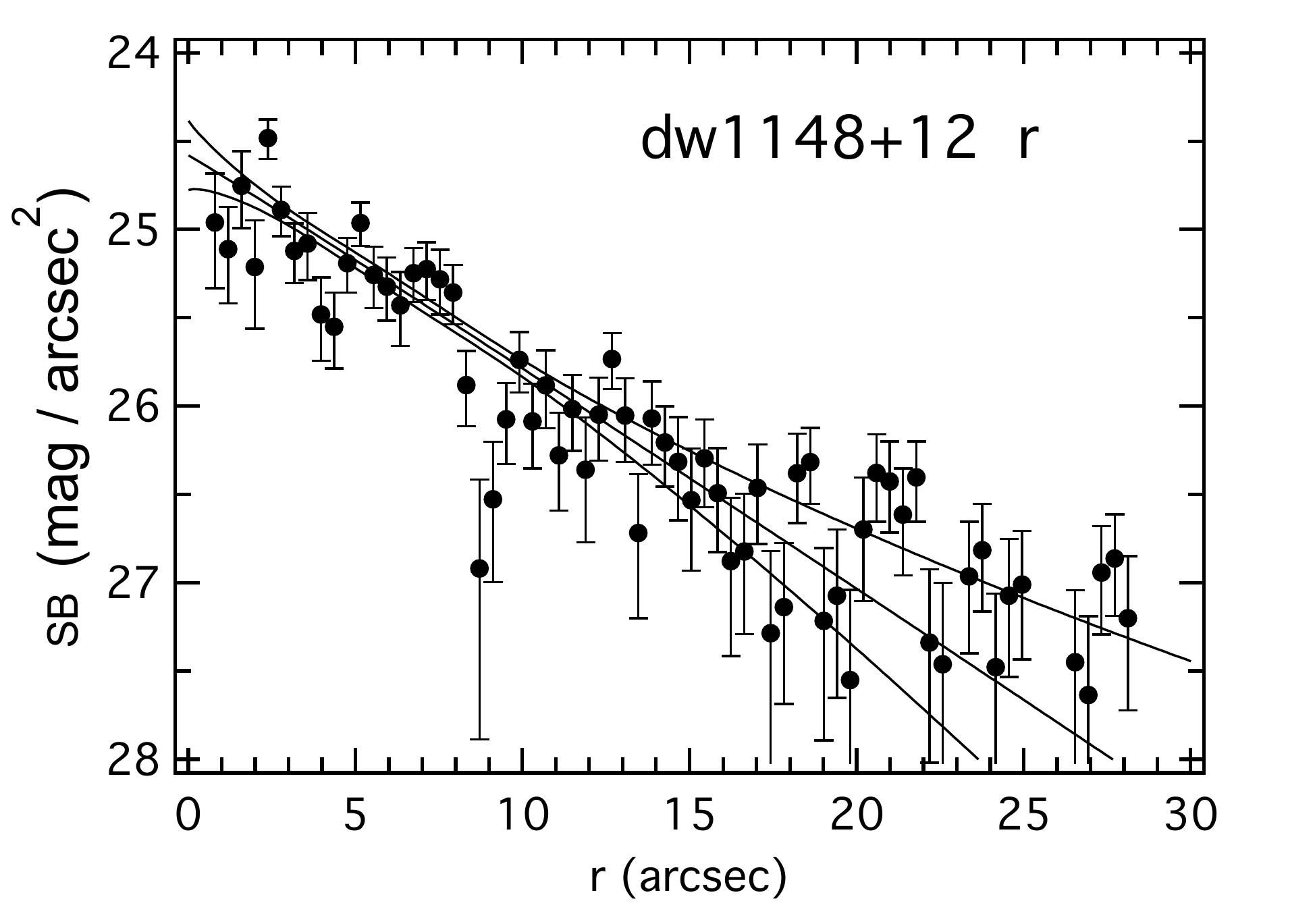}
 \includegraphics[width=3.6cm]{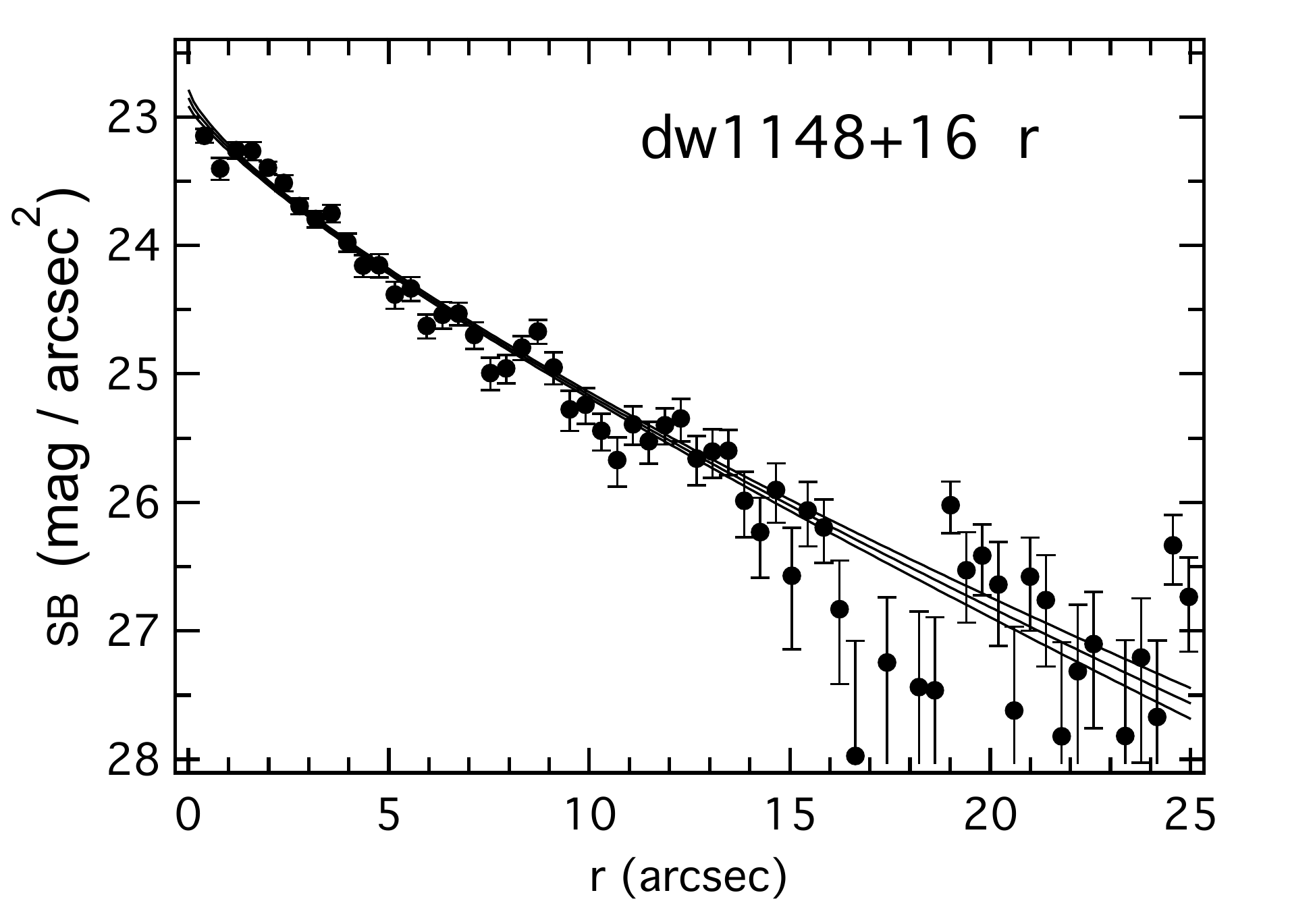}\\
 \includegraphics[width=3.6cm]{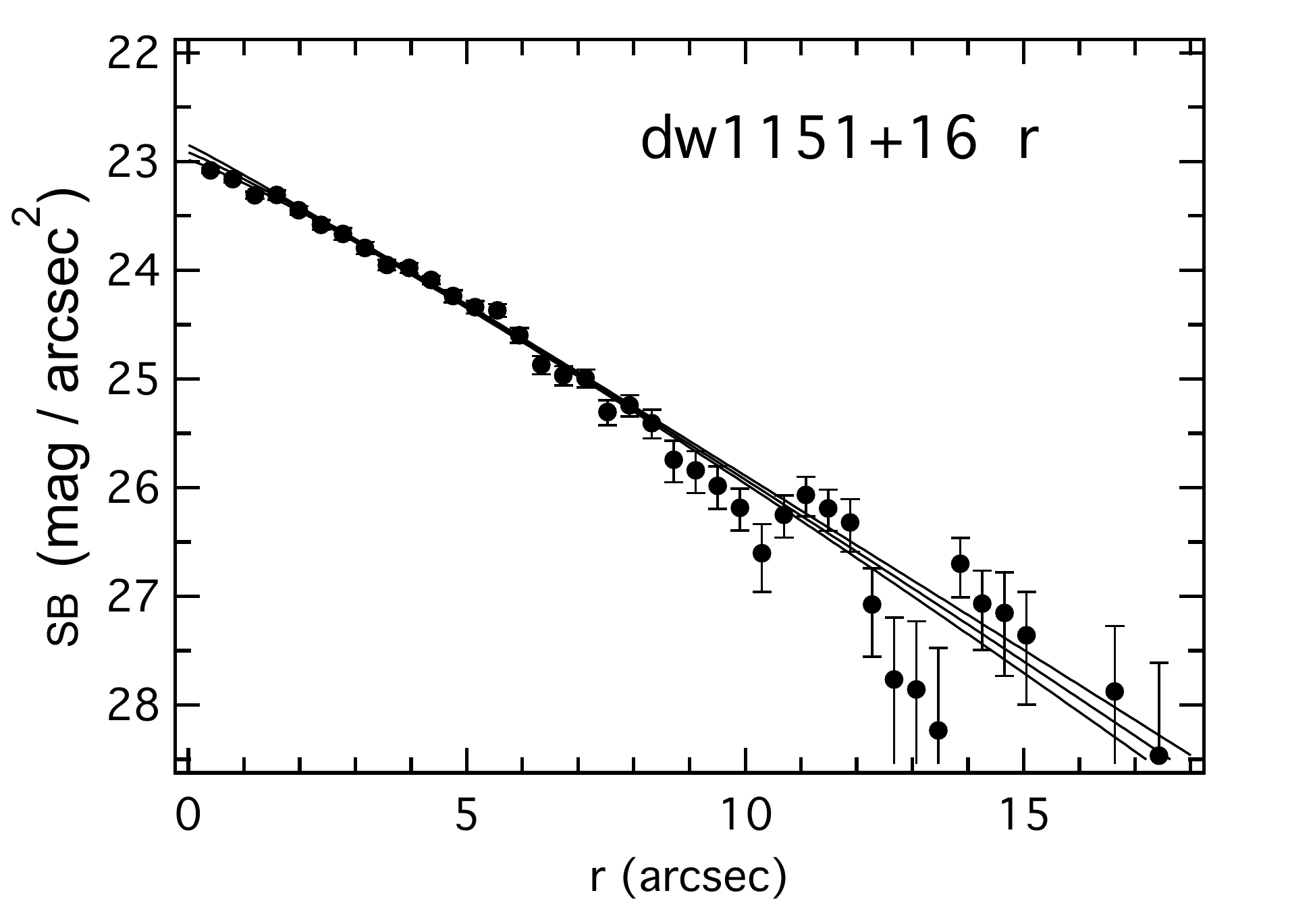}

 \caption{Surface brightness profiles of all new dwarf galaxy candidates in $r$ and the best-fitting S\'ersic profiles with $1 \sigma$ confidence intervals.}
 \label{sbp}
 \end{figure*}

We computed the total apparent magnitude $m$, the mean effective surface 
brightness $\langle \mu\rangle_{eff} $, and the effective radius $r_{eff}$ in $gr$ bands for each dwarf galaxy candidate, as well as for already known group members as many of them do not have accurate photometry. To measure the surface brightness profiles we used a circular aperture (step size of $0\farcs396$ corresponding to 1 pixel).
S\'ersic profiles  \citep{1968adga.book.....S} were fitted at the derived profiles using the equation
$$\mu_{sersic}(r)= \mu_0+1.0857\cdot\left(\frac{r}{r_0}\right)^{n},$$
where $\mu_0$ is the S\'ersic central surface brightness, $r_0$ the S\'ersic scale length, and $n$ the S\'ersic curvature index. The total extinction corrected absolute magnitude $M$ is calculated with a distance modulus of $m-M=30.06$\,mag, corresponding to $D=10.4$\,Mpc, as is used for Leo-I members with unknown distance estimates in the LV catalog.
See Fig.\,\ref{sbp} for all surface brightness profiles in the $r$ band and the associated S\'ersic fits.
In Table\,\ref{table2} we provide the derived photometry for the new candidates, in Tables\,\ref{table3} and \ref{table4} for the previously known (dwarf) members of the Leo-I group.

The magnitude uncertainties are estimated to be around $\approx$\,0.3\,mag \citep{2017A&A...602A.119M}. The main contributions to the error budget are from the uncertainties related to foreground star removal ($\approx$\,0.2\,mag) and sky background estimation ($\approx$\,0.2\,mag). The uncertainties for  $\langle \mu\rangle_{eff} $ are driven by the uncertainties in the measured total apparent magnitude ($\approx$\,0.3\,mag \,arcsec$^{-2}$). The error for $r_{eff}$ ($\approx$\,1.3\,arcsec) is given by the determination of the growth curve. Numerical uncertainties for the S\'ersic parameters are provided in the corresponding table.%\ref{table:2}.

\begin{table*}[ht]
\caption{Photometric and structural parameters of the new dwarf candidates in the surveyed region of the Leo-I group.}
\small
\centering
\setlength{\tabcolsep}{3pt}
\begin{tabular}{lccrrrrcrccrr}
\hline\hline 
Name & ${g_{tot}}$ & ${r_{tot}}$  & $A_g$ & $A_r$ & $M_{r}$ & $(g-r)_{0,tot}$ & $\mu_{0,r}$ & $r_{0,r}$ & $n_r$ & $\langle\mu\rangle_{eff,r}$ &  $r_{eff,r}$ & $\log r_{eff,r}$\\ 
 & mag & mag & mag & mag &  mag & mag & mag arcsec$^{-2}$ &arcsec & & mag arcsec$^{-2}$ &arcsec & $\log$ pc \\ 
 (1)& (2) & (3) &  (4) & (5) &  (6) & (7) &(8) & (9) & (10) & (11) & (12) & (13)\\ 
\hline \\
M\,96 subgroup & & & & & & & & &\\
dw1037+09 & 17.61 & 17.09 & 0.080 & 0.056 & -13.04 & 0.496 & $24.04 \pm 0.05 $  & $9.96 \pm 0.38 $  & $1.68 \pm 0.14 $  & 25.44 & 18.7 & 2.97\\
dw1044+11 & 19.39 & 19.17 & 0.088 & 0.061 & -10.97 & 0.200 & $25.42 \pm 0.45 $  & $10.87 \pm 5.89 $  & $0.87 \pm 0.40 $  & 26.59 & 12.1 & 2.78\\
dw1045+14a & 19.02 & 18.68 & 0.097 & 0.067 & -11.46 & 0.313 & $22.79 \pm 0.63 $  & $1.25 \pm 1.14 $  & $0.56 \pm 0.16 $  & 24.87 & 6.92 & 2.54\\
dw1045+14b & 19.79 & 19.29 & 0.094 & 0.065 & -10.85 & 0.470 & $24.50 \pm 0.17 $  & $5.24 \pm 0.83 $  & $1.13 \pm 0.18 $  & 25.39 & 6.61 & 2.52\\
dw1045+16 & 18.50 & 17.62 & 0.085 & 0.059 & -12.51 & 0.854 & $24.59 \pm 0.30 $  & $8.39 \pm 3.22 $  & $0.81 \pm 0.24 $  & 26.43 & 23.0 & 3.06\\
dw1045+13 & 18.81 & 18.08 & 0.110 & 0.076 & -12.07 & 0.696 & $24.96 \pm 0.15 $  & $7.41 \pm 0.70 $  & $1.92 \pm 0.52 $  & 26.59 & 20.0 & 3.00\\
dw1047+16 & 18.07 & 17.91 & 0.091 & 0.063 & -12.23 & 0.128 & $22.77 \pm 0.55 $  & $1.32 \pm 1.18 $  & $0.49 \pm 0.13 $  & 25.17 & 11.2 & 2.75\\
dw1048+13 & 19.83 & 18.67 & 0.111 & 0.077 & -11.48 & 1.126 & $25.53 \pm 0.10 $  & $13.63 \pm 0.64 $  & $2.94 \pm 0.78 $  & 26.19 & 12.6 & 2.80\\
dw1049+12a & 19.39 & 18.98 & 0.088 & 0.061 & -11.16 & 0.386 & $23.78 \pm 0.49 $  & $2.92 \pm 1.91 $  & $0.67 \pm 0.19 $  & 25.54 & 8.18 & 2.61\\
dw1049+15 & 18.56 & 17.88 & 0.088 & 0.061 & -12.26 & 0.655 & $24.30 \pm 0.12 $  & $9.38 \pm 0.90 $  & $1.42 \pm 0.23 $  & 25.08 & 10.9 & 2.74\\

dw1049+12b & 19.10 & 18.05 & 0.085 & 0.059 & -12.08 & 1.020 & $24.74 \pm 0.28 $  & $9.76 \pm 3.03 $  & $0.96 \pm 0.36 $  & 26.04 & 15.7 & 2.90\\
dw1051+11 & 17.85 & 16.95 & 0.092 & 0.063 & -13.19 & 0.872 & $25.34 \pm 0.07 $  & $16.76 \pm 0.63 $  & $4.15 \pm 1.20 $  & 26.20 & 28.2 & 3.15\\
dw1055+11 & 17.59 & 16.40 & 0.066 & 0.046 & -13.72 & 1.169 & $24.88 \pm 0.28 $  & $18.86 \pm 3.90 $  & $0.97 \pm 0.54 $  & 26.18 & 36.0 & 3.25\\
dw1059+11 & 18.98 & 18.60 & 0.060 & 0.041 & -11.51 & 0.359 & $24.61 \pm 0.11 $  & $9.73 \pm 0.72 $  & $1.68 \pm 0.26 $  & 25.02 & 7.65 & 2.58\\
dw1101+11 & 19.47 & 19.45 & 0.058 & 0.040 & -10.66 & 0.005 & $23.33 \pm 1.62 $  & $1.16 \pm 2.78 $  & $0.50 \pm 0.32 $  & 25.56 & 6.64 & 2.52\\
\\
Leo Triplet & & & & & & & & &\\
%dw1113+14 & 16.01 & 15.29 & 0.073 & 0.050 & -14.84 & 0.696 & $23.77 \pm 0.01 $  & $27.13 \pm 0.27 $  & $1.84 \pm 0.04 $  & 24.10 & 23.0 & 3.06\\
dw1116+14 & 20.33 & 19.57 & 0.071 & 0.049 & -10.56 & 0.742 & $25.67 \pm 0.13 $  & $10.63 \pm 0.58 $  & $3.28 \pm 1.11 $  & 25.81 & 7.08 & 2.55\\
dw1116+15a & 20.26 & 19.80 & 0.076 & 0.052 & -10.33 & 0.437 & $25.11 \pm 0.32 $  & $6.79 \pm 2.43 $  & $0.95 \pm 0.28 $  & 25.98 & 6.88 & 2.54\\
dw1116+15b & 20.42 & 19.31 & 0.068 & 0.047 & -10.81 & 1.091 & $25.33 \pm 0.26 $  & $8.85 \pm 2.61 $  & $1.00 \pm 0.32 $  & 27.02 & 13.8 & 2.84\\
dw1117+15 & 17.56 & 17.25 & 0.082 & 0.057 & -12.88 & 0.280 & $25.57 \pm 0.07 $  & $17.11 \pm 0.51 $  & $3.79 \pm 0.92 $  & 27.31 & 40.9 & 3.31\\
dw1117+12 & 21.22 & 19.87 & 0.073 & 0.050 & -10.25 & 1.322 & $25.24 \pm 0.52 $  & $7.29 \pm 4.29 $  & $0.93 \pm 0.51 $  & 26.10 & 7.02 & 2.54\\
dw1118+13a & 19.49 & 19.59 & 0.077 & 0.053 & -10.54 & -0.11 & $25.88 \pm 0.24 $  & $14.36 \pm 1.94 $  & $1.74 \pm 1.02 $  & 26.36 & 9.04 & 2.65\\
dw1118+13b & 18.15 & 17.78 & 0.069 & 0.047 & -12.34 & 0.341 & $25.33 \pm 0.13 $  & $15.09 \pm 1.66 $  & $1.35 \pm 0.24 $  & 26.45 & 21.5 & 3.03\\
dw1123+13 & 19.62 & 19.08 & 0.079 & 0.054 & -11.05 & 0.513 & $24.95 \pm 0.16 $  & $8.74 \pm 1.05 $  & $1.55 \pm 0.31 $  & 25.38 & 7.26 & 2.56\\
dw1127+13 & 19.76 & 18.85 & 0.093 & 0.064 & -11.28 & 0.872 & $25.51 \pm 0.13 $  & $12.85 \pm 0.95 $  & $1.72 \pm 0.52 $  & 26.10 & 11.2 & 2.75\\
\\
Field & & & & & & & & &\\
dw1013+18 & 18.02 & 17.65 & 0.106 & 0.073 & -12.50 & 0.340 & $22.67 \pm 0.10 $  & $3.29 \pm 0.38 $  & $0.76 \pm 0.04 $  & 24.16 & 7.99 & 2.60\\
dw1040+06 & 17.96 & 18.22 & 0.120 & 0.083 & -11.94 & -0.29 & $24.36 \pm 0.12 $  & $8.18 \pm 0.88 $  & $1.20 \pm 0.17 $  & 25.37 & 10.7 & 2.73\\
dw1109+18 & 17.73 & 17.18 & 0.077 & 0.054 & -12.95 & 0.523 & $23.25 \pm 0.06 $  & $7.35 \pm 0.41 $  & $1.07 \pm 0.06 $  & 24.17 & 9.94 & 2.70\\
dw1110+18 & 18.00 & 17.39 & 0.077 & 0.053 & -12.74 & 0.587 & $24.30 \pm 0.11 $  & $12.15 \pm 1.16 $  & $1.20 \pm 0.20 $  & 25.15 & 14.2 & 2.85\\

dw1130+20 & 17.53 & 17.28 & 0.068 & 0.047 & -12.84 & 0.220 & $22.63 \pm 0.14 $  & $2.75 \pm 0.58 $  & $0.61 \pm 0.05 $  & 24.49 & 10.9 & 2.74\\
dw1131+14 & 19.51 & 18.87 & 0.171 & 0.118 & -11.32 & 0.581 & $24.59 \pm 0.09 $  & $7.39 \pm 0.42 $  & $1.97 \pm 0.24 $  & 25.22 & 7.43 & 2.57\\
dw1137+16 & 17.32 & 16.77 & 0.097 & 0.067 & -13.37 & 0.523 & $24.49 \pm 0.12 $  & $14.19 \pm 1.99 $  & $0.89 \pm 0.10 $  & 25.89 & 26.6 & 3.12\\
dw1140+17 & 18.54 & 17.89 & 0.098 & 0.068 & -12.25 & 0.623 & $24.85 \pm 0.20 $  & $13.96 \pm 3.30 $  & $0.87 \pm 0.23 $  & 25.67 & 14.3 & 2.85\\
dw1145+14 & 19.86 & 19.20 & 0.147 & 0.101 & -10.97 & 0.613 & $24.02 \pm 0.12 $  & $4.94 \pm 0.49 $  & $1.39 \pm 0.17 $  & 24.65 & 4.89 & 2.39\\
dw1148+12 & 17.95 & 17.91 & 0.119 & 0.082 & -12.25 & 0.002 & $24.58 \pm 0.20 $  & $9.02 \pm 1.75 $  & $1.02 \pm 0.29 $  & 25.78 & 14.9 & 2.87\\
dw1148+16 & 17.49 & 17.34 & 0.150 & 0.104 & -12.84 & 0.109 & $22.85 \pm 0.06 $  & $3.77 \pm 0.32 $  & $0.78 \pm 0.04 $  & 24.61 & 11.3 & 2.75\\
dw1151+16 & 18.04 & 18.27 & 0.109 & 0.075 & -11.88 & -0.25 & $22.92 \pm 0.07 $  & $3.92 \pm 0.25 $  & $1.09 \pm 0.05 $  & 23.86 & 5.24 & 2.42\\
\hline\hline
\end{tabular}
\tablefoot{The quantities listed are as follows:
(1) name of candidate;
(2+3) total apparent magnitude in the $g$ and $r$ bands;
(4+5) galactic extinction in the $g$ and $r$ bands \citep{2011ApJ...737..103S};
(6) extinction corrected absolute $r$ band magnitude, using a distance modulus of $M-m=30.06$\,mag;
(7) integrated and extinction corrected $g-r$ color;
(8) S\'ersic central surface brightness in the $r$ band;
(9) S\'ersic scale length in the $r$ band;
(10) S\'ersic curvature index in the $r$ band;
(11) mean effective surface brightness in the $r$ band;
(12) effective radius in the $r$ band; 
(13) the logarithm of the effective radius in the $r$ band, converted to pc with a distance modulus of $M-m=30.06$\,mag. }
\label{table2}
\end{table*}

\section{Discussion}
\label{disc}
In the following we discuss the membership of the candidates based on their morphological parameters, the contamination of the field by nearby background galaxies, and the potential discovery of ultra diffuse galaxies (UDG).
\subsection{Membership estimation}
\begin{figure*}[ht]
\centering
\includegraphics[width=9cm]{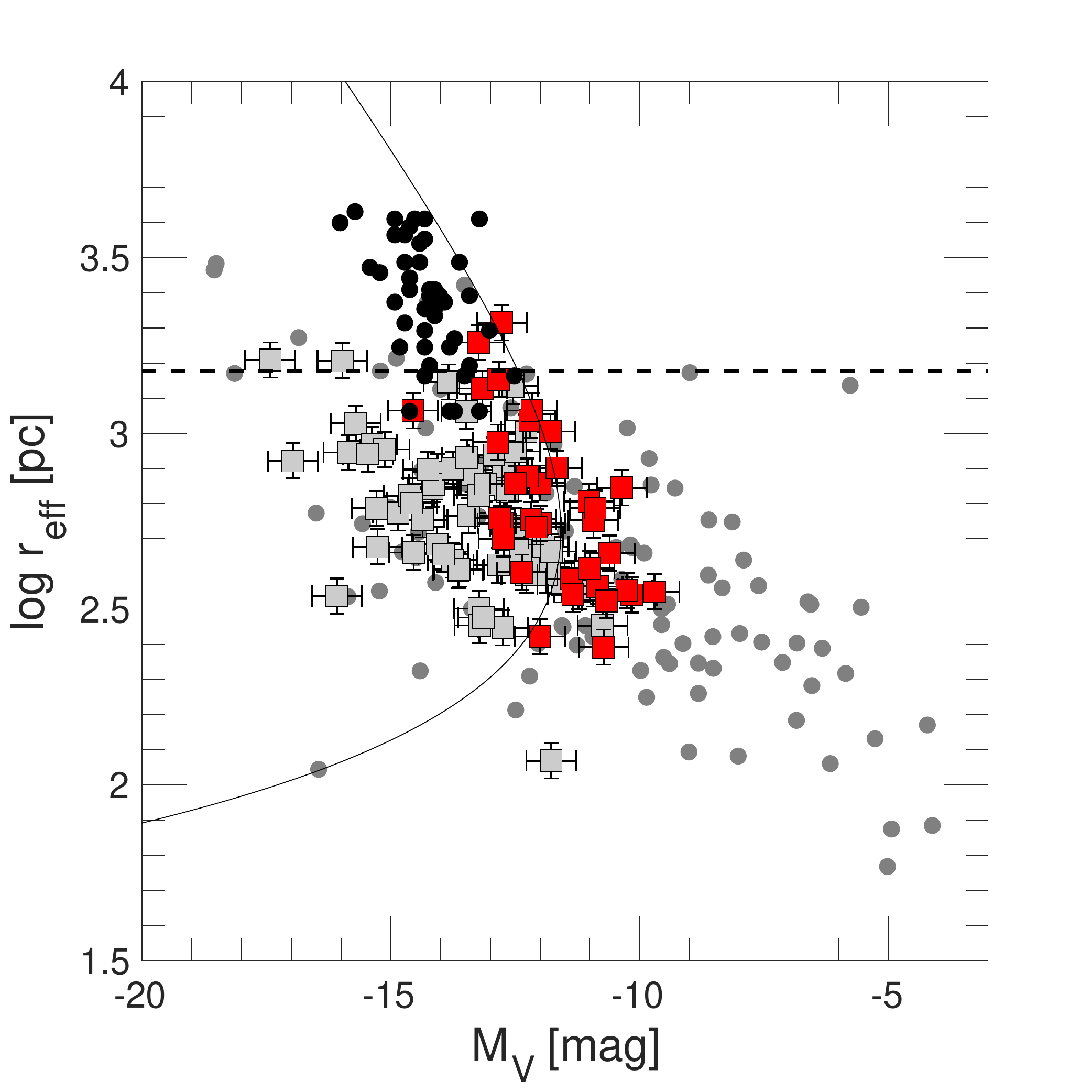}
\includegraphics[width=9cm]{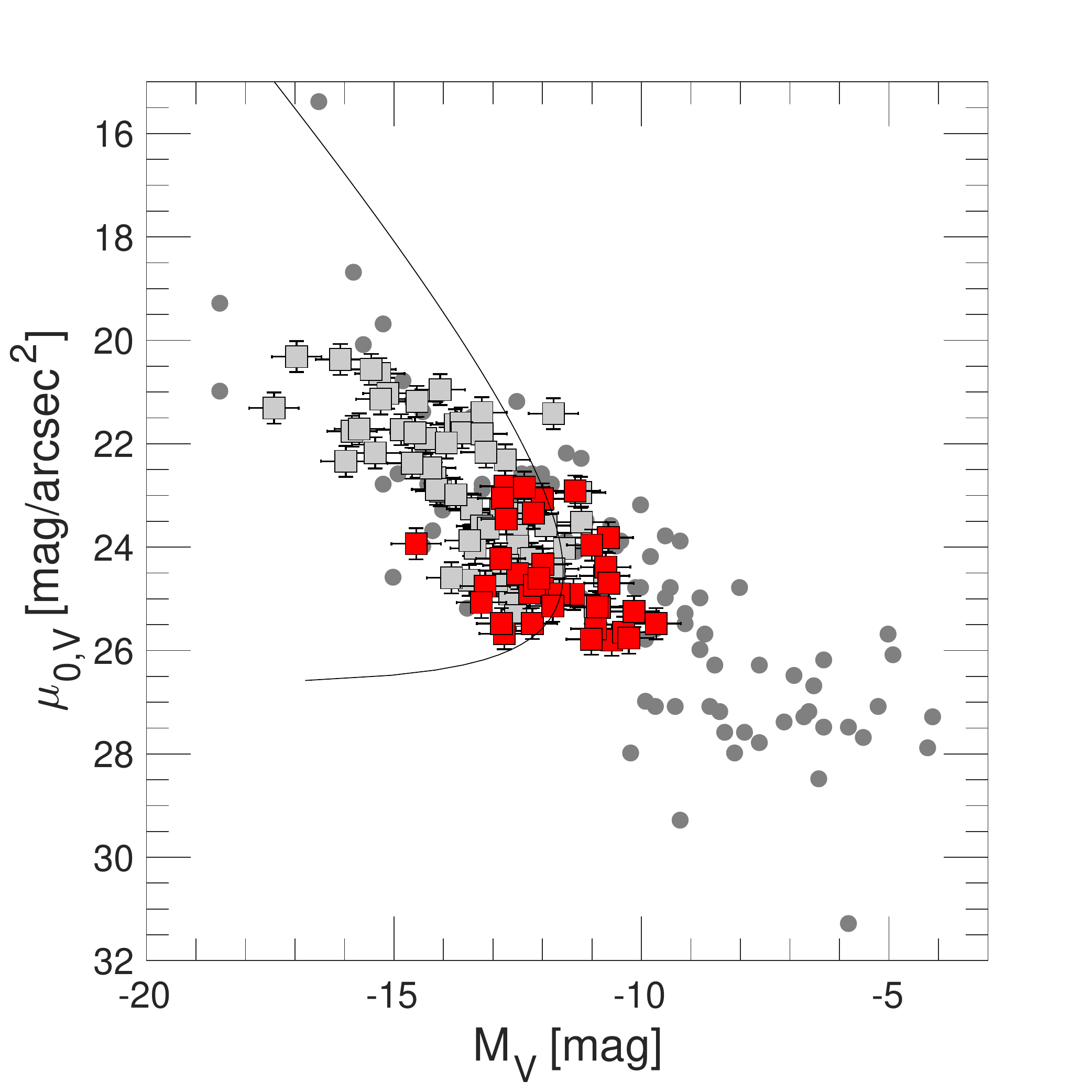}
 \caption{The ($r_{eff}$ -- $M$) and ($\mu_0$ -- $M$) scaling relations for the newly discovered dwarf candidates (red squares), previously discovered dwarf members (gray squares), and the Local Group dwarf galaxy population (gray dots). The estimated \om{conservative} completeness limit\om{, as derived in \citet{2017A&A...602A.119M},} is indicated with the line. The UDG candidates discovered in Coma \citep{2015ApJ...798L..45V} are overlaid as black dots in the ($r_{eff}$ -- $M$) diagram\om{, as well as the size cut (dashed line) of 1.5\,kpc for UDGs}. }
 \label{relations}
\end{figure*}

The standard approach to establish membership based on morphological properties is to compare the structural parameters of the candidates with known dwarf galaxies \citep[e.g.,][]{2000AJ....119..593J,2009AJ....137.3009C,2014ApJ...787L..37M,2017A&A...597A...7M,2017A&A...602A.119M}. If the objects fit into the ($\langle \mu\rangle_{eff} $ -- $M$), ($r_{eff}$ -- $M$), ($\mu_0$ -- $M$), and ($n$ -- $M$) scaling relations defined by the known dwarf galaxies in the local Universe it is reasonable to consider them as dwarf galaxy candidates. The ($\langle \mu\rangle_{eff} $ -- $M$) and ($\mu_0$ -- $M$) are especially crucial because the surface brightness is independent of the assumed distance of the object, therefore making it possible to assess the membership at a certain distance (see \citealt{2017A&A...597A...7M}, Fig.\,11 for what happens to galaxies with unreasonable distance estimates in those relations).
To transform our $gr$ photometry to the Johnson system we used the following equations \citep{SloanConv}:
$$V = g - 0.5784\cdot(g - r)_0 - 0.0038 $$
$$B = r + 1.3130\cdot(g - r)_0 + 0.2271$$
The structural parameters of the newly found dwarf candidates, together with the previously discovered Leo-I members and the Local Group dwarf population are plotted in Fig.\,\ref{relations}. 
\om{The structural parameters of the dwarf candidates fall into the relations defined by the Local Group dwarfs, thus we can assume that the candidates are indeed dwarf members of the Leo-I group.}
Additionally, we show the 44 ultra-diffuse galaxy (UDG) candidates in the Coma Cluster discovered by \citet{2015ApJ...798L..45V} (only $g$ band photometry is given, therefore we assume a color index of $(g - r) = 0.6$\,mag to transform them into $V$-band magnitudes). UDGs have typically an effective radius larger then $r_{eff}>1.5$\,kpc and a fainter central surface brightness than $\mu_g>24.0$\,mag\,arcsec$^{-2}$ \citep{2015ApJ...798L..45V}.

\begin{figure}[H]
\centering
\includegraphics[width=9cm]{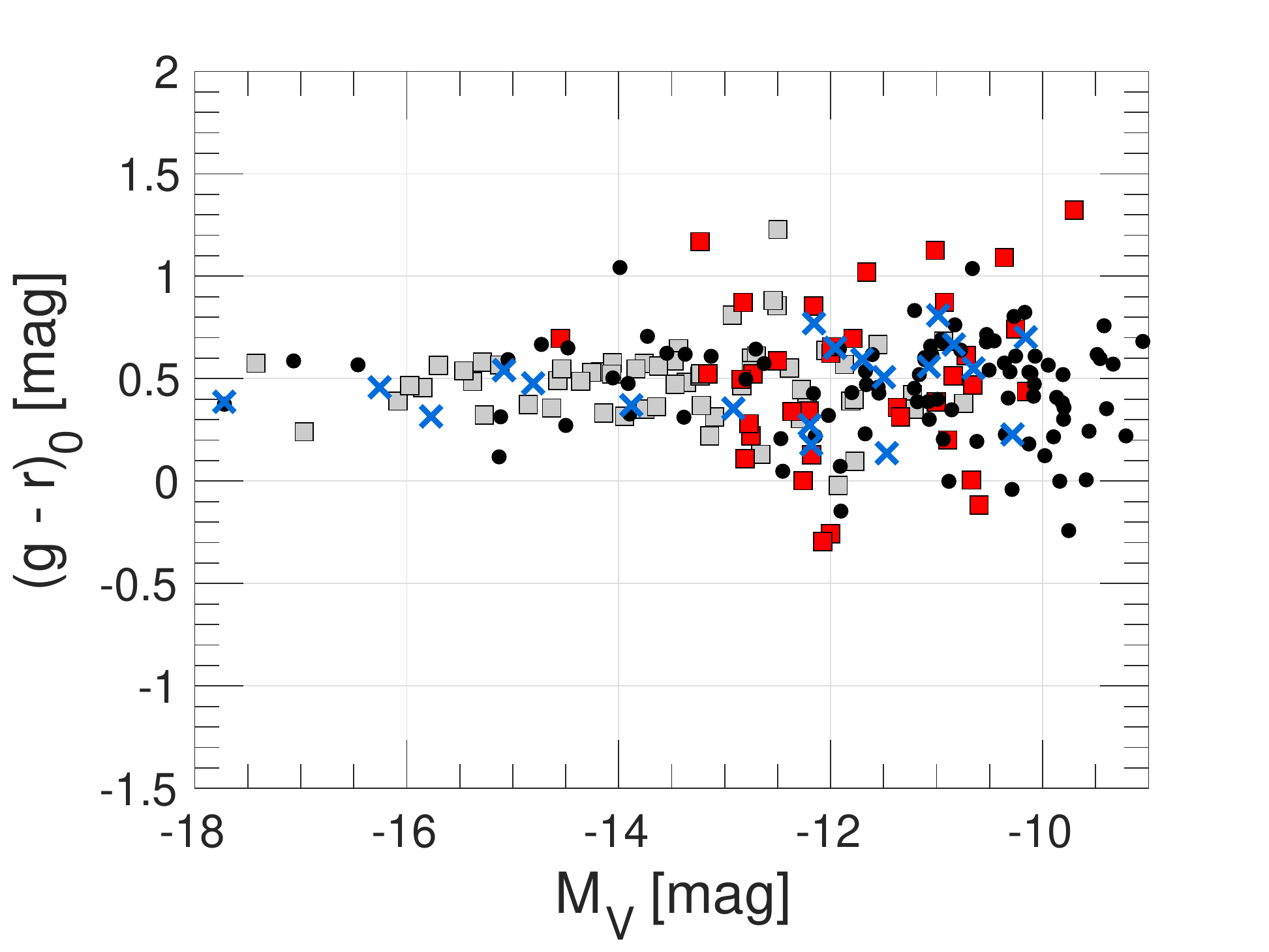}
 \caption{Color-magnitude relation for the previously known Leo-I dwarf members (gray squares), the new Leo-I members (red squares), the Centaurus group members \citep[black dots,][]{2015A&A...583A..79M,2017A&A...597A...7M},  and the M\,101 group members \citep[blue crosses,][]{2017A&A...602A.119M}. Both early and late type dwarf galaxies were considered. }
 \label{colors}
\end{figure}

Dwarf galaxies can also be characterized by their color using the color-magnitude relation \citep[e.g.][]{2008AJ....135..380L,2017A&A...608A.142V}. Here we compare the $(g-r)_0$ colors of the Leo-I group dwarfs with other well studied systems in the LV where $gr$ photometry is available, i.e. the Centaurus group \citep{2015A&A...583A..79M,2017A&A...597A...7M} and the M101 group complex \citep{2017A&A...602A.119M}. The calculated mean $(g-r)_0$ color and standard deviation for the three group populations are: $(g-r)_{0,Leo-I}=0.491\pm0.282$\,mag, $(g-r)_{0,Cen\,A}=0.463\pm0.258$\,mag, and $(g-r)_{0,M\,101}=0.472\pm0.190$\,mag. In Fig.\,\ref{colors} we show the color distribution as a function of total absolute $V$-magnitude for these different groups. \om{The dwarfs in the different galaxy groups follow a similar distribution in their colors. We note that the extreme blueish colors ($g-r<0$) of some objects -- uncommon for dwarf galaxies -- as well as the scatter at the faint-end of the scale, can arise from the photometric uncertainty.}

In the following we discuss some individual candidates which have apparently interesting features.\\
\textbf{dw1037+09:} This candidate has several knots within and around the galaxy, which could either be bright giant stars or globular clusters (GC).\\
\textbf{dw1110+18:} As for dw1037+09 there are several knots sprinkled among the object which could be bright giant stars or GCs. \\
%\textbf{dw1113+14:} While this galaxy has a smooth surface brightness profile in the outer region, it has a bright feature with some offset from the center. To check if this candidate is not an transient object we inspected the digitized sky survey (DSS) data at given position. Indeed, we find a corresponding high surface brightness feature at the same coordinates within an image of a different epoch, which rules out the possibility for a transient object. This candidate could be a blue compact dwarf \citep{1996A&AS..120..207P}.\\
\textbf{dw1130+20:} This galaxy has some bright knots, which could correspond to HII regions.

\begin{figure}%[H]
\centering
\includegraphics[width=9cm]{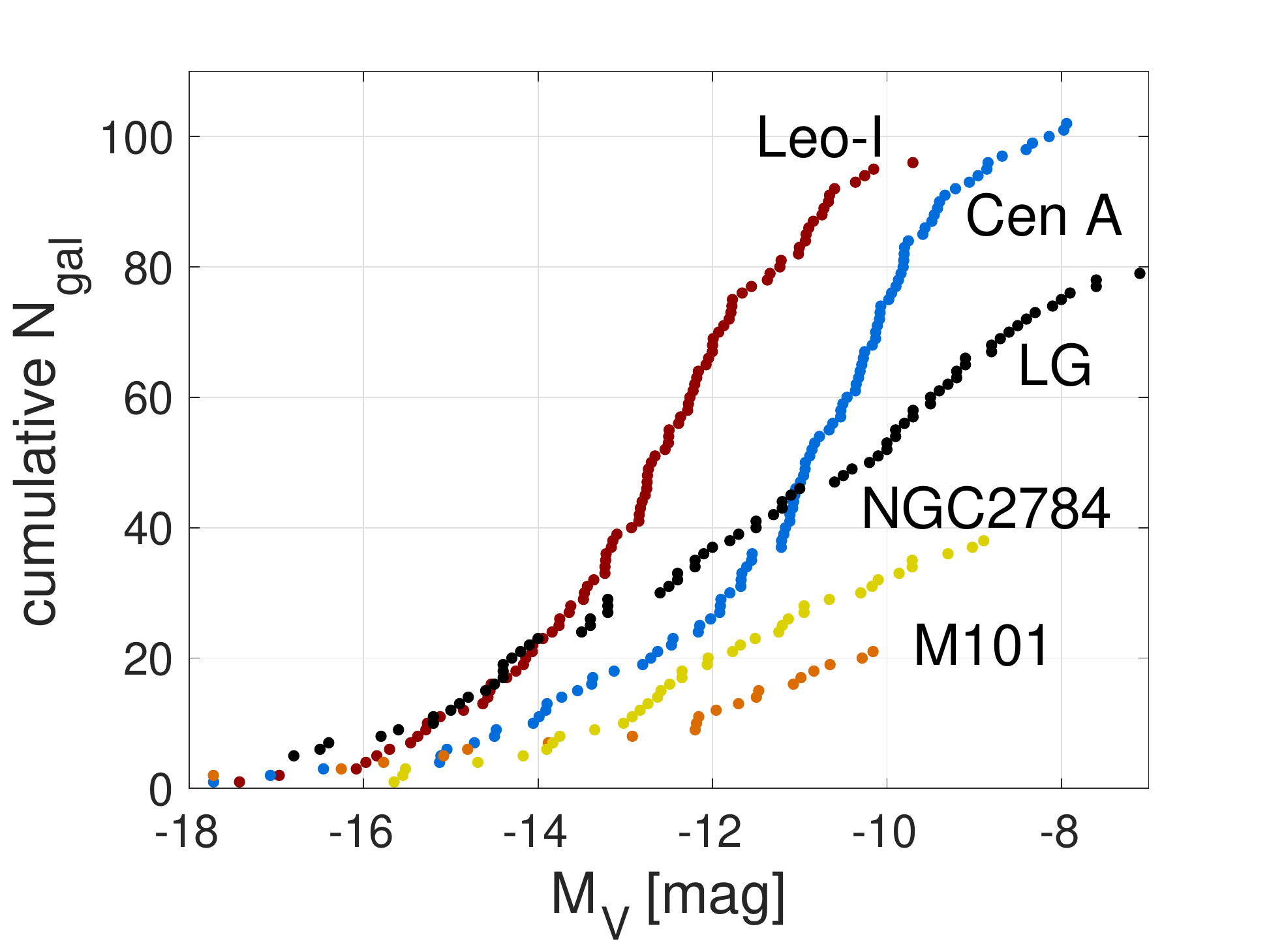}
 \caption{Cumulative galaxy luminosity functions 
 %\citep{2002MNRAS.335..712T} 
 for different galaxy groups in the Local Volume. Data taken from: Leo-I (this work), Centaurus group \citep{2015A&A...583A..79M,2017A&A...597A...7M}, LG \citep{2012AJ....144....4M}, NGC\,2784 group \citep{2017ApJ...848...19P}, and M\,101 group \citep{1999A&AS..137..337B,2017A&A...602A.119M}.
 }
 \label{LF}
\end{figure}

Under the assumption that all candidates are members of the Leo-I group, we can determine the galaxy luminosity function (see Fig.\,\ref{LF})
%\citep[LF,][]{2002MNRAS.335..712T}
and compare it to other nearby galaxy group environments, i.e. the Centaurus group \citep{2015A&A...583A..79M,2017A&A...597A...7M}, the LG \citep{2012AJ....144....4M}, the M101 group \citep{1999A&AS..137..337B,2017A&A...602A.119M}, and the NGC2784 group \citep{2017ApJ...848...19P}. Among these five groups, the Leo-I group is the richest galaxy aggregate with approximately 100 galaxies up to an absolute magnitude of $M_V=-10$, this is, if all candidates are confirmed as members. \om{The Leo-I group has approximately twice as many dwarfs as the LG and a steeper faint-end slope of the LF, comparable to the one of Cen\,A  The M\,101 and NGC\,2784 groups have shallower faint-end slopes. This indicates that galaxy groups with massive hosts have steeper faint-ends of the LF. While the  faint-end slopes of Leo-I and Cen\,A are comparable, the Leo-I group contains more brighter galaxies in the range of -16 to -14\,mag in $V$-bands, making it more rich (up to $M_V=-10$). In this range (-16 to -14\,$V$ mag), the LF of Leo-I is comparable to the one of the LG. 
}

\subsection{Background contamination}
\label{back}
One fundamental challenge when searching for new dwarf galaxies is given by the fact that survey fields are almost always contaminated by galaxy groups in the background. A prime example for such a confusion is the massive elliptical galaxy NGC\,5485 
with its many dwarf companions \citep{2011MNRAS.412.2498M} situated $\approx$20\,Mpc behind the Local Volume galaxy M\,101 (7\,Mpc, \citealt{2015MNRAS.449.1171N}).  Fig.\,8 in \citet{2016ApJ...833..168M} shows M\,101, the background elliptical NGC\,5485, and former M\,101 dwarf candidates \citep{2014ApJ...787L..37M} that  actually belong to the background galaxy population. Out of the seven reported dwarf candidates by \citet{2014ApJ...787L..37M} only three were confirmed to be M\,101 members with HST follow-up observations \citep{2017ApJ...837..136D}. Recently, more new dwarf candidates were reported around M\,101 \citep{2017ApJ...850..109B,2017A&A...602A.119M}, now awaiting confirmation as members by means of distance or velocity measurements.  Some will potentially be associated to the background elliptical NGC\,5485.\\
The possibility of contamination prompted us to study the background of the Leo-I group in more detail. In \citet{2017A&A...597A...7M} we used the Cosmicflows-2 catalog \citep{2013AJ....146...86T} to determine the background contamination of the Centaurus group. Here we query the Cosmicflows-2 catalog for bright galaxies with absolute magnitudes $M_B$<-19 and with radial velocities $v_{rad}$<2000\,km\,s$^{-1}$ within our survey footprint. Excluding the Leo-I galaxies this search resulted in 24 bright host galaxies potentially contaminating our survey.

\begin{figure*}[ht]
\centering
\hspace*{-0.7cm}
\includegraphics[width=20cm]{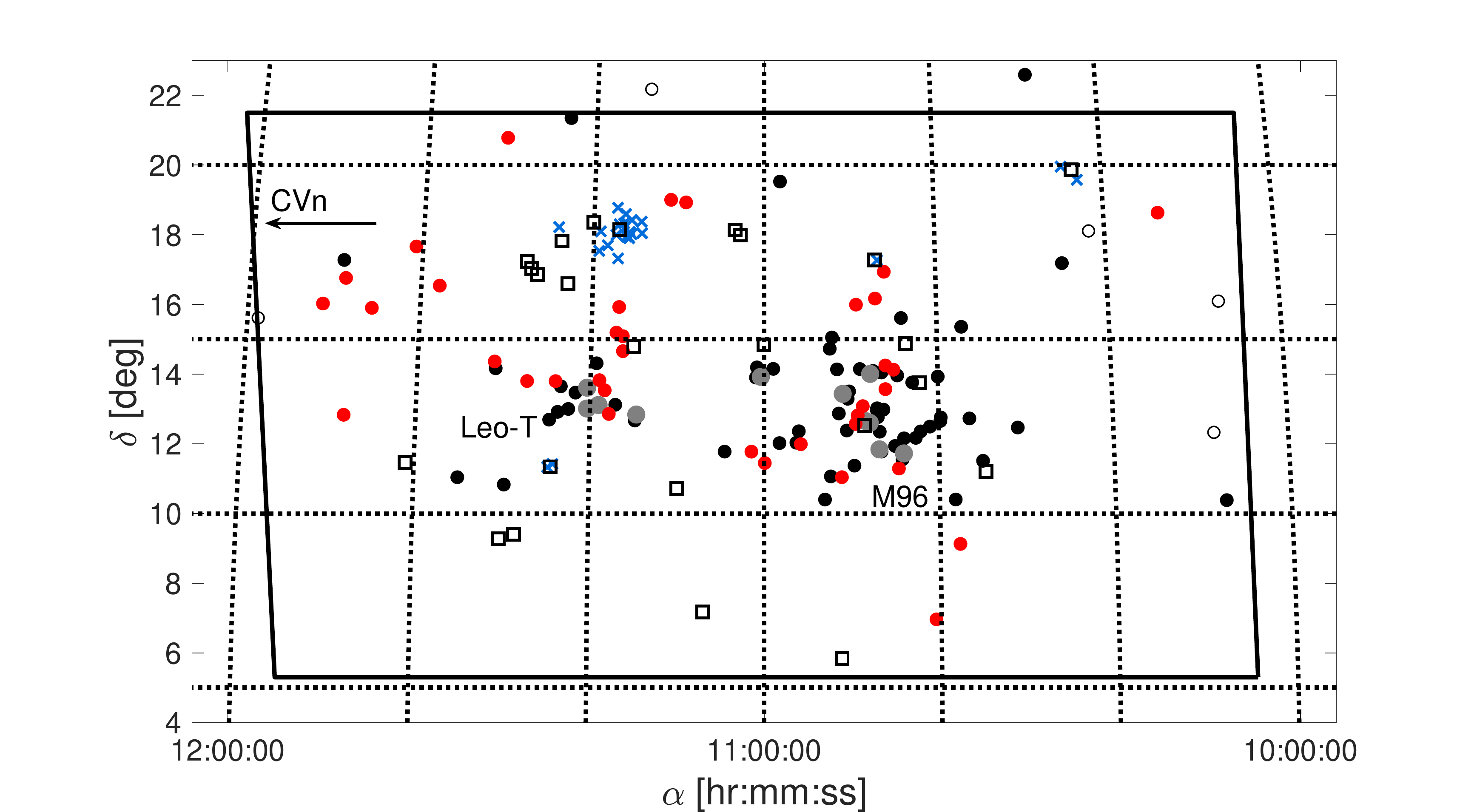}
 \caption{\om{Same as Fig.\,\ref{fieldImage} but with background host galaxies (black squares) and background dwarf galaxies (blue crosses) which are clustered around these hosts , however would be considered Leo-I dwarfs due to their morphology if not for their apparent closeness to these background hosts. See Section \ref{back} for details.} 
 }
 \label{fieldImageBack}
\end{figure*}

To test how these background galaxies will pollute our detections we surveyed for dwarf galaxies within 300\,kpc of each such host (approximatively the virial radius) with the same methods as used in our search for Leo-I dwarfs, but without removing candidates which are near to a background galaxy. Essentially, we search for the candidates we rejected as background sources. In Table\,\ref{table:app} we compiled the coordinates for the objects which would be considered as dwarf candidates based on their morphology. In total we found 26 additional dwarf candidates, of which 20 are clustered around NGC\,3607 at a distance of $\sim20$\,Mpc. This indicates that (a) it is not feasible to include every object in the survey footprint as Leo-I dwarf, and (b) that there probably will be some confusion between foreground and background, either by rejecting a foreground dwarf or including a background dwarf.\\
\om{Some Leo-I dwarf candidates are both near to a background host and a Leo-I host. In such a case we added a note the Table\,\ref{table:1}. To the north to the Leo Triplet there are four Leo-I candidates (dw1116+14, dw1116+15a, dw1116+15b, and dw1117+15) clustered around NGC\,3596 (15\,Mpc). See Fig.\,\ref{fieldImageBack} for the distribution of the background dwarf galaxies. Distance and velocity measurements will be crucial to distinguish their memberships.  Until then, the faint-end of the LF will be affected by these uncertain cases.}

\begin{table}[ht]
\centering
\setlength{\tabcolsep}{3pt}
\caption{Coordinates of the possible background dwarf galaxy in our survey footprint around bright host galaxies with v<2000\,km\,s$^{2}$.}
\label{table:app}
\begin{tabular}{lcc}
\hline\hline
& $\alpha$ & $\delta$  \\ 
Name & (J2000) & (J2000) \\ 
\hline \\[-2mm]
NGC\,3227\_1&10:22:53 &$+$19:34:36 \\%411
NGC\,3227\_2 &10:24:43 &$+$19:57:16\\%440
NGC\,3227\_3& 10:25:50&$+$19:43:22\\%441
%NGC\,3377\_1& 10:45:56 &$+$14:13:36 \\%272 dw1045+16
%NGC\,3377\_1& 10:46:24& $+$14:01:26\\%272 KK93
%NGC\,3338\_1& 10:39:55 &$+$13:54:28 \\%241 LeG6
NGC\,3666\_1&11:24:45 &$+$11:20:04 \\%195
NGC\,3666\_2&11:24:10 &$+$11:25:12 \\%195
%NGC\,3666\_3&11:24:34 &$+$12:40:20 \\%224 AGC 215415
NGC\,3370\_1&10:46:47 &$+$17:16:18\\%359
%NGC\,3370\_2& 10:45:56&$+$16:54:57 \\%359 dw1045+14b
%NGC\,3370\_3&10:47:01 &$+$16:08:47 \\%330 &dw1047+16
NGC\,3607\_1 & 11:14:22&$+$18:02:38\\%395
NGC\,3607\_2 &11:14:26 &$+$18:22:30 \\%395
NGC\,3607\_3 &11:15:35 &$+$18:25:21\\%395
NGC\,3607\_4 & 11:15:36& $+$18:01:04\\%395
NGC\,3607\_5&11:15:48 & $+$18:04:40\\%395
NGC\,3607\_6&11:15:52 &$+$17:54:04 \\%395
NGC\,3607\_7&11:15:57 & $+$17:56:25\\%395
NGC\,3607\_8&11:16:11 &$+$17:57:04 \\%395
NGC\,3607\_9&11:16:18 &$+$18:35:39 \\%395
NGC\,3607\_10&11:16:28 &$+$18:11:35 \\%395
NGC\,3607\_11& 11:16:30&$+$18:19:27 \\%395
NGC\,3607\_12& 11:17:01 & $+$18:18:07\\%396
NGC\,3607\_13& 11:17:07&$+$17:19:09 \\%367
NGC\,3607\_14&11:17:16 &$+$18:46:27 \\%425
NGC\,3607\_15&11:17:22 & $+$17:59:50\\%396
NGC\,3607\_16& 11:18:21&$+$17:41:50 \\%396
NGC\,3607\_17&11:19:13 &$+$18:05:47 \\%396
NGC\,3607\_18&11:19:21 &$+$17:32:09 \\%367
NGC\,3607\_19&11:24:08 &$+$18:13:16\\%397
NGC\,3607\_20&11:31:01 &$+$15:54:48 \\%341
\hline\hline
\end{tabular}
\end{table}

%\subsection{Tidal alignment}
%\citep{2014ApJ...786..144N}.
\subsection{UDG candidates}
Originally discovered by \citet{1984AJ.....89..919S} and described as ``a new type of very large diameter (10,000 pc), low central surface brightness (>25 B mag/arcsec$^2$) galaxy, that comes in both early (i.e., dE) and late (i.e., Im V) types", this class of galaxies has now been renamed as ultra diffuse galaxies \citep{2015ApJ...798L..45V} and was found in many different environments \citep{2016A&A...590A..20V}, i.e. in clusters \citep{2015ApJ...798L..45V,2015ApJ...807L...2K}, and in groups \citep{2016ApJ...833..168M}. Different possible formation scenarios have been proposed \citep[e.g.][]{2016MNRAS.459L..51A,2017MNRAS.466L...1D} and are under intense debate. \citet{2015ApJ...798L..45V} suggested to classify dwarf galaxies with $r_{eff}>1.5$\,kpc and a fainter central surface brightness than $\mu_g>24.0$\,mag\,arcsec$^{-2}$ as UDG, however, this boundary is rather arbitrary and should be considered more as a guideline.\\ Studying the properties of the Leo-I members we consider dw1055+11, dw1117+15, dw1051+11, KK\,96, and ACG\,215415 as UDG candidates. With $r_{eff}=1.3$\,kpc dw1137+16 is still considerably large and could be an UDG type. Better photometry is needed to derive the structural parameters more accurately.
However, we note that if these objects are more in the foreground (e.g. in the Canes Venatici-I cloud), they would be closer to our point of view and therefore would have smaller intrinsic sizes, making them common-sized dwarf galaxies.\\
\om{The UDG candidates are distributed in the outskirts of the aggregates and not in the central parts of the group. This is similar to what is found in galaxy clusters: in galaxy clusters the UDG density drops nearly to zero in the central regions because they cannot survive the tidal forces inflicted on them \citep{2016A&A...590A..20V}. We note that it is not feasible to assess the UDG distribution in Leo-I with only 5-6 candidates.}
\section{Conclusion}
\label{concl}
We have surveyed 500 square degrees of $gr$ images taken from SDSS within the extended region of the Leo-I group and found 36 new dwarf galaxy candidates. For every known member and new candidate we derived surface brightness photometry. Based on a comparison of their structural properties with other known dwarf galaxies in the nearby universe and their morphology we consider these candidates member of the Leo-I group, either lying in the vicinity of the M\,96 subgroup, the Leo Triplet, or in the nearby field.
To confirm their membership follow-ups are required to either measure their radial velocity, their distances, or both. Some of the candidates are exceptionally large with low surface brightness, a characteristic of ultra diffuse galaxies. If these UDGs are confirmed to be Leo-I members, those would be some of the closest UDGs from Earth and valuable targets to improve our understanding of this type of galaxies.

\begin{acknowledgements}
OM and BB are grateful to the Swiss National Science Foundation for financial support. HJ acknowledges the support of the Australian Research Council through Discovery Project DP150100862. 
\end{acknowledgements}

% \begin{figure*}[ht]
% \centering
% \includegraphics[width=17cm]{LeoT.png}
%  \caption{Leo Triplett.}
%  \label{leoT}
% \end{figure*}
\begin{sidewaystable*}
\caption{Photometric and structural parameters of the previously known Leo-I members.}
\small
\centering
\setlength{\tabcolsep}{3pt}
\begin{tabular}{lccccrrrrcrccrrlll}
\hline\hline 
Name & && ${g_{tot}}$ & ${r_{tot}}$  & $A_g$ & $A_r$ & $M_{r}$ & $(g-r)_{0,tot}$ & $\mu_{0,r}$ & $r_{0,r}$ & $n_r$ & $\langle\mu\rangle_{eff,r}$ &  $r_{eff,r}$ & $\log r_{eff,r}$ & Ref & $D$ & $v$\\ 
 & && mag & mag & mag & mag &  mag & mag & mag arcsec$^{-2}$ &arcsec & & mag arcsec$^{-2}$ &arcsec & $\log$ pc & & Mpc & km\,s$^{-1}$ \\ 
 (1)& && (2) & (3) &  (4) & (5) &  (6) & (7) &(8) & (9) & (10) & (11) & (12) & (13) & (14) & (15) & (16)\\ 
\hline \\
%UGC06881 &11:54:45&$+$20:03:25& 15.60 & 15.11 & 0.113 & 0.078 & -16.03 & 0.456 & $21.58 \pm 0.03 $  & $5.32 \pm 0.22 $  & $0.70 \pm 0.01 $  & 23.41 & 18.1 & 3.16 & & 16.40 (i) & 601 (I)\\ D=17Mpc
M\,96 subgroup & & & & & & & & &\\
AGC\,205156 & 10:30:53 & $+$12:26:48 & 18.65 & 18.22 & 0.102 & 0.071 & -11.94 & 0.398 & $21.22 \pm 0.12 $  & $1.80 \pm 0.17 $  & $1.12 \pm 0.08 $  & 22.13 & 2.41 & 2.08 & (AA) & 10.43 (LV) & 915 (AA) \\
AGC\,202248& 10:34:56 & $+$11:29:31 & 17.17 & 16.92 & 0.103 & 0.071 & -13.23 & 0.220 & $22.10 \pm 0.09 $  & $5.15 \pm 0.36 $  & $1.28 \pm 0.07 $  & 22.87 & 6.17 & 2.49 & (AA) &  & 1177 (AA) \\
NGC\,3299 & 10:36:24 & $+$12:42:25 & 13.08 & 12.48 & 0.082 & 0.057 & -17.66 & 0.575 & $21.06 \pm 0.02 $  & $22.48 \pm 0.55 $  & $0.77 \pm 0.02 $  & 22.09 & 33.3 & 3.22 &  &  & 604 (AA)\\
AGC\,205165 & 10:37:05&$+$15:20:13 & 15.92 & 15.39 & 0.123 & 0.085 & -14.77 & 0.490 & $21.85 \pm 0.02 $  & $7.83 \pm 0.16 $  & $0.97 \pm 0.01 $  & 22.97 & 13.0 & 2.81 & (AA) &  & 724 (AA)\\
AGC\,200499 & 10:38:08 & $+$10:22:52 & 14.32 & 13.90 & 0.094 & 0.065 & -16.24 & 0.390 & $20.25 \pm 0.05 $  & $5.35 \pm 0.19 $  & $0.98 \pm 0.02 $  & 20.15 & 7.10 & 2.55 & (AA) &  & 1175 (AA) \\
LeG\,04 &10:39:40&$+$12:44:07& 18.10 & 17.53 & 0.090 & 0.062 & -12.61 & 0.551 & $23.10 \pm 0.11 $  & $4.97 \pm 0.59 $  & $0.90 \pm 0.07 $  & 24.43 & 9.58 & 2.68  & (LV) & & \\
FS\,01 (LeG\,05) & 10:39:43	&$+$12:38:04 &16.74 & 16.14 & 0.084 & 0.058 & -13.99 & 0.573 & $21.50 \pm 0.03 $  & $4.04 \pm 0.15 $  & $0.82 \pm 0.02 $  & 22.91 & 8.99 & 2.65  & (FS) & & 780 (AA) \\
LeG\,06 &10:39:56&$+$13:54:33 & 17.12 & 16.60 & 0.117 & 0.081 & -13.56 & 0.481 & $23.90 \pm 0.03 $  & $14.32 \pm 0.36 $  & $1.50 \pm 0.06 $  & 24.59 & 15.7 & 2.90 & (LV) & & 1007 (AA) \\
UGC\,05812 &10:40:56&$+$12:28:21 & 15.07 & 14.56 & 0.080 & 0.056 & -15.58 & 0.486 & $21.97 \pm 0.02 $  & $13.08 \pm 0.22 $  & $1.17 \pm 0.03 $  & 22.97 & 19.1 & 2.98 & & & 1008 (AA)\\
FS\,04 &10:42:00&$+$12:20:05 &15.53 & 15.13 & 0.084 & 0.058 & -15.00 & 0.373 & $21.55 \pm 0.06 $  & $7.96 \pm 0.42 $  & $1.01 \pm 0.03 $  & 22.57 & 12.2 & 2.79 & (FS) & & 772 (AA)\\
LeG\,09 & 10:42:34&$+$12:09:02&17.03 & 16.41 & 0.082 & 0.057 & -13.72 & 0.587 & $24.38 \pm 0.08 $  & $18.37 \pm 1.46 $  & $1.05 \pm 0.09 $  & 25.29 & 23.8 & 3.07 & (LV) & & \\
LeG\,10 &10:43:55&$+$12:08:00 & 19.16 & 18.78 & 0.089 & 0.061 & -11.35 & 0.352 & $23.85 \pm 0.45 $  & $3.86 \pm 2.31 $  & $0.71 \pm 0.22 $  & 25.10 & 7.30 & 2.56 & (LV) & & \\
LeG\,11 & 10:44:02&$+$15:35:21 & 18.18 & 17.78 & 0.104 & 0.072 & -12.37 & 0.366 & $24.13 \pm 0.08 $  & $10.22 \pm 0.57 $  & $1.63 \pm 0.18 $  & 24.53 & 8.92 & 2.65 & (LV) & & \\
LeG\,12 & 10:44:07&$+$11:32:03 &19.20 & 18.75 & 0.098 & 0.068 & -11.40 & 0.421 & $23.60 \pm 0.75 $  & $2.07 \pm 2.36 $  & $0.57 \pm 0.25 $  & 25.07 & 7.33 & 2.56 & (LV) & & \\
AGC\,205445 & 10:44:35&$+$13:56:22 &16.11 & 15.59 & 0.103 & 0.072 & -14.56 & 0.487 & $21.71 \pm 0.03 $  & $6.48 \pm 0.20 $  & $0.92 \pm 0.02 $  & 22.93 & 11.7 & 2.77 & (AA) & & 633 (LV) \\
FS\,09 (LeG\,13) & 10:44:57&$+$11:55:00 & 17.59 & 17.10 & 0.073 & 0.050 & -13.03 & 0.466 & $23.13 \pm 0.11 $  & $8.16 \pm 0.67 $  & $1.39 \pm 0.15 $  & 23.79 & 8.70 & 2.64 & (FS) & & 871 (AA)\\
FS\,13 (LeG\,14) &10:46:14&$+$12:57:38 &18.49 & 17.83 & 0.081 & 0.056 & -12.31 & 0.638 & $24.25 \pm 0.05 $  & $9.57 \pm 0.34 $  & $1.70 \pm 0.13 $  & 24.92 & 10.4 & 2.72 & (FS) & & \\
FS\,14 (KK\,93) & 10:46:25&$+$14:01:25 &16.99 & 16.48 & 0.098 & 0.068 & -13.66 & 0.471 & $23.59 \pm 0.05 $  & $11.06 \pm 0.55 $  & $1.00 \pm 0.04 $  & 24.71 & 17.5 & 2.94 & (FS)& & \\
FS\,15 (LeG\,16) & 10:46:30&$+$11:45:21 &18.63 & 18.03 & 0.085 & 0.058 & -12.10 & 0.569 & $24.35 \pm 0.14 $  & $10.64 \pm 0.75 $  & $2.00 \pm 0.49 $  & 24.59 & 8.16 & 2.61 & (FS) & & \\
FS\,17 (LeG\,17) & 10:46:41&$+$12:19:37 &16.44 & 15.84 & 0.080 & 0.055 & -14.30 & 0.578 & $22.62 \pm 0.03 $  & $8.74 \pm 0.30 $  & $1.02 \pm 0.03 $  & 23.86 & 16.0 & 2.90 & (FS) & & 1030 (S+) \\
LeG\,18 &10:46:53&$+$12:44:26& 18.05 & 17.72 & 0.073 & 0.051 & -12.40 & 0.304 & $24.44 \pm 0.52 $  & $4.65 \pm 4.62 $  & $0.47 \pm 0.16 $  & 26.31 & 20.8 & 3.02 & (TT) & & 636 (AA) \\
FS\,20 (LeG\,19) &10:46:55&$+$12:47:19 &18.37 & 17.12 & 0.074 & 0.052 & -13.01 & 1.228 & $23.67 \pm 0.11 $  & $7.00 \pm 0.73 $  & $1.11 \pm 0.11 $  & 25.43 & 18.3 & 2.96 & (FS) & & \\
FS\,21 (KK94) & 10:46:57&$+$12:59:54 & 17.72 & 16.88 & 0.098 & 0.068 & -13.26 & 0.809 & $24.47 \pm 0.05 $  & $17.14 \pm 0.70 $  & $1.37 \pm 0.09 $  & 25.14 & 17.9 & 2.95 & (FS) & & 832 (H+)\\
Le\,G21 &10:47:01&$+$12:57:39& 18.60 & 18.18 & 0.096 & 0.066 & -11.96 & 0.391 & $24.46 \pm 0.07 $  & $10.03 \pm 0.46 $  & $1.86 \pm 0.23 $  & 24.90 & 8.81 & 2.64 & (TT) & & 843 (AA) \\
DDO\,088 & 10:47:22&$+$14:04:15 &13.85 & 13.35 & 0.114 & 0.079 & -16.16 & 0.465 & $22.07 \pm 0.02 $  & $25.52 \pm 0.36 $  & $1.14 \pm 0.02 $  & 22.95 & 33.1 & 3.09 & & 7.73 (LV) & 573 (H+) \\
CGCG\,066-026 &10:48:54&$+$14:07:28 &15.27 & 14.64 & 0.132 & 0.092 & -15.52 & 0.580 & $20.35 \pm 0.02 $  & $3.76 \pm 0.09 $  & $0.73 \pm 0.01 $  & 22.15 & 12.6 & 2.80 & & & 541 (LV) \\ %=PGC032348 
FS\,40 (LeG\,22) &10:49:37&$+$11:21:06 &17.79 & 17.16 & 0.102 & 0.071 & -12.99 & 0.599 & $24.40 \pm 0.09 $  & $13.55 \pm 1.03 $  & $1.27 \pm 0.14 $  & 25.20 & 16.2 & 2.91 & (FS) & & \\
LeG\,23 & 10:50:09&$+$13:29:02&19.66 & 19.25 & 0.104 & 0.072 & -10.89 & 0.379 & $24.37 \pm 0.19 $  & $5.35 \pm 0.76 $  & $1.44 \pm 0.31 $  & 25.09 & 5.85 & 2.47 & (LV) & & \\
UGC\,05944 &10:50:19&$+$13:16:18 &14.94 & 14.35 & 0.099 & 0.068 & -15.93 & 0.565 & $21.50 \pm 0.01 $  & $8.49 \pm 0.11 $  & $0.79 \pm 0.02 $  & 23.06 & 22.0 & 3.07 & & 11.07 (R+) & 1073 (LV) \\
KK\,96 &10:50:27&$+$12:21:34 &16.65 & 16.07 & 0.084 & 0.058 & -14.06 & 0.547 & $24.32 \pm 0.06 $  & $19.46 \pm 1.24 $  & $1.05 \pm 0.09 $  & 25.37 & 28.8 & 3.16 & (LV) & & \\
LeG\,26 & 10:51:21&$+$12:50:56&16.30 & 15.75 & 0.079 & 0.054 & -14.38 & 0.534 & $22.48 \pm 0.03 $  & $9.52 \pm 0.25 $  & $1.05 \pm 0.02 $  & 23.52 & 14.3 & 2.85 & (LV) & & 630 (LV)\\
AGC\,205540& 10:51:31&$+$14:06:53 & 17.75 & 17.18 & 0.107 & 0.074 & -12.97 & 0.540 & $22.06 \pm 0.04 $  & $4.27 \pm 0.15 $  & $1.24 \pm 0.04 $  & 22.99 & 5.77 & 2.46 & (AA) & & 832 (LV)\\
AGC\,205544 & 10:52:05&$+$15:01:50 & 16.86 & 16.27 & 0.072 & 0.050 & -13.85 & 0.561 & $21.53 \pm 0.02 $  & $3.94 \pm 0.11 $  & $0.85 \pm 0.01 $  & 22.91 & 8.48 & 2.63 & (AA) & & 828 (LV)\\
AGC\,202456 & 10:52:19&$+$11:02:35&15.94 & 15.37 & 0.078 & 0.054 & -14.76 & 0.547 & $20.95 \pm 0.02 $  & $4.88 \pm 0.09 $  & $0.90 \pm 0.01 $  & 22.24 & 9.44 & 2.67 & (AA) & &824 (LV) \\
LeG\,27 & 10:52:20&$+$14:42:26&18.15 & 17.27 & 0.073 & 0.050 & -12.86 & 0.856 & $23.40 \pm 0.05 $  & $7.29 \pm 0.33 $  & $1.38 \pm 0.08 $  & 24.52 & 11.2 & 2.75 & (LV) & & \\
LeG\,28 &10:53:01&$+$10:22:43 &17.11 & 16.43 & 0.083 & 0.057 & -13.70 & 0.646 & $23.03 \pm 0.04 $  & $10.15 \pm 0.34 $  & $1.17 \pm 0.04 $  & 23.83 & 12.0 & 2.78 & (LV) & & \\
LSBCD\,640-12 &10:55:56&$+$12:20:22 &17.21 & 16.68 & 0.060 & 0.042 & -13.44 & 0.512 & $23.51 \pm 0.08 $  & $11.14 \pm 0.81 $  & $1.15 \pm 0.09 $  & 24.37 & 13.7 & 2.84 & (S+) & & 847 (AA) \\
LSBCD\,640-13 &10:56:14&$+$12:00:35 & 16.20 & 15.85 & 0.065 & 0.045 & -14.27 & 0.332 & $22.79 \pm 0.03 $  & $11.42 \pm 0.26 $  & $1.59 \pm 0.06 $  & 23.69 & 14.7 & 2.87 & (S+) & & 989 (S+) \\
LSBCD\,640-14 &10:58:10&$+$11:59:53& 17.79 & 17.16 & 0.055 & 0.038 & -12.95 & 0.611 & $24.13 \pm 0.06 $  & $11.72 \pm 0.48 $  & $1.68 \pm 0.17 $  & 24.95 & 14.3 & 2.86 & (S+) & & \\
AGC\,205278 &10:58:52&$+$14:07:47& 16.66 & 16.32 & 0.060 & 0.042 & -14.07 & 0.317 & $21.87 \pm 0.03 $  & $4.62 \pm 0.14 $  & $0.87 \pm 0.02 $  & 23.17 & 9.35 & 2.72 & (AA) & & 686 (AA)\\
LeG\,33 &11:00:45&$+$14:10:21& 18.98 & 18.29 & 0.061 & 0.042 & -11.83 & 0.667 & $23.78 \pm 0.13 $  & $5.69 \pm 0.65 $  & $1.19 \pm 0.16 $  & 24.70 & 7.64 & 2.58 & (LV) & & \\
LSBCD\,640-08 &11:00:52&$+$13:52:53& 16.19 & 15.64 & 0.053 & 0.037 & -14.47 & 0.530 & $22.25 \pm 0.04 $  & $7.25 \pm 0.31 $  & $0.87 \pm 0.02 $  & 23.70 & 16.2 & 2.91 & (S+) & & \\
CGCG\,066-109 & 11:04:26&$+$11:45:20& 15.69 & 15.31 & 0.050 & 0.034 & -14.77 & 0.356 & $22.26 \pm 0.05 $  & $11.88 \pm 0.43 $  & $1.26 \pm 0.04 $  & 23.00 & 13.7 & 2.83 & & & 777 (AA) \\
\hline\hline
\end{tabular}
\tablefoot{The quantities listed are as follows:
(1) name of candidate;
(2+3) total apparent magnitude in the $g$ and $r$ bands;
(4+5) galactic extinction in the $g$ and $r$ bands \citep{2011ApJ...737..103S};
(6) extinction corrected absolute $r$ band magnitude, using a distance modulus of $M-m=30.06$\,mag;
(7) integrated and extinction corrected $g-r$ color;
(8) S\'ersic central surface brightness in the $r$ band;
(9) S\'ersic scale length in the $r$ band;
(10) S\'ersic curvature index in the $r$ band;
(11) mean effective surface brightness in the $r$ band;
(12) effective radius in the $r$ band; 
(13) the logarithm of the effective radius in the $r$ band, converted to pc with a distance modulus of $M-m=30.06$\,mag;
(14+15+16) reference for original discovery, distance measurement, and velocity measurement: (AA) \citep{2011AJ....142..170H}, (LV) \citep{2004ARep...48..267K,2004AJ....127.2031K,2013AJ....145..101K}, (FS) \citep{1990AJ....100....1F}, (S+) \citep{1997ApJS..111..233S}, (R+) \citep{2005A&A...437..823R},
 (S+) \citep{1992MNRAS.258..334S}, (TT) \citep{2002MNRAS.335..712T}, and (H+) \citep{2003A&A...401..483H}
} %(HIPASS) \citep{2006MNRAS.371.1855W}
\label{table3}
\end{sidewaystable*}

\begin{sidewaystable*}

\caption{Table\,\ref{table3} continued.}
\small
\centering
\setlength{\tabcolsep}{3pt}
\begin{tabular}{lccccrrrrcrccrrlll}
\hline\hline 
Name & && ${g_{tot}}$ & ${r_{tot}}$  & $A_g$ & $A_r$ & $M_{r}$ & $(g-r)_{0,tot}$ & $\mu_{0,r}$ & $r_{0,r}$ & $n_r$ & $\langle\mu\rangle_{eff,r}$ &  $r_{eff,r}$ & $\log r_{eff,r}$ & Ref & $D$ & $v$\\ 
 & && mag & mag & mag & mag &  mag & mag & mag arcsec$^{-2}$ &arcsec & & mag arcsec$^{-2}$ &arcsec & $\log$ pc & & Mpc & km\,s$^{-1}$ \\ 
 (1)& && (2) & (3) &  (4) & (5) &  (6) & (7) &(8) & (9) & (10) & (11) & (12) & (13) & (14) & (15) & (16)\\ 
\hline \\
Leo Triplet & & & & & & & & &\\
AGC\,202256 & 11:14:45&$+$12:38:52&  17.26 & 16.87 & 0.063 & 0.044 & -13.37 & 0.368 & $21.73 \pm 0.12 $  & $4.07 \pm 0.38 $  & $1.08 \pm 0.06 $  & 22.71 & 5.86 & 2.49 & (AA) & 11.0 (LV) & 630 (AA) \\
IC\,2684 & 11:17:01&$+$13:05:57&  15.38 & 14.79 & 0.091 & 0.063 & -15.35 & 0.566 & $20.91 \pm 0.04 $  & $3.75 \pm 0.18 $  & $0.65 \pm 0.01 $  & 23.12 & 18.5 & 2.97 & & & 588 (AA) \\
AGC\,215354 &11:19:16&$+$14:17:24& 17.23 & 16.69 & 0.073 & 0.051 & -13.44 & 0.522 & $21.33 \pm 0.04 $  & $2.92 \pm 0.13 $  & $0.85 \pm 0.02 $  & 22.76 & 6.54 & 2.51 & (AA) & & 790 (LV)\\
DGSAT-1$^*$ & 11:21:37&$+$13:26:50 & 19.63 & 18.92 & 0.085 & 0.059 & -11.21 & 0.683 & $24.81 \pm 0.18 $  & $8.52 \pm 1.23 $  & $1.36 \pm 0.38 $  & 25.27 & 7.43 & 2.57 & (J+) & &\\
AGC\,213436 & 11:22:24&$+$12:58:46&  16.40 & 15.85 & 0.082 & 0.057 & -14.28 & 0.522 & $20.78 \pm 0.03 $  & $2.05 \pm 0.08 $  & $0.63 \pm 0.01 $  & 22.85 & 9.98 & 2.70 & (AA) & & 626 (LV) \\
IC\,2787 & 11:23:19&$+$13:37:47& 15.02 & 14.45 & 0.085 & 0.058 & -15.68 & 0.538 & $20.49 \pm 0.02 $  & $3.11 \pm 0.09 $  & $0.60 \pm 0.01 $  & 22.73 & 18.0 & 2.95 & & & 708 (LV) \\
IC\,2791 & 11:23:38&$+$12:53:46&  16.74 & 16.35 & 0.088 & 0.061 & -13.79 & 0.363 & $21.51 \pm 0.06 $  & $3.73 \pm 0.23 $  & $0.87 \pm 0.03 $  & 22.97 & 8.40 & 2.62 && & 666 (AA)\\
AGC\,215415 &11:24:34&$+$12:40:30& 18.18 & 17.26 & 0.125 & 0.087 & -12.91 & 0.881 & $25.63 \pm 0.08 $  & $21.10 \pm 0.88 $  & $3.59 \pm 1.06 $  & 26.49 & 28.0 & 3.15 & (AA) & & 1002 (AA)\\
KKH\,68 &11:30:53&$+$14:08:46&16.22 & 15.83 & 0.126 & 0.087 & -13.89 & 0.351 & $22.84 \pm 0.07 $  & $11.03 \pm 0.64 $  & $1.07 \pm 0.05 $  & 23.89 & 16.3 & 2.82 & (KKH) &8.50 (LV) & 880 (AA) \\
\\
Field & & & & & & & & &\\
UGC\,05456 &10:07:19&$+$10:21:48& 13.41 & 13.13 & 0.133 & 0.092 & -17.06 & 0.241 & $20.31 \pm 0.04 $  & $10.54 \pm 0.51 $  & $0.88 \pm 0.04 $  & 21.31 & 17.2 & 2.94 & &  10.52 (LV) & 536 (AA)\\
CGCG\,095-078 &10:58:02&$+$19:30:19& 15.34 & 14.99 & 0.084 & 0.058 & -15.40 & 0.322 & $21.03 \pm 0.02 $  & $6.87 \pm 0.12 $  & $1.12 \pm 0.01 $  & 21.95 & 9.80 & 2.74 &  & 11.70 (LV) &652 (LV)\\
KKH\,67 & 11:23:03&$+$21:19:18& 17.49 & 17.34 & 0.077 & 0.053 & -12.70 & 0.130 & $24.75 \pm 0.04 $  & $18.15 \pm 0.48 $  & $1.97 \pm 0.13 $  & 25.08 & 14.1 & 2.83 & (KKH) & & \\
AGC\,213091 &11:29:35&$+$10:48:34&17.68 & 17.19 & 0.119 & 0.083 & -12.46 & 0.447 & $23.10 \pm 0.03 $  & $7.71 \pm 0.18 $  & $1.49 \pm 0.04 $  & 23.75 & 8.15 & 2.51 & (AA) & 8.22 (LV) &743 (LV) \\
KKH\,69 &11:34:53&$+$11:01:07& 16.51 & 16.17 & 0.082 & 0.057 & -13.22 & 0.313 & $23.54 \pm 0.02 $  & $16.55 \pm 0.27 $  & $1.77 \pm 0.05 $  & 24.02 & 14.8 & 2.72 & (KKH) & 7.4 (LV) & 881 (LV)\\
LV\,J1149+1715 &11:49:06&$+$17:15:20& 17.44 & 17.43 & 0.127 & 0.088 & -11.91 & -0.02 & $23.37 \pm 0.10 $  & $7.03 \pm 0.65 $  & $1.13 \pm 0.10 $  & 24.38 & 9.82 & 2.52 & (LV) & 7.11 (LV) & 623 (LV) \\
AGC\,215145 &11:54:12&$+$12:26:04& 17.95 & 17.83 & 0.101 & 0.070 & -11.80 & 0.096 & $24.39 \pm 0.05 $  & $11.36 \pm 0.30 $  & $2.42 \pm 0.20 $  & 24.71 & 9.51 & 2.57 & (HI) & 8.20 (i) & 1004 (AA)\\
\hline\hline
\end{tabular}
\tablefoot{
*: full name: NGC3628-DGSAT-1.
The quantities listed are as follows:
(1) name of candidate;
(2+3) total apparent magnitude in the $g$ and $r$ bands;
(4+5) galactic extinction in the $g$ and $r$ bands \citep{2011ApJ...737..103S};
(6) extinction corrected absolute $r$ band magnitude, using a distance modulus of $M-m=30.06$\,mag;
(7) integrated and extinction corrected $g-r$ color;
(8) S\'ersic central surface brightness in the $r$ band;
(9) S\'ersic scale length in the $r$ band;
(10) S\'ersic curvature index in the $r$ band;
(11) mean effective surface brightness in the $r$ band;
(12) effective radius in the $r$ band; 
(13) the logarithm of the effective radius in the $r$ band, converted to pc with a distance modulus of $M-m=30.06$\,mag;
(14+15+16) reference for original discovery, distance measurement, and velocity measurement: (J+) \citep{2016A&A...588A..89J}, (AA) \citep{2011AJ....142..170H}, (LV) \citep{2004ARep...48..267K,2004AJ....127.2031K,2013AJ....145..101K}, (KKH) \citep{2001A&A...366..428K}, and (HI) \citep{2006MNRAS.371.1855W}.
}
\label{table4}
\end{sidewaystable*}

\bibliographystyle{aa}
\bibliography{bibliographie}

%\section*{Appendix}

\end{document}